\documentclass[12pt]{article}
\usepackage{amsmath,amssymb,amsthm,amsxtra,overpic,bbm,bm,epsfig,subfigure}
\usepackage{hyperref}
\usepackage{mathrsfs}
\usepackage{enumitem}
\usepackage{graphicx}
\usepackage{color}
\usepackage[table]{xcolor}
\usepackage{comment}
\usepackage{epstopdf}
\usepackage{float}
\usepackage{cite}
\usepackage{slashed,stmaryrd}
\usepackage{youngtab}
\usepackage{makecell}

\usepackage[title]{appendix}

\allowdisplaybreaks[1]

\def\thefootnote{\fnsymbol{footnote}}

\usepackage{multirow}

\newcommand{\Op}{\mathcal{O}}
\newcommand{\Opr}{\mathcal{R}}
\newcommand{\loopf}{\frac{1}{16\pi^2 \varepsilon}}
\newcommand{\tr}[1]{{\rm Tr}\left(#1\right)}

\newcommand{\rmi}{{\rm i}}

\newcommand{\Dlr}{\mbox{$\raisebox{2mm}{\boldmath ${}^\leftrightarrow$}\hspace{-4mm}D^{}_\mu$}}
\newcommand{\Dilr}{\mbox{$\raisebox{2mm}{\boldmath ${}^\leftrightarrow$}\hspace{-4mm}D^I_\mu$}}
\newcommand{\Dilrs}{\mbox{$\raisebox{2mm}{\boldmath ${}^\leftrightarrow$}\hspace{-4mm}\slashed{D}^I$}}
\newcommand{\Dlrs}{\mbox{$\raisebox{2mm}{\boldmath ${}^\leftrightarrow$}\hspace{-4mm}\slashed{D}$}}
\newcommand{\Dl}{\mbox{$\raisebox{2mm}{\boldmath ${}^\leftarrow$}\hspace{-4mm}D^{}_\mu$}}

\newcommand{\Dlrn}{\mbox{$\raisebox{2mm}{\boldmath ${}^\leftrightarrow$}\hspace{-4mm}D^\nu$}}

\newcommand{\befune}[1]{ \mu \frac{{\rm d}#1}{{\rm d}\mu}}
\newcommand{\befun}{ \dot{C}}

\usepackage{ulem}

\begin{document}
	
	{\normalsize \flushright TUM-HEP 1460/23\\}
	\vspace{0.2cm}
	
%

\begin{center}
{\Large\bf Renormalization Group Equations for the SMEFT Operators up to Dimension Seven}
\end{center}

\vspace{0.2cm}

\begin{center}
{\bf  Di Zhang}~\footnote{E-mail: di1.zhang@tum.de} 
\\
\vspace{0.2cm}
{\small
Physik-Department, Technische Universität München, James-Franck-Straße, 85748 Garching, Germany}
\end{center}

\vspace{1.5cm}

\begin{abstract}
	In this paper, we propose a Green's basis and also a new physical basis for dimension-seven (dim-7) operators, which are suitable for the matching of ultraviolet models onto the Standard Model effective field theory (SMEFT) and the deviation of renormalization group equations (RGEs) for dim-7 operators in the SMEFT. The reduction relations to convert operators in the Green's basis to those in the physical basis are achieved as well, where some redundant dim-6 operators in the Green's basis are involved if the dim-5 operator exists. Working in these two bases for dim-7 operators and with the help of the reduction relations, we work out the one-loop RGEs resulting from the mixing among different dimensional operators for the dim-5 and dim-7 operators up to $\mathcal{O} \left( \Lambda^{-3} \right)$ in the SMEFT. These new results complete the previous results for RGEs of the dim-5 and dim-7 operators and hence can be used for a consistent one-loop analysis of the SMEFT at $\mathcal{O} \left( \Lambda^{-3} \right)$.
\end{abstract}


\def\thefootnote{\arabic{footnote}}
\setcounter{footnote}{0}
\newpage

\section{Introduction}

The Standard Model (SM) of particle physics is our best understanding of strong, weak and electromagnetic interactions among fundamental particles, and has successfully passed the plethora of precision tests, especially in the electroweak sector~\cite{ParticleDataGroup:2022pth}. However, its extraordinary power is irremediably gone when it comes to neutrinos and some compelling cosmological observations~\cite{ParticleDataGroup:2022pth,Xing:2020ijf}. The non-zero neutrino masses and the existence of dark matter  cannot be accommodated in the SM. Meanwhile, the SM can  not provide a successful explanation for the matter-antimatter asymmetry of the Universe. Thus, the SM is believed to be incomplete and serves as an effective field theory (EFT) at low-energy scales. Due to the unknown dynamics of new physics, the SM effective field theory (SMEFT)~\cite{Buchmuller:1985jz,Grzadkowski:2010es} (see, e.g., Refs.~\cite{Brivio:2017vri,Isidori:2023pyp} for the latest review) valid below a cut-off scale $\Lambda$ is extensively used to discuss indirect consequences of new physics in a model-independent way. The SMEFT is composed of the SM Lagrangian and higher-mass-dimensional operators constituted by the SM fields and preserving the SM gauge symmetry, in which new physics effects are entirely embedded in the Wilson coefficients of those non-renormalizable operators. On the one hand, one can map the Wilson coefficients of those higher-dimensional operators in the SMEFT onto low-energy observables to search for new physics indirectly or constrain the size of relevant Wilson coefficients. On the other hand, all ultraviolet (UV) models extending the SM by introducing heavy degrees of freedom can be matched onto the SMEFT at the cut-off scale $\Lambda$, and then connected to low-energy observables with the help of the renormalization group equations (RGEs) and the mapping of the SMEFT~\cite{Henning:2014wua}. Apart from matching between UV models and the SMEFT, RG running and mapping are intrinsic to the SMEFT and hence only need to be done once even for different UV models. Moreover, as an EFT should be, the SMEFT can improve the perturbative convergence involved in loop calculations for multi-scale UV models by means of the matching and running~\cite{Weinberg:1980wa}. More discussions about the advantages and usages of the SMEFT can be found in Ref.~\cite{Henning:2014wua}.


An operator basis is basic to working within the SMEFT. There are two different types of bases for operators, i.e., the physical basis and the Green's basis~\cite{Jiang:2018pbd}. The former is related to $S$-matrices, in which all operators are independent under integral by part, Fierz transformations, algebraic relations and field redefinitions [equivalent to field's equation of motion (EoM) at linear order in the perturbation~\cite{Criado:2018sdb}]. The latter is directly associated with one-particle-irreducible (1PI) off-shell Green's functions, and operators in this basis are independent under integral by part, Fierz transformations and algebraic relations but redundant under field redefinitions. Thus, they can be converted to operators in the physical basis by exploiting field redefinitions. The Green's basis is quite useful to deal with the matching of UV models onto the SMEFT and to derive RGEs of the SMEFT, especially with the Feynman diagrammatic approach. For dimension-five (dim-5) operators, the physical and Green's bases are the same and consist of the unique Weinberg operator~\cite{Weinberg:1979sa}, whereas for higher-dimensional operators, the physical basis is usually a proper subset of the Green's basis. The number of operators in those bases increases rapidly with the increase of mass dimension~\cite{Henning:2015alf,Criado:2019ugp,Fonseca:2019yya}. So far, the physical bases for operators up to dimension twelve have been constructed in a flood of literature~\cite{Buchmuller:1985jz,Grzadkowski:2010es,Ma:2019gtx,Aoude:2019tzn,Lehman:2014jma,Liao:2016hru,Liao:2019tep,Murphy:2020rsh,Li:2020gnx,Durieux:2019eor,AccettulliHuber:2021uoa,Liao:2020jmn,Li:2020xlh,Harlander:2023psl}, and the Green's bases for dim-6 and dim-8 operators have been presented in Refs.~\cite{Gherardi:2020det,Chala:2021cgt,Ren:2022tvi}. Moreover, the reduction relations to covert all dim-6 operators and dim-8 bosonic operators in the Green's basis to those in the physical basis can be found  in Refs.~\cite{Gherardi:2020det,Chala:2021cgt}. On the other hand, as a significant bridge to connect parameters at high- and low-energy scales, the RGEs in the SMEFT have been widely studied and derived for the dim-5~\cite{Babu:1993qv,Chankowski:1993tx,Antusch:2001ck}, dim-6~\cite{Jenkins:2013zja,Jenkins:2013wua,Alonso:2013hga,Alonso:2014zka,Davidson:2018zuo,Wang:2023bdw}, dim-7~\cite{Liao:2016hru,Liao:2019tep,Chala:2021juk} and dim-8~\cite{AccettulliHuber:2021uoa,Chala:2021pll,DasBakshi:2022mwk,Helset:2022pde,DasBakshi:2023htx,Chala:2023jyx} operators. However, the RGEs for dim-7 and dim-8 operators are not complete~\cite{DasBakshi:2023htx}. Contributions from dim-5 and dim-6 operators to the RGEs of dim-7 operators have not been fully achieved so far~\footnote{Such contributions to the dim-7 Weinberg-like operator $\Op^{}_{\ell H}$ have been obtained with some approximations~\cite{Chala:2021juk}. Meanwhile, contributions from dim-6 and dim-7 operators to the RGEs of the dim-5 operator have also been partially achieved with the same approximation.}, and the RGEs for dim-8 fermionic operators are still lacking. One may refer to Table~3 in Ref.~\cite{DasBakshi:2023htx} for a summary of the status of the SMEFT RGEs but note that contributions from the dim-5 operator to those of dim-6 operators have been slightly revised in Ref.~\cite{Wang:2023bdw}.

In this work, we mainly focus on the bases and RGEs for dim-7 operators, but the RGE for the dim-5 operator up to $\mathcal{O} \left(\Lambda^{-3} \right)$ is also concerned. A physical basis for dim-7 operators has been put forward in Ref.~\cite{Liao:2019tep}, but to get rid of redundant degrees of freedom induced by some flavor relations, the flavor indices of operators in this basis are picked with restrictions and hence can not run over all flavors. Therefore, this basis is awfully inconvenient to match UV models onto the SMEFT and to derive RGEs of dim-7 operators. To avoid such an inconvenience, we propose a new physical basis for dim-7 operators by means of tensor decompositions of ${\rm SU(}n{\rm )}$ group with $n$ being the number of fermion generation. In this new basis, there is no limitation on flavor indices of operators, thus they can run over all flavors and the redundant degrees of freedom are automatically removed. Furthermore, we present a Green's basis for dim-7 operators and derive the reduction relations between the Green's and physical bases for the first time. It should be emphasized that some redundant dim-6 operators in the Green's basis are involved in those reduction relations for dim-7 operators due to the existence of the dim-5 operator. These two bases together with the general reduction relations for dim-7 operators are applicable both to matching and to deriving RGEs and can be incorporated into the package {\sf Matchmakereft} aiming at automatically dealing with these two issues in EFTs~\cite{Carmona:2021xtq}. Then, working in these two bases and with the help of the reduction relations, we calculate all RGEs of the dim-5 and dim-7 operators resulting from the mixing among operators of different dimensions up to $\mathcal{O}\left( \Lambda^{-3} \right)$ with the off-shell scheme. These results complete the RGEs for the dim-5 and dim-7 operators in the previous works~\cite{Babu:1993qv,Chankowski:1993tx,Antusch:2001ck,Liao:2016hru,Liao:2019tep,Chala:2021juk} and are important for a consistent one-loop analysis of the SMEFT.

The rest of this paper is organized as follows. In Sec.~\ref{sec:framework}, we construct the Green's and physical bases for dim-7 operators and establish the reduction relations to convert operators in the Green's basis to those in the physical basis. In Sec.~\ref{sec:derivation}, all counterterms for the dim-5 and dim-7 operators are calculated first in the Green's basis and then reduced into those in the physical basis, from which the RGEs of the dim-5 and dim-7 operators are derived. The main conclusions are summarized in
Sec.~\ref{sec:conclusions}. Some details are given in a series of appendices. The general reduction relations for dim-7 operators are presented in Appendix~\ref{app:conversion}. All counterterms for the dim-5 and dim-7 operators in the Green's basis and as well as those for relevant dim-6 operators are listed in Appendix~\ref{app:counterterm-green}. Finally, all counterterms in the physical basis are collected in Appendix~\ref{app:counterterm-phys}.

\section{The SMEFT up to $\mathcal{O} \left( \Lambda^{-3} \right)$}\label{sec:framework}

The Lagrangian of the SMEFT up to $\mathcal{O} \left( \Lambda^{-3} \right)$ is given by
\begin{eqnarray}\label{eq:LSMEFT}
	\mathcal{L}^{}_{\rm SMEFT} = \mathcal{L}^{}_{\rm SM} + \frac{1}{2} \left( C^{\alpha\beta}_5 \Op^{(5)}_{\alpha\beta} + {\rm h.c.} \right) + \sum_i C^i_6 \Op^{(6)}_i + \sum_j C^j_7 \Op^{(7)}_j \;,
\end{eqnarray}
where $\Op^{(6)}_{i}$ and $\Op^{(7)}_{j}$ are independent dim-6 and dim-7 operators in the physical basis, and $C^i_6$ and $C^j_7$ are the corresponding Wilson coefficients and suppressed by $\Lambda^{-2}$ and $\Lambda^{-3}$, respectively. Here $i$ and $j$ run over all independent operators including hermitian conjugates of non-hermitian operators in the physical basis. The unique dim-5 operator $\Op^{(5)}$, also called Weinberg operator~\cite{Weinberg:1979sa}, is given by $\Op^{(5)}_{\alpha\beta} = \overline{\ell^{}_{\alpha\rm L}} \widetilde{H} \widetilde{H}^{\rm T} \ell^{\rm c}_{\beta \rm L}$ with $\ell^{\rm c}_{\rm L} \equiv {\sf C} \overline{\ell^{}_{\rm L}}^{\rm T}$ and ${\sf C}$ being the charge-conjugate matrix, and $C^{\alpha\beta}_5$ is the corresponding Wilson coefficient. All dim-6 and dim-7 operators in the physical basis can be found in Refs.~\cite{Buchmuller:1985jz,Grzadkowski:2010es} and Refs.~\cite{Lehman:2014jma,Liao:2016hru,Liao:2019tep}, respectively. Throughout this work, we adopt the Warsaw basis~\cite{Grzadkowski:2010es} for dim-6 operators, and a modified basis based on that in Ref. \cite{Liao:2019tep} for dim-7 operators. The SM Lagrangian is
\begin{eqnarray}\label{eq:LSM}
	\mathcal{L}^{}_{\rm SM} &=& - \frac{1}{4} G^{A}_{\mu\nu} G^{A\mu\nu} - \frac{1}{4} W^I_{\mu\nu} W^{I\mu\nu} - \frac{1}{4} B^{}_{\mu\nu} B^{\mu\nu} + \left( D^{}_\mu H \right)^\dagger \left( D^\mu H \right) - m^2 H^\dagger H - \lambda \left( H^\dagger H \right)^2 
	\nonumber
	\\
	&& + \sum^{}_f \overline{f} \rmi \slashed{D} f - \left[ \overline{Q^{}_{\alpha \rm L}} \left(Y^{}_{\rm u} \right)^{}_{\alpha \beta} \widetilde{H} U^{}_{\beta \rm R} +  \overline{Q^{}_{\alpha \rm L}}  (Y^{}_{\rm d})^{}_{\alpha \beta} H D^{}_{\beta \rm R}+ \overline{\ell^{}_{\alpha \rm L}}(Y^{}_l)^{}_{\alpha \beta} H E^{}_{\beta \rm R}+ {\rm h.c.} \right]\;,
\end{eqnarray}
in which $f= Q^{}_{\rm L}, U^{}_{\rm R}, D^{}_{\rm R}, \ell^{}_{\rm L}, E^{}_{\rm R}$, the gauge-fixing and Faddeev-Popov ghost terms stemming from the standard procedure of quantization are suppressed, and the covariant derivative $D^{}_\mu \equiv \partial^{}_\mu - \rmi g^{}_1 Y B^{}_\mu - \rmi g^{}_2 T^I W^I_\mu - \rmi g^{}_s T^A G^A_\mu$ is defined as usual. 

In the SMEFT, field redefinitions are often applied for getting rid of redundant operators or converting them to non-redundant ones, such as during the construction of a physical operator basis and the one-loop matching of UV models onto the SMEFT. However, it is quite complicated to directly conduct field redefinitions. Fortunately, imposing EoMs is equivalent to applying field redefinitions at linear order in the perturbation~\cite{Criado:2018sdb} but much easier, and in most cases, EoMs are enough for the purpose. In this work, the EoMs including contributions from the dim-5 operator are sufficient to convert redundant dim-6 and dim-7 operators to non-redundant dim-7 operators, i.e., up to $\mathcal{O} \left( \Lambda^{-3} \right)$, because the differences between imposing EoMs and field redefinitions to reduce redundant dim-6 and dim-7 operators only lead to higher dimensional (at least dim-8) operators (see, e.g., Ref.~\cite{Criado:2018sdb} for more details). Therefore, we will make use of the SMEFT EoMs instead of field redefinitions. Starting with Eqs.~\eqref{eq:LSMEFT} and \eqref{eq:LSM}, one can easily obtain the EoMs of all SM fields up to $\mathcal{O} \left( \Lambda^{-1} \right)$, namely (see also Ref.~\cite{Barzinji:2018xvu})
\begin{eqnarray}\label{eq:eom}
    \rmi \slashed{D} \ell^a_{\alpha \rm L} &=&  \left( Y^{}_l \right)^{}_{\alpha\beta} H^a E^{}_{\beta\rm R} - C^{\alpha\beta}_5 \widetilde{H}^a \widetilde{H}^{\rm T} \ell^{\rm c}_{\beta \rm L} \;,
    \nonumber
    \\
    \rmi \slashed{D} E^{}_{\alpha \rm R} &=& \left( Y^\dagger_l \right)^{}_{\alpha\beta} H^\dagger \ell^{}_{\beta\rm L} \;,
    \nonumber
    \\
    \rmi \slashed{D} Q^{a}_{\alpha\rm L} &=& \left( Y^{}_{\rm d} \right)^{}_{\alpha\beta} H^a D^{}_{\beta\rm R} + \left( Y^{}_{\rm u} \right)^{}_{\alpha\beta} \widetilde{H}^a U^{}_{\beta\rm R} \;,
    \nonumber
    \\
    \rmi \slashed{D} D^{}_{\alpha\rm R} &=& \left( Y^\dagger_{\rm d} \right)^{}_{\alpha\beta} H^\dagger Q^{}_{\beta \rm L} \;,
    \nonumber
    \\
    \rmi \slashed{D} U^{}_{\alpha\rm R} &=& \left( Y^\dagger_{\rm u} \right)^{}_{\alpha\beta} \widetilde{H}^\dagger Q^{}_{\beta\rm L} \;,
    \nonumber
    \\
    D^\nu B^{}_{\mu\nu} &=& \frac{1}{2} g^{}_1 \left( H^\dagger \rmi \Dlr H + 2 \sum^{}_f Y (f) \overline{f} \gamma^{}_\mu f \right) \;,
    \nonumber
    \\
    \left( D^\nu W^{}_{\mu\nu} \right)^I &=& \frac{1}{2} g^{}_2 \left( H^\dagger \rmi \Dilr H + \overline{Q^{}_{\rm L}} \sigma^I \gamma^{}_\mu Q^{}_{\rm L} + \overline{\ell^{}_{\rm L}} \sigma^I \gamma^{}_\mu \ell^{}_{\rm L} \right) \;,
    \nonumber
    \\
    \left( D^2 H \right)^a &=& - m^2 H^a - 2\lambda H^a \left( H^\dagger H \right) - \left( \overline{E^{}_{\rm R}} Y^\dagger_l \ell^a_{\rm L} + \overline{D^{}_{\rm R}} Y^\dagger_{\rm d} Q^a_{\rm L} - \epsilon^{ab} \overline{Q^b_{\rm L}} Y^{}_{\rm u} U^{}_{\rm R} \right) 
    \nonumber
    \\
    && - \frac{1}{2} \epsilon^{ab} C^{\alpha\beta}_5 \left( \overline{\ell^b_{\alpha\rm L}} \widetilde{H}^{\rm T} \ell^{\rm c}_{\beta\rm L}  + \overline{\ell^{}_{\alpha\rm L}} \widetilde{H} \ell^{b \rm c}_{\beta \rm L} \right) \;,
\end{eqnarray}
where $\Dlr \equiv D^{}_\mu - \Dl$ and $\Dilr \equiv \sigma^I D^{}_\mu - \Dl \sigma^I$ with $\Dl$ acting on the left, $Y(f)$ is  the hypercharge for the fermionic fields $f= Q^{}_{\rm L}, U^{}_{\rm R}, D^{}_{\rm R}, \ell^{}_{\rm L}, E^{}_{\rm R}$, $\sigma^I$ with $I=1,2,3$ are the Pauli matrices, and $a,b=1,2$ are the weak isospin indices. As shown in Eq.~\eqref{eq:eom}, only the EoMs of lepton and Higgs doublets contain contributions from the dim-5 operator. This indicates that redundant dim-6 operators removed by imposing the EoMs of lepton and Higgs doublets may also be reduced to non-redundant dim-7 operators, besides non-redundant dim-6 operators.

\subsection{The Green's basis for dim-7 operators}

A Green's basis is  very convenient and helpful to perform one-loop matching and calculate RGEs in EFTs with the off-shell scheme and can make calculations clearer and more intuitive. On the one hand, as a minimal but complete pre-set basis, the Green's basis provides specific gauge invariant forms of all possible amplitudes directly from 1PI diagrams beforehand. Then the corresponding Wilson coefficients or counterterms can be easily determined by matching for one-loop matching or requiring them to cancel out all divergences for calculations of RGEs. During this procedure, we do not need to perform the nontrivial restoration of amplitudes' gauge structures, like those in Ref.~\cite{Bilenky:1993bt}, and less 1PI diagrams are involved. On the other hand, it provides us with simple criteria to choose a set of 1PI diagrams enough to gain all matching conditions or counterterms, i.e., those whose external lines run over the field ingredients (excluding gauge fields in the covariant derivative) of each class of operators in the Green's basis. Generally, this set of 1PI diagrams is not the optimal one but is very easy to be determined. Therefore, it offers an intuitive way to the final results, where each step is general and clear, i.e., generating 1PI diagrams, calculating in the Green's basis, and converting results to those in the physical basis. This is mainly how {\sf Matchmakereft} package~\cite{Carmona:2021xtq} is programmed. We will adopt this strategy to calculate all counterterms and derive the concerned RGEs.

\begin{table}
	\centering
	\resizebox{\textwidth}{!}{
		\begin{tabular}{l|c|c|l|c|c}
			\hline\hline
			\multicolumn{3}{c|}{$\psi^2 H^4$} & \multicolumn{3}{c}{$\psi^2 H^3 D$} \\
			\hline
			\cellcolor{gray!30}{$\Op^{}_{\ell H}$} & \cellcolor{gray!30}{$\epsilon^{ab} \epsilon^{de}  \left( \ell^a_{\rm L} C \ell^d_{\rm L} \right) H^b H^e \left( H^\dagger H \right)$} & \cellcolor{gray!30}{$\frac{1}{2} n(n+1)$} & $\Op^{}_{\ell e H D}$ & $\epsilon^{ab} \epsilon^{de} \left( \ell^a_{\rm L} C \gamma^{}_\mu E^{}_{\rm R} \right) H^b H^d {\rm i} D^\mu H^e$ & $n^2$ \\
			\hline
			\multicolumn{3}{c|}{$\psi^2 H^2 D^2$} & \multicolumn{3}{c}{$\psi^4 H$} \\
			\hline
			$\Op^{}_{\ell H D1}$ & $\epsilon^{ab} \epsilon^{de} \left( \ell^a_{\rm L} C D^\mu \ell^b_{\rm L} \right) H^d D^{}_\mu H^e$ & $n^2$ & \cellcolor{gray!30}{$\Op^{}_{\overline{e} \ell\ell\ell H}$} & \cellcolor{gray!30}{$\epsilon^{ab} \epsilon^{de} \left( \overline{E^{}_{\rm R}} \ell^a_{\rm L}\right) \left( \ell^b_{\rm L} C \ell^d_{\rm L} \right) H^e$} & \cellcolor{gray!30}{$\frac{1}{3} n^2 \left( 2n^2 +1 \right)$} \\
			$\Op^{}_{\ell HD2}$ & $\epsilon^{ad} \epsilon^{be} \left( \ell^a_{\rm L} C D^\mu \ell^b_{\rm L} \right) H^d D^{}_\mu H^e$ & $n^2$ & $\Op^{}_{\overline{d} \ell q \ell H 1}$ & $\epsilon^{ab} \epsilon^{de} \left( \overline{D^{}_{\rm R}} \ell^a_{\rm L} \right) \left( Q^b_{\rm L} C \ell^d_{\rm L} \right) H^e $ & $n^4$ \\
			\cellcolor{gray!30}{$\Opr^{}_{\ell HD3}$} & \cellcolor{gray!30}{$\epsilon^{ad} \epsilon^{be} \left( \ell^a_{\rm L} C  \ell^b_{\rm L} \right) D^\mu H^d D^{}_\mu H^e$} & \cellcolor{gray!30}{$\frac{1}{2} n(n+1)$} & $\Op^{}_{\overline{d} \ell q \ell H 2}$ & $\epsilon^{ad} \epsilon^{be} \left( \overline{D^{}_{\rm R}} \ell^a_{\rm L} \right) \left( Q^b_{\rm L} C \ell^d_{\rm L} \right) H^e$ & $n^4$ \\
			\cellcolor{gray!30}{$\Opr^{}_{\ell HD4}$} & \cellcolor{gray!30}{$\epsilon^{ad} \epsilon^{be} \left( D^\mu \ell^a_{\rm L} C D^{}_\mu \ell^b_{\rm L} \right) H^d H^e$} & \cellcolor{gray!30}{$\frac{1}{2} n(n+1) $} & $\Op^{}_{\overline{d} \ell u e H}$ & $\epsilon^{ab} \left( \overline{D^{}_{\rm R}} \ell^a_{\rm L} \right) \left( U^{}_{\rm R} C E^{}_{\rm R} \right) H^b$ & $n^4$ \\
			$\Opr^{}_{\ell HD5}$ & $\epsilon^{ab} \epsilon^{de} \left( \ell^a_{\rm L} C \sigma^{}_{\mu\nu} D^\mu \ell^b_{\rm L} \right) H^d D^\nu H^e $ & $n^2$ & $\Op^{}_{\overline{q} u \ell \ell H}$ & $\epsilon^{ab} \left( \overline{Q^{}_{\rm L}} U^{}_{\rm R} \right) \left( \ell^{}_{\rm L} C \ell^a_{\rm L} \right) H^b$ & $n^4$ \\
			\cellcolor{gray!30}{$\Opr^{}_{\ell HD6}$} & \cellcolor{gray!30}{$\epsilon^{ad} \epsilon^{be} \left( D^\mu \ell^a_{\rm L} C \sigma^{}_{\mu\nu} D^\nu \ell^b_{\rm L} \right) H^d H^e$} & \cellcolor{gray!30}{$\frac{1}{2} n(n+1)$} & $\Op^{}_{\overline{\ell} dud \widetilde{H}}$ & $\left( \overline{\ell^{}_{\rm L}} D^{}_{\rm R} \right) \left( U^{}_{\rm R} C D^{}_{\rm R} \right) \widetilde{H} $ & $ n^4 $ \\
			\cline{1-3}
			\multicolumn{3}{c|}{$\psi^2 H^2 X$} & \cellcolor{gray!30}{$\Op^{}_{\overline{\ell}dddH}$} & \cellcolor{gray!30}{$\left( \overline{\ell^{}_{\rm L}} D^{}_{\rm R} \right) \left( D^{}_{\rm R} C D^{}_{\rm R} \right) H$} & \cellcolor{gray!30}{$\frac{1}{3} n^2 \left( n^2 - 1 \right)$} \\
			\cline{1-3}
			\cellcolor{gray!30}{$\Op^{}_{\ell HB}$} & \cellcolor{gray!30}{$\epsilon^{ab} \epsilon^{de} \left( \ell^a_{\rm L} C \sigma^{}_{\mu\nu} \ell^d_{\rm L} \right) H^b H^e B^{\mu\nu}$} & \cellcolor{gray!30}{$\frac{1}{2} n(n-1)$} & \cellcolor{gray!30}{$\Op^{}_{\overline{e}qdd\widetilde{H}}$} & \cellcolor{gray!30}{$\epsilon^{ab} \left( \overline{ E^{}_{\rm R}} Q^a_{\rm L} \right) \left( D^{}_{\rm R} C D^{}_{\rm R} \right) \widetilde{H}^b$} & \cellcolor{gray!30}{$\frac{1}{2} n^3 (n-1)$} \\
			$\Op^{}_{\ell HW}$ & $\epsilon^{ab} \left( \epsilon \sigma^I \right)^{de} \left( \ell^a_{\rm L} C \sigma^{}_{\mu\nu} \ell^d_{\rm L} \right) H^b H^e W^{I \mu\nu}$ & $n^2$ & $\Op^{}_{\overline{\ell} d qq \widetilde{H}}$ & $\epsilon^{ab} \left( \overline{\ell^{}_{\rm L}} D^{}_{\rm R} \right) \left( Q^{}_{\rm L} C Q^a_{\rm L} \right) \widetilde{H}^b$ & $ n^4 $ \\[0.1cm]
			\hline
			\multicolumn{6}{c}{$\psi^4 D$} \\
			\hline
			$\Op^{}_{\overline{e}dddD}$ & $\left(\overline{E^{}_{\rm R}} \gamma^{}_\mu D^{}_{\rm R} \right) \left( D^{}_{\rm R} C {\rm i} D^\mu D^{}_{\rm R} \right) $ & $n^4$ & $\Op^{}_{\overline{\ell}qddD}$ & $\left( \overline{\ell^{}_{\rm L}} \gamma^{}_\mu Q^{}_{\rm L} \right) \left( D^{}_{\rm R} C {\rm i} D^\mu D^{}_{\rm  R} \right)$ & $n^4$  \\
			$\Op^{}_{\overline{d}u \ell \ell D}$ & $\epsilon^{ab} \left( \overline{D^{}_{\rm R}} \gamma^{}_\mu U^{}_{\rm R} \right) \left( \ell^a_{\rm L} C {\rm i} D^\mu \ell^b_{\rm L} \right)$ & $n^4$ & $\Opr^{}_{\overline{\ell}dDqd}$ & $\left( \overline{\ell^{}_{\rm L}} D^{}_{\rm R} \right) \left( {\rm i}D^\mu Q^{}_{\rm L} C \gamma^{}_\mu D^{}_{\rm R} \right)$ & $n^4$ \\
			$\Opr^{}_{\overline{d}\ell\ell D u}$ & $\epsilon^{ab} \left( \overline{D^{}_{\rm R}} \ell^a_{\rm L} \right) \left( \ell^b_{\rm L }  C \gamma^{}_\mu {\rm i}D^\mu U^{}_{\rm R } \right) $ & $n^4$ & $\Opr^{}_{\overline{\ell} d q D d}$ & $\left( \overline{\ell^{}_{\rm L}} D^{}_{\rm R} \right) \left( Q^{}_{\rm L} C \gamma^{}_\mu {\rm i}D^\mu D^{}_{\rm R} \right)$ & $n^4$ \\
			$\Opr^{}_{\overline{d} D \ell\ell u}$ & $\epsilon^{ab} \left( \overline{D^{}_{\rm R}} {\rm i}D^\mu \ell^a_{\rm L} \right) \left( \ell^b_{\rm L} C \gamma^{}_\mu U^{}_{\rm R} \right)$ & $n^4$ & & & \\[0.1cm]
			\hline
			\hline
		\end{tabular}
	}
	\caption{Dimension-7 operators in the Green's basis, where operators in the grey cells should be replaced with those in Eqs.~\eqref{eq:decomposing-1}, \eqref{eq:decomposing-2} and \eqref{eq:decomposing-3}. The expressions in the third and sixth columns are the corresponding number of independent operators in this Green's basis with $n$ being the number of fermion generation.}
	\label{tab:green-basis}
\end{table}

The Green's bases for dim-6 and dim-8 operators have been put forward in Refs.~\cite{Jiang:2018pbd,Gherardi:2020det} (also see Ref.~\cite{Carmona:2021xtq}, where dim-6 evanescent operators are included as well) and in Refs.~\cite{Chala:2021cgt,Ren:2022tvi}, respectively, but that for dim-7 operators is still lacking. Following the procedure and analysis in Ref.~\cite{Lehman:2014jma}, we work out all dim-7 operators in the Green's basis for the first time, which are listed in Table~\ref{tab:green-basis} but with the operators in the grey cells replaced with those in Eqs.~\eqref{eq:decomposing-1}, \eqref{eq:decomposing-2} and \eqref{eq:decomposing-3}. More explanation is given in what follows. We adopt most notations in Ref.~\cite{Liao:2016hru} for dim-7 operators, where fermion fields in an operator and their flavors are identically ordered, the weak isospin indices of two $\rm SU(2)^{}_L$ doublets are contracted if they are not explicitly given, the color indices in the operators containing two quark fields are contracted directly, while for those consisting of three quark fields, the color indices are in the same order as the quark fields and contracted with the 3-rank antisymmetric tensor, and $\psi^{}_1 C\psi^{}_2 \equiv \overline{\psi^{\rm c}_1} \psi^{}_2$ is used to avoid too many indices on the fields. For instance, the explicit forms for $\Op^{}_{\overline{q}u\ell\ell H}$ and $\Op^{}_{\overline{\ell}qddD}$ in Table~\ref{tab:green-basis} are given by
\begin{eqnarray}
	\Op^{\alpha\beta\gamma\lambda}_{\overline{q}u\ell\ell H} &=& \epsilon^{ab} \left( \overline{Q^{id}_{\alpha\rm L}} U^{i}_{\beta\rm R}\right) \left( \overline{\ell^{d\rm c}_{\gamma\rm L}} \ell^{a}_{\lambda\rm L} \right) H^b \;,
	\nonumber
	\\
	\Op^{\alpha\beta\gamma\lambda}_{\overline{\ell}qddD} &=& \epsilon^{ijk} \left( \overline{\ell^a_{\alpha\rm L}} \gamma^{}_\mu Q^{ia}_{\beta\rm L} \right) \left( \overline{D^{j \rm c}_{\gamma\rm R}} \rmi D^\mu D^{k}_{\lambda\rm R} \right) \;,
\end{eqnarray}
in which $\{i,j,k\}$, $\{a,b,d,e\}$, $\{\alpha,\beta,\gamma,\lambda,\dots\}$ are color, weak isospin and flavor indices, respectively. In Table~\ref{tab:green-basis}, $\Opr^{}_{\dots}$ denote operators that can be totally converted to those in the physical basis via field redefinitions or equivalently EoMs but are independent in the Green's basis. There are eight such operators (barring flavor structures and hermitian conjugates) in the Green's basis for dim-7 operators. In the third and sixth columns of Table~\ref{tab:green-basis}, the number of the corresponding operators is also given with $n$ being the number of fermion generation, where their hermitian conjugates are not included. The total number of the operators with the same field ingredients is consistent with that given by two independent-operator-counting packages, i.e., {\sf Basisgen}~\cite{Criado:2019ugp} and {\sf Sym2Int}~\cite{Fonseca:2019yya}, with the off-shell option. If there are no symmetries in the flavor structure of an operator, the number of the operator is naively governed by $n^{\#}$ with $\#$ being the number of fermion fields in the operator. However, it is not such a naive case for all operators as shown in Table~\ref{tab:green-basis}, and there are some symmetries in the flavor structures of the operators in gray cells, i.e.,~\footnote{For instance, $\mathcal{R}^{\beta\alpha}_{\ell HD3} = \epsilon^{ad}\epsilon^{be} \left( \ell^a_{\beta\rm L} C \ell^b_{\alpha\rm L} \right) D^\mu H^d D^{}_\mu H^e = \epsilon^{ad}\epsilon^{be} \left( \ell^b_{\alpha\rm L} C \ell^a_{\beta\rm L} \right) D^\mu H^d D^{}_\mu H^e = \epsilon^{ad} \epsilon^{be} \left( \ell^a_{\alpha\rm L} C \ell^b_{\beta\rm L} \right) $ $\times D^\mu H^d D^{}_\mu H^e = \mathcal{R}^{\alpha\beta}_{\ell HD3} $ holds, where $\ell^a_{\beta\rm L} C \ell^b_{\alpha\rm L} = \ell^b_{\alpha\rm L} C \ell^a_{\beta\rm L}$ has been used in the first equality, and in the third equality, we have exchanged $\rm SU(2)$ indices, i.e., $a \leftrightarrow b$ and $ d \leftrightarrow e$. }
\begin{eqnarray}\label{eq:symmetry}
	&&\Op^{\alpha\beta}_{\ell H} - \Op^{\beta\alpha}_{\ell H} = 0 \;,
	\nonumber
	\\
	&& \Op^{\alpha\beta}_{\ell HB} + \Op^{\beta\alpha}_{\ell HB} = 0 \;,
	\nonumber
	\\
	&& \Op^{\alpha\beta\gamma\lambda}_{\overline{e}qdd\widetilde{H}} + \Op^{\alpha\beta\lambda\gamma}_{\overline{e}qdd\widetilde{H}} =0 \;,
	\nonumber
	\\
	&&\Op^{\alpha\beta\gamma\lambda}_{\overline{e}\ell\ell\ell H} + \Op^{\alpha\lambda\gamma\beta}_{\overline{e}\ell\ell\ell H} - \Op^{\alpha\lambda\beta\gamma}_{\overline{e}\ell\ell\ell H} - \Op^{\alpha\gamma\beta\lambda}_{\overline{e}\ell\ell\ell H} = 0 \;,
	\nonumber
	\\	
	&& \Op^{\alpha\beta\gamma\lambda}_{\overline{\ell}dddH} + \Op^{\alpha\beta\lambda\gamma}_{\overline{\ell}dddH} = 0 \;,\quad  \Op^{\alpha\beta\gamma\lambda}_{\overline{\ell}dddH} + \Op^{\alpha\gamma\lambda\beta}_{\overline{\ell}dddH} + \Op^{\alpha\lambda\beta\gamma}_{\overline{\ell}dddH} = 0 \;,
	\nonumber
	\\
	&& \Opr^{\alpha\beta}_{\ell HD3} - \Opr^{\beta\alpha}_{\ell HD3} = 0 \;,
	\nonumber
	\\
	&& \Opr^{\alpha\beta}_{\ell HD4} - \Opr^{\beta\alpha}_{\ell HD4} = 0 \;,
	\nonumber
	\\
	&& \Opr^{\alpha\beta}_{\ell HD6} - \Opr^{\beta\alpha}_{\ell HD6} = 0 \;.
\end{eqnarray}
This indicates that the operators $\Op^{}_{\ell HB}$, $\Op^{}_{\overline{e}qdd\widetilde{H}}$ and $\Op^{}_{\overline{\ell} dddH}$ would not exist if there is only one generation of fermions, and not all degrees of freedom of operators in the flavor space are independent. One needs to remove those redundant degrees of freedom from the operators involved in Eq.~\eqref{eq:symmetry}. For symmetries among two flavor indices, it is quite trivial and the operators are symmetric or antisymmetric with respect to their flavor indices, and so are their Wilson coefficients. While for symmetries among more than two flavor indices, it becomes more complicated and it is not transparent to get rid of the redundant degrees of freedom, e.g., those in $\Op^{}_{\overline{e}\ell\ell\ell H}$ and $\Op^{}_{\overline{\ell}dddH}$.

To count the independent degrees of freedom and remove the redundant ones, one can decompose the involved operators into combinations with explicit symmetries by means of ${\rm SU}(n)$ tensor decomposition. More specifically, the identical fermion fields in operators transform under the same ${\rm SU}(n)$ flavor symmetry, namely, $\Yvcentermath1  \yng(1) \otimes \yng(1) $  and $\Yvcentermath1  \yng(1) \otimes \yng(1) \otimes \yng(1) $ for two and three identical fermion fields, respectively. Therefore, they can be decomposed as
\begin{eqnarray}\label{eq:yangt}
	&&\Yvcentermath1 \yng(1) \otimes \yng(1) = \yng(2) \oplus \yng(1,1) \;,
	\nonumber
	\\
	&& \Yvcentermath1 \yng(1) \otimes \yng(1)  \otimes \yng(1) = \yng(3) \oplus \yng(2,1) \oplus \yng(2,1) \oplus \yng(1,1,1) \;.
\end{eqnarray}
In other words, the operators with two identical fermion fields can be decomposed into symmetric and antisymmetric combinations. For example,
\begin{eqnarray}
	\Op^{\alpha\beta}_{\ell H} &=& \Op^{(S)\alpha\beta}_{\ell H} + \Op^{(A)\alpha\beta}_{\ell H}
\end{eqnarray}
with
\begin{eqnarray}\label{eq:decomposing-1}
	\Op^{(S)\alpha\beta}_{\ell H} = \frac{1}{2} \left( \Op^{\alpha\beta}_{\ell H} + \Op^{\beta\alpha}_{\ell H} \right) \;,\quad \Op^{(A)\alpha\beta}_{\ell H} = \frac{1}{2} \left( \Op^{\alpha\beta}_{\ell H} - \Op^{\beta\alpha}_{\ell H} \right) \;,
\end{eqnarray}
where the superscripts $``(S)"$ and $``(A)"$ stand for symmetric and antisymmetric combinations, respectively. As a result, their Wilson coefficients can be accordingly written in  symmetric and antisymmetric forms, that is
\begin{eqnarray}\label{eq:decomposing-WC-1}
	G^{(S)\alpha\beta}_{\ell H} = \frac{1}{2} \left( G^{\alpha\beta}_{\ell H} + G^{\beta\alpha}_{\ell H} \right) \;,\quad G^{(A)\alpha\beta}_{\ell H} = \frac{1}{2} \left( G^{\alpha\beta}_{\ell H} - G^{\beta\alpha}_{\ell H} \right) \;,
\end{eqnarray}
and as expected, they satisfy $G^{\alpha\beta}_{\ell H} \Op^{\alpha\beta}_{\ell H} = G^{(S)\alpha\beta}_{\ell H} \Op^{(S)\alpha\beta}_{\ell H} + G^{(A)\alpha\beta}_{\ell H} \Op^{(A)\alpha\beta}_{\ell H}$ with $``G^{\dots}_{\dots}"$ denoting the Wilson coefficient in the Green's basis. One can easily get the number of independent operators from Eq.~\eqref{eq:yangt}, i.e., $n(n+1)/2$ for $\Op^{(S)}_{\ell H}$ and $n(n-1)/2$ for $\Op^{(A)}_{\ell H}$ if there is no symmetry in the flavor structure of the operator. However, as shown in Eq.~\eqref{eq:symmetry}, the antisymmetric part $\Op^{(A)}_{\ell H}$ vanishes automatically and only the symmetric part $\Op^{(S)}_{\ell H}$ is left. For the operators with three identical fermion fields, besides the totally symmetric and antisymmetric combinations, there are two mixed-symmetry combinations. Taking $\Op^{}_{\overline{e}\ell\ell\ell H}$ as an example, one can decompose it into
\begin{eqnarray}
	\Op^{\alpha\beta\gamma\lambda}_{\overline{e}\ell\ell\ell H} &=& 	\Op^{(S)\alpha\beta\gamma\lambda}_{\overline{e}\ell\ell\ell H} + 	\Op^{(A)\alpha\beta\gamma\lambda}_{\overline{e}\ell\ell\ell H} + 	\Op^{(M)\alpha\beta\gamma\lambda}_{\overline{e}\ell\ell\ell H} + 	\Op^{(M^\prime)\alpha\beta\gamma\lambda}_{\overline{e}\ell\ell\ell H} 
\end{eqnarray}
with
\begin{eqnarray}\label{eq:decomposing-2}
	\Op^{(S)\alpha\beta\gamma\lambda}_{\overline{e}\ell\ell\ell H} &=& \frac{1}{6} \left( \Op^{\alpha\beta\gamma\lambda}_{\overline{e}\ell\ell\ell H} + \Op^{\alpha\lambda\beta\gamma}_{\overline{e}\ell\ell\ell H} +  \Op^{\alpha\gamma\lambda\beta}_{\overline{e}\ell\ell\ell H} + \Op^{\alpha\beta\lambda\gamma}_{\overline{e}\ell\ell\ell H} + \Op^{\alpha\gamma\beta\lambda}_{\overline{e}\ell\ell\ell H} + \Op^{\alpha\lambda\gamma\beta}_{\overline{e}\ell\ell\ell H} \right) \;,
	\nonumber
	\\
	\Op^{(A)\alpha\beta\gamma\lambda}_{\overline{e}\ell\ell\ell H} &=& \frac{1}{6} \left( \Op^{\alpha\beta\gamma\lambda}_{\overline{e}\ell\ell\ell H} + \Op^{\alpha\lambda\beta\gamma}_{\overline{e}\ell\ell\ell H} +  \Op^{\alpha\gamma\lambda\beta}_{\overline{e}\ell\ell\ell H} - \Op^{\alpha\beta\lambda\gamma}_{\overline{e}\ell\ell\ell H} - \Op^{\alpha\gamma\beta\lambda}_{\overline{e}\ell\ell\ell H} - \Op^{\alpha\lambda\gamma\beta}_{\overline{e}\ell\ell\ell H} \right) \;,
	\nonumber
	\\
	\Op^{(M)\alpha\beta\gamma\lambda}_{\overline{e}\ell\ell\ell H} &=& \frac{1}{3} \left( \Op^{\alpha\beta\gamma\lambda}_{\overline{e}\ell\ell\ell H} + \Op^{\alpha\gamma\beta\lambda}_{\overline{e}\ell\ell\ell H} -  \Op^{\alpha\lambda\gamma\beta}_{\overline{e}\ell\ell\ell H} - \Op^{\alpha\gamma\lambda\beta}_{\overline{e}\ell\ell\ell H} \right) \;,
	\nonumber
	\\
	\Op^{(M^\prime)\alpha\beta\gamma\lambda}_{\overline{e}\ell\ell\ell H} &=& \frac{1}{3} \left( \Op^{\alpha\beta\gamma\lambda}_{\overline{e}\ell\ell\ell H} + \Op^{\alpha\lambda\gamma\beta}_{\overline{e}\ell\ell\ell H} -  \Op^{\alpha\lambda\beta\gamma}_{\overline{e}\ell\ell\ell H} - \Op^{\alpha\gamma\beta\lambda}_{\overline{e}\ell\ell\ell H} \right) \;,
\end{eqnarray}
in which the first two are totally symmetric and antisymmetric operators, and the last two have mixed symmetry. Similarly, the corresponding Wilson coefficients can be taken to be 
\begin{eqnarray}\label{eq:decomposing-WC-2}
	G^{(S)\alpha\beta\gamma\lambda}_{\overline{e}\ell\ell\ell H} &=& \frac{1}{6} \left( G^{\alpha\beta\gamma\lambda}_{\overline{e}\ell\ell\ell H} + G^{\alpha\lambda\beta\gamma}_{\overline{e}\ell\ell\ell H} +  G^{\alpha\gamma\lambda\beta}_{\overline{e}\ell\ell\ell H} + G^{\alpha\beta\lambda\gamma}_{\overline{e}\ell\ell\ell H} + G^{\alpha\gamma\beta\lambda}_{\overline{e}\ell\ell\ell H} + G^{\alpha\lambda\gamma\beta}_{\overline{e}\ell\ell\ell H} \right) \;,
	\nonumber
	\\
	G^{(A)\alpha\beta\gamma\lambda}_{\overline{e}\ell\ell\ell H} &=& \frac{1}{6} \left( G^{\alpha\beta\gamma\lambda}_{\overline{e}\ell\ell\ell H} + G^{\alpha\lambda\beta\gamma}_{\overline{e}\ell\ell\ell H} +  G^{\alpha\gamma\lambda\beta}_{\overline{e}\ell\ell\ell H} - G^{\alpha\beta\lambda\gamma}_{\overline{e}\ell\ell\ell H} - G^{\alpha\gamma\beta\lambda}_{\overline{e}\ell\ell\ell H} - G^{\alpha\lambda\gamma\beta}_{\overline{e}\ell\ell\ell H} \right) \;,
	\nonumber
	\\
	G^{(M)\alpha\beta\gamma\lambda}_{\overline{e}\ell\ell\ell H} &=& \frac{1}{6} \left( 2G^{\alpha\beta\gamma\lambda}_{\overline{e}\ell\ell\ell H} - G^{\alpha\lambda\beta\gamma}_{\overline{e}\ell\ell\ell H} - G^{\alpha\gamma\lambda\beta}_{\overline{e}\ell\ell\ell H} + G^{\alpha\beta\lambda\gamma}_{\overline{e}\ell\ell\ell H} + G^{\alpha\gamma\beta\lambda}_{\overline{e}\ell\ell\ell H} - 2G^{\alpha\lambda\gamma\beta}_{\overline{e}\ell\ell\ell H} \right) \;,
	\nonumber
	\\
	G^{(M^\prime)\alpha\beta\gamma\lambda}_{\overline{e}\ell\ell\ell H} &=& \frac{1}{6} \left( 2G^{\alpha\beta\gamma\lambda}_{\overline{e}\ell\ell\ell H} - G^{\alpha\lambda\beta\gamma}_{\overline{e}\ell\ell\ell H} - G^{\alpha\gamma\lambda\beta}_{\overline{e}\ell\ell\ell H} + G^{\alpha\beta\lambda\gamma}_{\overline{e}\ell\ell\ell H} - 2G^{\alpha\gamma\beta\lambda}_{\overline{e}\ell\ell\ell H} + G^{\alpha\lambda\gamma\beta}_{\overline{e}\ell\ell\ell H} \right) \;,
\end{eqnarray}
and satisfy $G^{\alpha\beta\gamma\lambda}_{\overline{e}\ell\ell\ell H} \Op^{\alpha\beta\gamma\lambda}_{\overline{e}\ell\ell\ell H} = \sum^{}\limits_{ I=S,A,M,M^\prime} G^{(I)\alpha\beta\gamma\lambda}_{\overline{e}\ell\ell\ell H} \Op^{(I)\alpha\beta\gamma\lambda}_{\overline{e}\ell\ell\ell H} $. Notice that $\Op^{(M^\prime)}_{\overline{e}\ell\ell\ell H}$ turns out to be vanishing due to the relation given in Eq.~\eqref{eq:symmetry} and hence one only needs to focus on $\Op^{(S),(A),(M)}_{\overline{e}\ell\ell\ell H}$. With the help of Eq.~\eqref{eq:yangt}, the number of independent operators is found to be $n^2(n+1)(n+2)/6$ for $\Op^{(S)}_{\overline{e}\ell\ell\ell H}$, $n^2(n-1)(n-2)/6$ for $\Op^{(A)}_{\overline{e}\ell\ell\ell H}$, and $n^2(n-1)(n+1)/3$ for $\Op^{(M)}_{\overline{e}\ell\ell\ell H}$, and their sum is in accordance with the total number of degrees of freedom, i.e., $n^2(2n^2+1)/3$. 

As can be seen from the above examples, after operators are decomposed into combinations with explicit symmetries, some of them are automatically vanishing due to the flavor relations shown in Eq.~\eqref{eq:symmetry}, and the rest is independent. Moreover, the number of independent degrees of freedom can be easily counted with the help of Young tableaus in Eq.~\eqref{eq:yangt}. Though the decompositions of operators with symmetries among two flavor indices are trivial and even not necessary, we still decompose them along with those having symmetries among three flavor indices. Decomposing the rest of operators involved in Eq.~\eqref{eq:symmetry} in the same way, one obtains
\begin{eqnarray}\label{eq:decomposing-3}
	\Op^{(A)\alpha\beta}_{\ell HB} &=& \frac{1}{2} \left( \Op^{\alpha\beta}_{\ell HB} - \Op^{\beta\alpha}_{\ell HB} \right) \;,
	\nonumber
	\\
	\Op^{(A)\alpha\beta\gamma\lambda}_{\overline{e}qdd\widetilde{H}} &=& \frac{1}{2} \left( \Op^{\alpha\beta\gamma\lambda}_{\overline{e}qdd\widetilde{H}} - \Op^{\alpha\beta\lambda\gamma}_{\overline{e}qdd\widetilde{H}} \right) \;,
	\nonumber
	\\
	\Op^{(M)\alpha\beta\gamma\lambda}_{\overline{\ell} dddH} &=& \frac{1}{3} \left( \Op^{\alpha\beta\gamma\lambda}_{\overline{\ell} dddH} + \Op^{\alpha\gamma\beta\lambda}_{\overline{\ell} dddH} -  \Op^{\alpha\beta\lambda\gamma}_{\overline{\ell} dddH} - \Op^{\alpha\lambda\beta\gamma}_{\overline{\ell} dddH} \right) \;,
	\nonumber
	\\
	\Opr^{(S)\alpha\beta}_{\ell HD3} &=& \frac{1}{2} \left( \Opr^{\alpha\beta}_{\ell HD3} + \Opr^{\beta\alpha}_{\ell HD3} \right) \;,
	\nonumber
	\\
	\Opr^{(S)\alpha\beta}_{\ell HD4} &=& \frac{1}{2} \left( \Opr^{\alpha\beta}_{\ell HD4} + \Opr^{\beta\alpha}_{\ell HD4} \right) \;,
	\nonumber
	\\
	\Opr^{(S)\alpha\beta}_{\ell HD6} &=& \frac{1}{2} \left( \Opr^{\alpha\beta}_{\ell HD6} + \Opr^{\beta\alpha}_{\ell HD6} \right) 
\end{eqnarray}
with their Wilson coefficients being
\begin{eqnarray}\label{eq:decomposing-WC-3}
	G^{(A)\alpha\beta}_{\ell HB} &=& \frac{1}{2} \left( G^{\alpha\beta}_{\ell HB} - G^{\beta\alpha}_{\ell HB} \right) \;,
	\nonumber
	\\
	G^{(A)\alpha\beta\gamma\lambda}_{\overline{e}qdd\widetilde{H}} &=& \frac{1}{2} \left( G^{\alpha\beta\gamma\lambda}_{\overline{e}qdd\widetilde{H}} - G^{\alpha\beta\lambda\gamma}_{\overline{e}qdd\widetilde{H}} \right) \;,
	\nonumber
	\\
	G^{(M)\alpha\beta\gamma\lambda}_{\overline{\ell} dddH} &=& \frac{1}{6} \left( 2G^{\alpha\beta\gamma\lambda}_{\overline{\ell} dddH} - G^{\alpha\lambda\beta\gamma}_{\overline{\ell} dddH} - G^{\alpha\gamma\lambda\beta}_{\overline{\ell} dddH}  - 2G^{\alpha\beta\lambda\gamma}_{\overline{\ell} dddH} + G^{\alpha\gamma\beta\lambda}_{\overline{\ell} dddH} + G^{\alpha\lambda\gamma\beta}_{\overline{\ell} dddH} \right) \;,
	\nonumber
	\\
	G^{(S)\alpha\beta}_{\ell HD3} &=& \frac{1}{2} \left(G^{\alpha\beta}_{\ell HD3} + G^{\beta\alpha}_{\ell HD3} \right) \;,
	\nonumber
	\\
	G^{(S)\alpha\beta}_{\ell HD4} &=& \frac{1}{2} \left( G^{\alpha\beta}_{\ell HD4} + G^{\beta\alpha}_{\ell HD4} \right) \;,
	\nonumber
	\\
	G^{(S)\alpha\beta}_{\ell HD6} &=& \frac{1}{2} \left( G^{\alpha\beta}_{\ell HD6} + G^{\beta\alpha}_{\ell HD6} \right) \;,
\end{eqnarray}
and other degrees of freedom associated with these operators are automatically vanishing. Then, one may substitute those operators with explicit symmetries in Eqs.~\eqref{eq:decomposing-1}, \eqref{eq:decomposing-2} and \eqref{eq:decomposing-3} for the corresponding operators in the grey cells of Table~\ref{tab:green-basis}, and their Wilson coefficients are given in Eqs.~\eqref{eq:decomposing-WC-1}, \eqref{eq:decomposing-WC-2} and \eqref{eq:decomposing-WC-3}. Together with the rest of operators in Table~\ref{tab:green-basis}, they constitute a Green's basis, where all operators and their Wilson coefficients only contain independent degrees of freedom.

\subsection{The physical basis for dim-7 operators}

With the help of field redefinitions or EoMs in the SMEFT, one can convert the operators in the Green's basis into those in a physical basis. But as pointed out in Ref.~\cite{Liao:2019tep}, there are some non-trivial flavor relations among operators $\Op^{}_{\dots}$ listed in Table~\ref{tab:green-basis} after field redefinitions or EoMs are applied and fermion flavors are considered. This new feature appears first at dimension seven and has to be taken into account when constructing a physical basis and also converting operators in the Green's basis to those in a physical basis. Those non-trivial flavor relations between dim-7 operators are found to be~\cite{Liao:2019tep}
\begin{eqnarray}\label{eq:flavor-relations}
	&&\left[ 4\Op^{\alpha\beta}_{\ell H D 2} + 2 \mathcal{K}^{\alpha\beta} + 2 \left( Y^{}_l \right)^{}_{\beta\gamma} \Op^{\alpha\gamma}_{\ell e HD} - \frac{1}{2} g^{}_2 \Op^{\alpha\beta}_{\ell H W} \right] - \alpha \leftrightarrow \beta = g^{}_1 \Op^{\alpha\beta}_{\ell HB} \;,
	\nonumber
	\\
	&&\left[ \Op^{\alpha\beta}_{\ell HD1} + \mathcal{K}^{\alpha\beta} \right] - \alpha \leftrightarrow \beta = 0 \;,
	\nonumber
	\\
	&&\left[ \Op^{\alpha\beta\gamma\lambda}_{\overline{d}u\ell\ell D} + \left( Y^{}_{\rm d} \right)^{}_{\rho \alpha} \Op^{\rho\beta\lambda\gamma}_{\overline{q}u\ell\ell H} - \left(Y^\dagger_{\rm u} \right)^{}_{\beta\rho} \Op^{\alpha\gamma\rho\lambda}_{\overline{d}\ell q \ell H 2} \right] - \gamma \leftrightarrow \lambda = 0 \;,
	\nonumber
	\\
	&&\left[ \Op^{\alpha\beta\gamma\lambda}_{\overline{\ell}qddD} + \left( Y^{}_{\rm u} \right)^{}_{\beta\rho} \Op^{\alpha\gamma\rho\lambda}_{\overline{\ell}dud\widetilde{H}} \right] - \gamma \leftrightarrow \lambda = - \left( Y^\dagger_l \right)^{}_{\rho\alpha} \Op^{\rho\beta\gamma\lambda}_{\overline{e}qdd\widetilde{H}} - \left( Y^{}_{\rm d} \right)^{}_{\beta\rho} \Op^{\alpha\rho\gamma\lambda}_{\overline{\ell}dddH} \;,
	\nonumber
	\\
	&&\Op^{\alpha\beta\gamma\lambda}_{\overline{e}dddD} - \Op^{\alpha\gamma\beta\lambda}_{\overline{e}dddD} = \left( Y^\dagger_{\rm d} \right)^{}_{\lambda\rho} \Op^{\alpha\rho\beta\gamma}_{\overline{e}qdd\widetilde{H}} \;,
	\nonumber
	\\
	&&\Op^{\alpha\beta\gamma\lambda}_{\overline{e}dddD} + \Op^{\alpha\lambda\gamma\beta}_{\overline{e}dddD} - \gamma \leftrightarrow \lambda = \left( Y^{}_l \right)^{}_{\rho\alpha} \Op^{\rho\beta\gamma\lambda}_{\overline{\ell}dddH} \;
\end{eqnarray}
with 
\begin{eqnarray}
	\mathcal{K}^{\alpha\beta} = \left( Y^{}_{\rm u} \right)^{}_{\gamma\lambda} \Op^{\gamma\lambda\alpha\beta}_{\overline{q}u\ell\ell H} - \left( Y^\dagger_{\rm d} \right)^{}_{\gamma\lambda} \Op^{\gamma\alpha\lambda\beta}_{\overline{d}\ell q \ell H 2} - \left( Y^\dagger_l \right)^{}_{\gamma\lambda} \Op^{\gamma\lambda\alpha\beta}_{\overline{e}\ell\ell\ell H} \;,
\end{eqnarray}
where $\alpha \leftrightarrow \beta$ and $\gamma \leftrightarrow \lambda$ indicate the exchange of flavor indices. Because of those flavor relations given in Eq.~\eqref{eq:flavor-relations}, there are some redundant degrees of freedom in $\Op^{}_{\dots}$, which need to be converted to the non-redundant ones. To avoid the inverse of small Yukawa couplings and automatically get rid of redundant degrees of freedom in operators, we propose a new physical basis for dim-7 operators. All operators in this basis are listed in Table~\ref{tab:physical-basis}, where $\Op^{}_{\ell DH1}$, $\Op^{}_{\ell DH2}$, $\Op^{}_{\overline{d}u\ell\ell D}$, $\Op^{}_{\overline{\ell}qddD}$ and $\Op^{}_{\overline{e}dddD}$ are also decomposed according to ${\rm SU}(n)$ tensor decomposition. But in contrast to operators in the grey cells of Table~\ref{tab:green-basis}, $\Op^{(A)}_{\ell HD1}$, $\Op^{(A)}_{\ell HD2}$, $\Op^{(A),(M),(M^\prime)}_{\overline{e}dddD}$, $\Op^{(A)}_{\overline{d}u\ell\ell D}$ and $\Op^{(A)}_{\overline{\ell}qddD}$ do not vanish and can be converted to other independent operators in Table~\ref{tab:physical-basis} due to the non-trivial flavor relations in Eq.~\eqref{eq:flavor-relations}. Here, we take $\Op^{(A),(M),(M^\prime)}_{\overline{e}dddD}$ as an example, i.e.,
\begin{table}
	\centering
	\renewcommand\arraystretch{1.3}
	\resizebox{\textwidth}{!}{
		\begin{tabular}{l|l|l}
			\hline\hline
			\multicolumn{3}{c}{$\psi^2H^4$} \\
			\hline
			$\Op^{(S)\alpha\beta}_{\ell H} = \frac{1}{2} \left( \Op^{\alpha\beta}_{\ell H} + \Op^{\beta\alpha}_{\ell H} \right) $ & $C^{(S)\alpha\beta}_{\ell H} = \frac{1}{2} \left( C^{\alpha\beta}_{\ell H} + C^{\beta\alpha}_{\ell H} \right) $ & $\frac{1}{2} n(n+1)$ \\
			\hline
			\multicolumn{3}{c}{$\psi^2H^3D$} \\
			\hline
			$\Op^{\alpha\beta}_{\ell eHD} $ & $ C^{\alpha\beta}_{\ell eHD} $  & $n^2$ \\
			\hline
			\multicolumn{3}{c}{$\psi^2H^2 D^2$} \\
			\hline
			$ \Op^{(S)\alpha\beta}_{\ell HD1} = \frac{1}{2} \left( \Op^{\alpha\beta}_{\ell HD1} + \Op^{\beta\alpha}_{\ell HD1} \right) $ & $ C^{(S)\alpha\beta}_{\ell HD1} = \frac{1}{2} \left( C^{\alpha\beta}_{\ell HD1} + C^{\beta\alpha}_{\ell HD1} \right) $ &  $\frac{1}{2} n(n+1)$ \\
			$ \Op^{(S)\alpha\beta}_{\ell HD2} = \frac{1}{2} \left( \Op^{\alpha\beta}_{\ell HD2} + \Op^{\beta\alpha}_{\ell HD2} \right) $ & $ C^{(S)\alpha\beta}_{\ell HD2} = \frac{1}{2} \left( C^{\alpha\beta}_{\ell HD2} + C^{\beta\alpha}_{\ell HD2} \right) $ &  $\frac{1}{2} n(n+1)$ \\
			\hline
			\multicolumn{3}{c}{$\psi^2H^2 X$} \\
			\hline
			$\Op^{(A)\alpha\beta}_{\ell HB} = \frac{1}{2} \left( \Op^{\alpha\beta}_{\ell HB} - \Op^{\beta\alpha}_{\ell HB} \right) $ & $C^{(A)\alpha\beta}_{\ell HB} = \frac{1}{2} \left( C^{\alpha\beta}_{\ell HB} - C^{\beta\alpha}_{\ell HB} \right) $ & $\frac{1}{2} n(n-1)$ \\
			$\Op^{\alpha\beta}_{\ell HW} $ & $C^{\alpha\beta}_{\ell HW} $ & $n^2$ \\
			\hline
			\multicolumn{3}{c}{$\psi^4H$} \\
			\hline
			\makecell[l]{$ \Op^{(S)\alpha\beta\gamma\lambda}_{\overline{e}\ell\ell\ell H} = \frac{1}{6} \left( \Op^{\alpha\beta\gamma\lambda}_{\overline{e}\ell\ell\ell H} + \Op^{\alpha\lambda\beta\gamma}_{\overline{e}\ell\ell\ell H} +  \Op^{\alpha\gamma\lambda\beta}_{\overline{e}\ell\ell\ell H} \right.$ \\  $\hphantom{\Op^{(S)\alpha\beta\gamma\lambda}_{\overline{e}\ell\ell\ell H} = \frac{1}{6} } \left.+ \Op^{\alpha\beta\lambda\gamma}_{\overline{e}\ell\ell\ell H} + \Op^{\alpha\gamma\beta\lambda}_{\overline{e}\ell\ell\ell H} + \Op^{\alpha\lambda\gamma\beta}_{\overline{e}\ell\ell\ell H} \right) $} & \makecell[l]{ $C^{(S)\alpha\beta\gamma\lambda}_{\overline{e}\ell\ell\ell H} = \frac{1}{6} \left( C^{\alpha\beta\gamma\lambda}_{\overline{e}\ell\ell\ell H} + C^{\alpha\lambda\beta\gamma}_{\overline{e}\ell\ell\ell H} +  C^{\alpha\gamma\lambda\beta}_{\overline{e}\ell\ell\ell H} \right.$ \\ $\hphantom{\Op^{(S)\alpha\beta\gamma\lambda}_{\overline{e}\ell\ell\ell H} = \frac{1}{6} } \left.+ C^{\alpha\beta\lambda\gamma}_{\overline{e}\ell\ell\ell H} + C^{\alpha\gamma\beta\lambda}_{\overline{e}\ell\ell\ell H} + C^{\alpha\lambda\gamma\beta}_{\overline{e}\ell\ell\ell H} \right)$} & $\frac{1}{6}n^2(n+1)(n+2)$ \\
			\makecell[l]{$\Op^{(A)\alpha\beta\gamma\lambda}_{\overline{e}\ell\ell\ell H} = \frac{1}{6} \left( \Op^{\alpha\beta\gamma\lambda}_{\overline{e}\ell\ell\ell H} + \Op^{\alpha\lambda\beta\gamma}_{\overline{e}\ell\ell\ell H} +  \Op^{\alpha\gamma\lambda\beta}_{\overline{e}\ell\ell\ell H} \right.$ \\ $\hphantom{\Op^{(S)\alpha\beta\gamma\lambda}_{\overline{e}\ell\ell\ell H} = \frac{1}{6} }\left. - \Op^{\alpha\beta\lambda\gamma}_{\overline{e}\ell\ell\ell H} - \Op^{\alpha\gamma\beta\lambda}_{\overline{e}\ell\ell\ell H} - \Op^{\alpha\lambda\gamma\beta}_{\overline{e}\ell\ell\ell H} \right)$ }& \makecell[l]{$C^{(A)\alpha\beta\gamma\lambda}_{\overline{e}\ell\ell\ell H} = \frac{1}{6} \left( C^{\alpha\beta\gamma\lambda}_{\overline{e}\ell\ell\ell H} + C^{\alpha\lambda\beta\gamma}_{\overline{e}\ell\ell\ell H} +  C^{\alpha\gamma\lambda\beta}_{\overline{e}\ell\ell\ell H} \right. $ \\ $ \hphantom{\Op^{(S)\alpha\beta\gamma\lambda}_{\overline{e}\ell\ell\ell H} = \frac{1}{6} }\left. - C^{\alpha\beta\lambda\gamma}_{\overline{e}\ell\ell\ell H} - C^{\alpha\gamma\beta\lambda}_{\overline{e}\ell\ell\ell H} - C^{\alpha\lambda\gamma\beta}_{\overline{e}\ell\ell\ell H} \right)$} & $\frac{1}{6} n^2(n-1)(n-2)$ \\
			\makecell[l]{ $\Op^{(M)\alpha\beta\gamma\lambda}_{\overline{e}\ell\ell\ell H} = \frac{1}{3} \left( \Op^{\alpha\beta\gamma\lambda}_{\overline{e}\ell\ell\ell H} + \Op^{\alpha\gamma\beta\lambda}_{\overline{e}\ell\ell\ell H} -  \Op^{\alpha\lambda\gamma\beta}_{\overline{e}\ell\ell\ell H} \right.$ \\ $\hphantom{\Op^{(M)\alpha\beta\gamma\lambda}_{\overline{e}\ell\ell\ell H} = \frac{1}{3} } \left. - \Op^{\alpha\gamma\lambda\beta}_{\overline{e}\ell\ell\ell H} \right)$} & \makecell[l]{$C^{(M)\alpha\beta\gamma\lambda}_{\overline{e}\ell\ell\ell H} = \frac{1}{6} \left( 2C^{\alpha\beta\gamma\lambda}_{\overline{e}\ell\ell\ell H} - C^{\alpha\lambda\beta\gamma}_{\overline{e}\ell\ell\ell H} -  C^{\alpha\gamma\lambda\beta}_{\overline{e}\ell\ell\ell H} \right. $ \\ $ \hphantom{\Op^{(S)\alpha\beta\gamma\lambda}_{\overline{e}\ell\ell\ell H} = \frac{1}{6} }\left. + C^{\alpha\beta\lambda\gamma}_{\overline{e}\ell\ell\ell H} + C^{\alpha\gamma\beta\lambda}_{\overline{e}\ell\ell\ell H} - 2 C^{\alpha\lambda\gamma\beta}_{\overline{e}\ell\ell\ell H} \right)$} & $\frac{1}{3} n^2(n-1)(n+1)$ \\
			$\Op^{\alpha\beta\gamma\lambda}_{\overline{d}\ell q \ell H1} $ & $C^{\alpha\beta\gamma\lambda}_{\overline{d}\ell q \ell H1} $ & $n^4$ \\ $\Op^{\alpha\beta\gamma\lambda}_{\overline{d}\ell q \ell H2} $ & $C^{\alpha\beta\gamma\lambda}_{\overline{d}\ell q \ell H2} $  & $n^4$ \\ $\Op^{\alpha\beta\gamma\lambda}_{\overline{d}\ell ueH} $ & $C^{\alpha\beta\gamma\lambda}_{\overline{d}\ell ueH} $ & $n^4$ \\
			$\Op^{\alpha\beta\gamma\lambda}_{\overline{q} u \ell \ell H} $ & $C^{\alpha\beta\gamma\lambda}_{\overline{q} u \ell \ell H} $ & $n^4$ \\
			$\Op^{\alpha\beta\gamma\lambda}_{\overline{\ell}dud\widetilde{H}} $ & $C^{\alpha\beta\gamma\lambda}_{\overline{\ell}dud\widetilde{H}} $ & $n^4$ \\
			\makecell[l]{$\Op^{(M)\alpha\beta\gamma\lambda}_{\overline{\ell} dddH} = \frac{1}{3} \left( \Op^{\alpha\beta\gamma\lambda}_{\overline{\ell} dddH} + \Op^{\alpha\gamma\beta\lambda}_{\overline{\ell} dddH} -  \Op^{\alpha\beta\lambda\gamma}_{\overline{\ell} dddH} \right.$ \\ $ \hphantom{\Op^{(M)\alpha\beta\gamma\lambda}_{\overline{\ell} dddH} = \frac{1}{3}} \left. - \Op^{\alpha\lambda\beta\gamma}_{\overline{\ell} dddH} \right)$}  &  \makecell[l]{$C^{(M)\alpha\beta\gamma\lambda}_{\overline{\ell} dddH} = \frac{1}{6} \left( 2C^{\alpha\beta\gamma\lambda}_{\overline{\ell} dddH} - C^{\alpha\lambda\beta\gamma}_{\overline{\ell} dddH} - C^{\alpha\gamma\lambda\beta}_{\overline{\ell} dddH} \right.$ \\ $ \hphantom{\Op^{(M)\alpha\beta\gamma\lambda}_{\overline{\ell} dddH} = \frac{1}{3}} \left. - 2C^{\alpha\beta\lambda\gamma}_{\overline{\ell} dddH} + C^{\alpha\gamma\beta\lambda}_{\overline{\ell} dddH} + C^{\alpha\lambda\gamma\beta}_{\overline{\ell} dddH} \right)$ }  & $\frac{1}{3} n^2(n-1)(n+1)$ \\
			$\Op^{(A)\alpha\beta\gamma\lambda}_{\overline{e}qdd\widetilde{H}} = \frac{1}{2} \left( \Op^{\alpha\beta\gamma\lambda}_{\overline{e}qdd\widetilde{H}} - \Op^{\alpha\beta\lambda\gamma}_{\overline{e}qdd\widetilde{H}}  \right) $ & $C^{(A)\alpha\beta\gamma\lambda}_{\overline{e}qdd\widetilde{H}} = \frac{1}{2} \left( C^{\alpha\beta\gamma\lambda}_{\overline{e}qdd\widetilde{H}} - C^{\alpha\beta\lambda\gamma}_{\overline{e}qdd\widetilde{H}}  \right) $ & $\frac{1}{2} n^3(n-1)$  \\
			$ \Op^{\alpha\beta\gamma\lambda}_{\overline{\ell}dqq\widetilde{H}}$ & $ C^{\alpha\beta\gamma\lambda}_{\overline{\ell}dqq\widetilde{H}}$ & $n^4$ \\
			\hline
			\multicolumn{3}{c}{$\psi^4D$} \\
			\hline
			\makecell[l]{$\Op^{(S)\alpha\beta\gamma\lambda}_{\overline{e} dddD} = \frac{1}{6} \left( \Op^{\alpha\beta\gamma\lambda}_{\overline{e}dddD} + \Op^{\alpha\lambda\beta\gamma}_{\overline{e}dddD} +  \Op^{\alpha\gamma\lambda\beta}_{\overline{e}dddD} \right.$ \\ $\hphantom{\Op^{(S)\alpha\beta\gamma\lambda}_{\overline{e}\ell\ell\ell H} = \frac{1}{6} } \left.+ \Op^{\alpha\beta\lambda\gamma}_{\overline{e}dddD} + \Op^{\alpha\gamma\beta\lambda}_{\overline{e}dddD} + \Op^{\alpha\lambda\gamma\beta}_{\overline{e}dddD} \right)$ }   &  \makecell[l]{$C^{(S)\alpha\beta\gamma\lambda}_{\overline{e} dddD} = \frac{1}{6} \left( C^{\alpha\beta\gamma\lambda}_{\overline{e}dddD} + C^{\alpha\lambda\beta\gamma}_{\overline{e}dddD} +  C^{\alpha\gamma\lambda\beta}_{\overline{e}dddD} \right.$ \\ $\hphantom{C^{(S)\alpha\beta\gamma\lambda}_{\overline{e}\ell\ell\ell H} = \frac{1}{6} } \left.+ C^{\alpha\beta\lambda\gamma}_{\overline{e}dddD} + C^{\alpha\gamma\beta\lambda}_{\overline{e}dddD} + C^{\alpha\lambda\gamma\beta}_{\overline{e}dddD} \right)$ }  & $\frac{1}{6}n^2(n+1)(n+2)$ \\
			$\Op^{(S)\alpha\beta\gamma\lambda}_{\overline{d}u\ell\ell D} = \frac{1}{2} \left( \Op^{\alpha\beta\gamma\lambda}_{\overline{d} u\ell\ell D} + \Op^{\alpha\beta\lambda\gamma}_{\overline{d} u\ell\ell D} \right) $ &  $C^{(S)\alpha\beta\gamma\lambda}_{\overline{d}u\ell\ell D} = \frac{1}{2} \left( C^{\alpha\beta\gamma\lambda}_{\overline{d} u\ell\ell D} + C^{\alpha\beta\lambda\gamma}_{\overline{d} u\ell\ell D} \right) $ & $\frac{1}{2} n^3(n+1)$ \\ $\Op^{(S)\alpha\beta\gamma\lambda}_{\overline{\ell}qddD} = \frac{1}{2} \left( \Op^{\alpha\beta\gamma\lambda}_{\overline{\ell} qdd D} + \Op^{\alpha\beta\lambda\gamma}_{\overline{\ell} qdd D} \right) $ & $C^{(S)\alpha\beta\gamma\lambda}_{\overline{\ell}qddD} = \frac{1}{2} \left( C^{\alpha\beta\gamma\lambda}_{\overline{\ell} qdd D} + C^{\alpha\beta\lambda\gamma}_{\overline{\ell} qdd D} \right) $ & $\frac{1}{2} n^3(n+1)$ \\
			\hline\hline
		\end{tabular}
	}
	\caption{A physical basis for dim-7 operators. Operators and their Wilson coefficients are listed in the first and second columns, respectively, and the explicit forms of operators can be found in Table~\ref{tab:green-basis}. The number of corresponding independent operators is given in the last column.}\label{tab:physical-basis}
\end{table}

 \begin{eqnarray}
	\Op^{(A)\alpha\beta\gamma\lambda}_{\overline{e} dddD} &=& \frac{1}{6} \left( \Op^{\alpha\beta\gamma\lambda}_{\overline{e}dddD} + \Op^{\alpha\lambda\beta\gamma}_{\overline{e}dddD} +  \Op^{\alpha\gamma\lambda\beta}_{\overline{e}dddD} - \Op^{\alpha\beta\lambda\gamma}_{\overline{e}dddD} - \Op^{\alpha\gamma\beta\lambda}_{\overline{e}dddD} - \Op^{\alpha\lambda\gamma\beta}_{\overline{e}dddD} \right)
	\nonumber
	\\
	&=&  \frac{1}{6} \left[ \left( Y^\dagger_{\rm d} \right)^{}_{\lambda\rho} \Op^{\alpha\rho\beta\gamma}_{\overline{e}qdd\widetilde{H}} +  \left( Y^\dagger_{\rm d} \right)^{}_{\gamma\rho} \Op^{\alpha\rho\lambda\beta}_{\overline{e}qdd\widetilde{H}} +  \left( Y^\dagger_{\rm d} \right)^{}_{\beta\rho} \Op^{\alpha\rho\gamma\lambda}_{\overline{e}qdd\widetilde{H}}  \right] \;,
	\nonumber
	\\
	\Op^{(M)\alpha\beta\gamma\lambda}_{\overline{e} dddD} &=& \frac{1}{3} \left( \Op^{\alpha\beta\gamma\lambda}_{\overline{e}dddD} + \Op^{\alpha\lambda\gamma\beta}_{\overline{e}dddD} - \Op^{\alpha\beta\lambda\gamma}_{\overline{e}dddD} - \Op^{\alpha\gamma\lambda\beta}_{\overline{e}dddD} \right) = \frac{1}{3} \left( Y^{}_l \right)^{}_{\rho\alpha} \Op^{\rho\beta\gamma\lambda}_{\overline{\ell}dddH} \;,
	\nonumber
	\\
	\Op^{(M^\prime)\alpha\beta\gamma\lambda}_{\overline{e} dddD} &=& \frac{1}{3} \left( \Op^{\alpha\beta\gamma\lambda}_{\overline{e}dddD} + \Op^{\alpha\beta\lambda\gamma}_{\overline{e}dddD} -  \Op^{\alpha\lambda\gamma\beta}_{\overline{e}dddD} - \Op^{\alpha\lambda\beta\gamma}_{\overline{e}dddD} \right) 
	\nonumber
	\\
	&=& \frac{1}{3} \left[  \left( \Op^{\alpha\lambda\beta\gamma}_{\overline{e}dddD} + \Op^{\alpha\gamma\beta\lambda}_{\overline{e}dddD} -  \Op^{\alpha\beta\lambda\gamma}_{\overline{e}dddD} - \Op^{\alpha\gamma\lambda\beta}_{\overline{e}dddD} \right) + \left( \Op^{\alpha\beta\gamma\lambda}_{\overline{e}dddD}  - \Op^{\alpha\gamma\beta\lambda}_{\overline{e}dddD}  \right) \right.
	\nonumber
	\\
	&& + \left. \left( \Op^{\alpha\gamma\lambda\beta}_{\overline{e}dddD} - \Op^{\alpha\lambda\gamma\beta}_{\overline{e}dddD}  \right) - 2 \left( \Op^{\alpha\lambda\beta\gamma}_{\overline{e}dddD} - \Op^{\alpha\beta\lambda\gamma}_{\overline{e}dddD}  \right) \right]
	\nonumber
	\\
	&=& \frac{1}{3} \left[ \left( Y^{}_l \right)^{}_{\rho\alpha} \Op^{\rho\gamma\beta\lambda}_{\overline{\ell}dddH} +  \left( Y^\dagger_{\rm d} \right)^{}_{\lambda\rho} \Op^{\alpha\rho\beta\gamma}_{\overline{e}qdd\widetilde{H}} + \left( Y^\dagger_{\rm d} \right)^{}_{\beta\rho} \Op^{\alpha\rho\gamma\lambda}_{\overline{e}qdd\widetilde{H}} - 2  \left( Y^\dagger_{\rm d} \right)^{}_{\gamma\rho} \Op^{\alpha\rho\lambda\beta}_{\overline{e}qdd\widetilde{H}}  \right] \;,
\end{eqnarray}
and then they can contribute to the Wilson coefficients of other independent operators, namely
\begin{eqnarray}
	C^{(A)\alpha\beta\gamma\lambda}_{\overline{e}qdd\widetilde{H}} &\supset& \frac{1}{2} \left( C^{(A)\alpha\rho\gamma\lambda}_{\overline{e}dddD} + 2 C^{(M^\prime)\alpha\gamma\rho\lambda}_{\overline{e}dddD}  \right)  \left( Y^\dagger_{\rm d} \right)^{}_{\rho\beta}  \;,
	\nonumber
	\\
	C^{(M)\alpha\beta\gamma\lambda}_{\overline{\ell}dddH} &\supset& \frac{1}{3} \left( C^{(M)\rho\beta\gamma\lambda}_{\overline{e}dddD} + C^{(M^\prime)\rho\gamma\beta\lambda}_{\overline{e}dddD}  \right) \left( Y^{}_l \right)^{}_{\alpha\rho} \;,
\end{eqnarray}
with
\begin{eqnarray}
	C^{(A)\alpha\beta\gamma\lambda}_{\overline{e}dddD} &=& \frac{1}{6} \left( C^{\alpha\beta\gamma\lambda}_{\overline{e}dddD} + C^{\alpha\lambda\beta\gamma}_{\overline{e}dddD} +  C^{\alpha\gamma\lambda\beta}_{\overline{e}dddD} - C^{\alpha\beta\lambda\gamma}_{\overline{e}dddD} - C^{\alpha\gamma\beta\lambda}_{\overline{e}dddD} - C^{\alpha\lambda\gamma\beta}_{\overline{e}dddD} \right) \;,
	\nonumber
	\\
	C^{(M)\alpha\beta\gamma\lambda}_{\overline{e}dddD} &=& \frac{1}{6} \left( 2C^{\alpha\beta\gamma\lambda}_{\overline{e}dddD} - C^{\alpha\lambda\beta\gamma}_{\overline{e}dddD} -  C^{\alpha\gamma\lambda\beta}_{\overline{e}dddD} - 2C^{\alpha\beta\lambda\gamma}_{\overline{e}dddD} + C^{\alpha\gamma\beta\lambda}_{\overline{e}dddD} + C^{\alpha\lambda\gamma\beta}_{\overline{e}dddD} \right) \;,
	\nonumber
	\\
	C^{(M^\prime)\alpha\beta\gamma\lambda}_{\overline{e}dddD} &=& \frac{1}{6} \left( 2C^{\alpha\beta\gamma\lambda}_{\overline{e}dddD} - C^{\alpha\lambda\beta\gamma}_{\overline{e}dddD} -  C^{\alpha\gamma\lambda\beta}_{\overline{e}dddD} + C^{\alpha\beta\lambda\gamma}_{\overline{e}dddD} + C^{\alpha\gamma\beta\lambda}_{\overline{e}dddD} - 2C^{\alpha\lambda\gamma\beta}_{\overline{e}dddD} \right) \;.
\end{eqnarray}
This indicates that when we derive the reduction relations between the Green's basis in Table~\ref{tab:green-basis} and the physical basis in Table~\ref{tab:physical-basis}, not only are operators $\Opr^{\dots}_{\dots}$ involved but also some combinations of operators $\Op^{\dots}_{\dots}$, i.e., $\{ \Op^{(A)}_{\ell HD1}, \Op^{(A)}_{\ell HD2}, \Op^{(A),(M),(M^\prime)}_{\overline{e}dddD}, \Op^{(A)}_{\overline{d}u\ell\ell D}, \Op^{(A)}_{\overline{\ell}qddD} \}$ need to be reduced into the physical ones via the EoMs given in Eq.~\eqref{eq:eom}. In Table~\ref{tab:physical-basis}, all Wilson coefficients are listed explicitly in the second column, and the number of independent operators is presented in the third column as well. The latter is consistent with those given by {\sf Basisgen}~\cite{Criado:2019ugp}, {\sf Sym2Int}~\cite{Fonseca:2019yya}, and Hilbert series~\cite{Henning:2015alf}. 
\\
\\
Before ending this section, some comments on the Green's basis and the physical basis for dim-7 operators are given in order.
\begin{itemize}
	\item The basis for dim-7 operators in Table~\ref{tab:green-basis} or Table~\ref{tab:physical-basis} is not unique. There is a different physical basis for dim-7 operators proposed in Ref.~\cite{Liao:2019tep}, where the authors picked independent degrees of freedom from $\Op^{}_{\overline{e}\ell\ell\ell H}$ and $\Op^{}_{\overline{\ell}dddH}$ by imposing limitations on flavor indices. Compared to the basis in Ref.~\cite{Liao:2019tep}, our basis in Table~\ref{tab:physical-basis} has at least three advantages. First, $\Op^{(S),(A),(M)}_{\overline{e}\ell\ell\ell H}$ and $\Op^{(M)}_{\overline{\ell}dddH}$ automatically give the right number of independent degrees of freedom without any restrictions on the flavor indices. Therefore, their flavor indices can run over all flavors. Moreover, the reduction relations between the Green's and physical bases can be established definitely and easily. Finally, the so-call flavor-blind basis~\cite{Liao:2019tep} is not necessary since one can directly make use of the non-redundant operators in Table~\ref{tab:physical-basis} to calculate relevant amplitudes.

	\item Working in the Green's basis in Table~\ref{tab:green-basis} and the physical basis in Table~\ref{tab:physical-basis}, one can achieve the reduction relations to convert redundant operators in the Green's basis to those in the physical basis with the help of EoMs in Eq.~\eqref{eq:eom}. However, this is not the whole story. Since we are working in the SMEFT up to $\mathcal{O}\left( \Lambda^{-3} \right)$, some redundant dim-6 operators can also be converted to the physical dim-7 operators thanks to the existence of the dim-5 operator. Considering that just the EoMs of lepton and Higgs doublets acquire contributions from the dim-5 operator, only the redundant dim-6 operators reduced by making use of the EoMs of lepton or Higgs doublets need to be taken into consideration, i.e., those listed in Table~\ref{tab:dim-6}. By applying the EoMs in Eq.~\eqref{eq:eom} to operators in Tables~\ref{tab:green-basis} and \ref{tab:dim-6}, we work out all reduction relations for dim-7 operators in the physics basis listed in Table~\ref{tab:physical-basis}, and present them in Appendix~\ref{app:conversion}.

	\begin{table}
		\centering
		\renewcommand\arraystretch{1.6}
		\resizebox{\textwidth}{!}{
			\begin{tabular}{lc|lc|lc}
				\hline
				\hline
				$\Opr^{}_{2W}$ & $-\frac{1}{2} \left( D^{}_\mu W^{I \mu\nu} \right) \left( D^\rho W^I_{\rho\nu} \right)$ &  $\Opr^{}_{WDH}$ & $D^{}_\nu W^{I \mu\nu} \left( H^\dagger \rmi \Dilr H \right)$ & $\Opr^{}_{DH}$ & $\left( D^{}_\mu D^\mu H \right)^\dagger \left( D^{}_\nu D^\nu H \right)$ 
				\\
				$\Opr^\prime_{HD}$ & $\left( H^\dagger H \right) \left( D^{}_\nu H \right)^\dagger \left( D^\mu H \right)$ & $\Opr^{\prime\prime}_{HD}$ & $\left( H^\dagger H \right) D^{}_\mu \left( H^\dagger \rmi \Dlr H \right)$ & $\Opr^{\alpha\beta}_{\ell D}$ & $\frac{\rmi}{2} \overline{\ell^{}_{\alpha\rm L}} \left\{ D^{}_\mu D^\mu, \slashed{D}  \right\} \ell^{}_{\beta \rm L}$ 
				\\
				$\Opr^{\alpha\beta}_{uHD1}$ & $\left( \overline{Q^{}_{\alpha\rm L}} U^{}_{\beta\rm R} \right) D^{}_\mu D^\mu \widetilde{H}$ & $\Opr^{\alpha\beta}_{uHD2}$ & $\left( \overline{Q^{}_{\alpha\rm L}} \rmi \sigma^{}_{\mu\nu} D^\mu U^{}_{\beta \rm R} \right) D^\nu \widetilde{H}$ &  $\Opr^{\alpha\beta}_{uHD4}$ & $\left( \overline{Q^{}_{\alpha\rm L}} D^\mu U^{}_{\beta \rm R} \right) D^\mu \widetilde{H}$ 
				\\
				$\Opr^{\alpha\beta}_{dHD1}$ & $\left( \overline{Q^{}_{\alpha\rm L}} D^{}_{\beta\rm R} \right) D^{}_\mu D^\mu H$ & $\Opr^{\alpha\beta}_{dHD2}$ & $\left( \overline{Q^{}_{\alpha\rm L}} \rmi \sigma^{}_{\mu\nu} D^\mu D^{}_{\beta \rm R} \right) D^\nu H$ & $\Opr^{\alpha\beta}_{dHD4}$ & $\left( \overline{Q^{}_{\alpha\rm L}} D^{}_\mu D^{}_{\beta\rm R} \right) D^\mu H$ 
				\\
				$\Opr^{\alpha\beta}_{eHD1}$ & $\left( \overline{\ell^{}_{\alpha\rm L}} E^{}_{\beta\rm R} \right) D^{}_\mu D^\mu H$ &	$\Opr^{\alpha\beta}_{eHD2}$ & $\left( \overline{\ell^{}_{\alpha\rm L}} \rmi \sigma^{}_{\mu\nu} D^\mu E^{}_{\beta \rm R} \right) D^\nu H$ & $\Opr^{\alpha\beta}_{eHD4}$ & $\left( \overline{\ell^{}_{\alpha\rm L}} D^{}_\mu E^{}_{\beta \rm R} \right) D^\mu H$ 
				\\
				$\Opr^{\prime\alpha\beta}_{W\ell}$ & $\frac{1}{2} \left( \overline{\ell^{}_{\alpha\rm L}} \sigma^I \gamma^\mu \rmi \Dlrn \ell^{}_{\beta \rm L} \right) W^I_{\mu\nu}$ & $\Opr^{\prime\alpha\beta}_{\widetilde{W}\ell}$ & $\frac{1}{2} \left( \overline{\ell^{}_{\alpha\rm L}} \sigma^I \gamma^\mu \rmi \Dlrn \ell^{}_{\beta \rm L} \right) \widetilde{W}^I_{\mu\nu}$ & $\Opr^{\prime\alpha\beta}_{B\ell}$ & $\frac{1}{2} \left( \overline{\ell^{}_{\alpha\rm L}} \gamma^\mu \rmi \Dlrn \ell^{}_{\beta \rm L} \right) B_{\mu\nu}$ 
				\\
				$\Opr^{\prime\alpha\beta}_{\widetilde{B}\ell}$ & $\frac{1}{2} \left( \overline{\ell^{}_{\alpha\rm L}} \gamma^\mu \rmi \Dlrn \ell^{}_{\beta \rm L} \right) \widetilde{B}_{\mu\nu}$ & $\Opr^{\prime(1)\alpha\beta}_{H \ell}$ & $\left( \overline{\ell^{}_{\alpha\rm L}} \rmi \Dlrs \ell^{}_{\beta \rm L} \right) \left( H^\dagger H \right)$ & $\Opr^{\prime\prime(1)\alpha\beta}_{H \ell}$ & $\left( \overline{\ell^{}_{\alpha\rm L}} \gamma^\mu \ell^{}_{\beta \rm L} \right) \partial^{}_\mu \left( H^\dagger H \right)$ 
				\\
				$\Opr^{\prime(3)\alpha\beta}_{H \ell}$ & $\left( \overline{\ell^{}_{\alpha\rm L}} \rmi \Dilrs \ell^{}_{\beta \rm L} \right) \left( H^\dagger \sigma^I H \right)$ & $\Opr^{\prime\prime(3)\alpha\beta}_{H \ell}$ & $\left( \overline{\ell^{}_{\alpha\rm L}} \sigma^I \gamma^\mu \ell^{}_{\beta \rm L} \right) D^{}_\mu \left( H^\dagger \sigma^I H \right)$ & &
				\\
				\hline
				\hline
			\end{tabular}
		}
		\caption{Dim-6 operators in the Green's basis converted to physical dim-7 operators with the help of EoMs of the lepton or Higgs doublets. The dual tensors are defined by $\widetilde{X}^{}_{\mu\nu} = \frac{1}{2} \epsilon^{}_{\mu\nu\rho\sigma} X^{\rho\sigma}$ with $\epsilon^{}_{0123} = +1$ and $X$ denoting $W^I$ and $B$.}
		\label{tab:dim-6}
	\end{table}

	\item The Green's basis in Table~\ref{tab:green-basis} and the physical basis in Table~\ref{tab:physical-basis} for dim-7 operators together with those reduction relations in Appendix~\ref{app:conversion} can not only be used to derive RGEs involving dim-7 operators in the SMEFT but also be applied for matching of some lepton-number-violating UV models onto the SMEFT up to dimension seven, such as seesaw models~\cite{Minkowski:1977sc,Yanagida:1979as,Gell-Mann:1979vob,Glashow:1979nm,Mohapatra:1979ia,Konetschny:1977bn,Magg:1980ut,Cheng:1980qt,Mohapatra:1980yp,Schechter:1980gr,Lazarides:1980nt,Foot:1988aq,Ma:1998dn}~\footnote{Recently, the complete one-loop matchings of seesaw models onto the SMEFT up to dimension six have been achieved~\cite{Zhang:2021tsq,Zhang:2021jdf,Coy:2021hyr,Ohlsson:2022hfl,Li:2022ipc,Du:2022vso,Zhang:2022osj}, which are renamed seesaw EFTs (SEFTs) to distinguish from the SMEFT. The scotogenic model~\cite{Ma:2006km} has been matched onto the SMEFT up to dimension seven at the one-loop level~\cite{Liao:2022cwh}.}. Moreover, it is possible to incorporate them into some EFT tools, such as {\sf Matchmakereft} package, which performs both one-loop matching and RGE calculations in EFTs by means of the Feynman diagrammatic approach.
\end{itemize}

\section{RGEs for the dim-5 and dim-7 operators up to $\mathcal{O}\left( \Lambda^{-3} \right)$}\label{sec:derivation}

\subsection{General structure of RGEs}\label{subsec:gsr}
The general structure of RGEs for the dim-5  and dim-7 operators can be formulated as
\begin{eqnarray}\label{eq:rge-structure}
	\befune{C^{}_5} &=& \gamma^{(5,5)} C^{}_5 + \hat{\gamma}^{(5,5)} C^{}_5 C^{}_5 C^{}_5+ \gamma^{(5,6)}_i C^{}_5 C^{i}_6 + \gamma^{(5,7)}_i C^{i}_7 \;,
	\nonumber
	\\
	\befune{C^{i}_7} &=& \gamma^{(7,7)}_{ij} C^{j}_7 + \gamma^{(7,5)}_i C^{}_5 C^{}_5 C^{}_5 + \gamma^{(7,6)}_{ij} C^{}_5 C^{j}_6 \;,
\end{eqnarray}
up to $\mathcal{O}{\left( \Lambda^{-3} \right)}$. The RGEs for dim-6 operators up to $\mathcal{O}{\left( \Lambda^{-3} \right)}$ are exactly the same as those up to $\mathcal{O}{\left( \Lambda^{-2} \right)}$, which have been achieved in Refs.~\cite{Jenkins:2013zja,Jenkins:2013wua,Alonso:2013hga,Alonso:2014zka,Davidson:2018zuo,Wang:2023bdw}. In Eq.~\eqref{eq:rge-structure}, $\gamma^{(m^{}_1,m^{}_2)}$ and $\hat{\gamma}^{(m^{}_1,m^{}_2)}$ stand for the anomalous dimension matrices or tensors for Wilson coefficients of dim-$m^{}_1$ operators resulting from dim-$m^{}_2$ operators (maybe combined with the dim-5 operator). The formulae in Eq. \eqref{eq:rge-structure} are written in a compact form with all flavor indices inexplicit, and they help a lot in understanding the general structure of RGEs and also in discussing the origination of contributions from different dimensional operators. As can be seen from Eq. \eqref{eq:rge-structure}, the RGE of the Wilson coefficient of the dim-5 operator may acquire contributions from triple insertions of the dim-5 operator (i.e., $\hat{\gamma}^{(5,5)}$), insertions of one dim-5 operator and one dim-6 operator (i.e., $\gamma^{(5,6)}$), and single insertions of dim-7 operators (i.e., $\gamma^{(5,7)}$) at $\mathcal{O} \left( \Lambda^{-3} \right)$, besides those at $\mathcal{O} \left( \Lambda^{-1} \right)$ from single insertions of the dim-5 operator (i.e., $\gamma^{(5,5)}$). Similarly, RGEs of dim-7 operators may receive contributions from triple insertions of the dim-5 operator (i.e., $\hat{\gamma}^{(7,5)}$), insertions of one dim-5 operator and one dim-6 operator (i.e., $\gamma^{(7,6)}$), in addition to those from single insertions of dim-7 operators (i.e., $\gamma^{(7,5)}$). More discussions and comments on the contributions to RGEs for the dim-5 and dim-7 operators are given below:
\begin{itemize}
	\item For the RGE of the dim-5 operator, $\gamma^{(5,5)}$ results from single insertions of the dim-5 operator and shows up at $\mathcal{O} \left( \Lambda^{-1} \right)$, while $\hat{\gamma}^{(5,5)}$, $\gamma^{(5,6)}$ and $\gamma^{(5,7)}$ appear only at $\mathcal{O} \left( \Lambda^{-3} \right)$. The former has been acquired in previous works~\cite{Babu:1993qv,Chankowski:1993tx,Antusch:2001ck} a long time ago, and the latter has been taken into  account in Ref.~\cite{Chala:2021juk} but with the down-type quark and lepton Yukawa couplings being neglected. For those of dim-7 operators, all contributions are at $\mathcal{O} \left( \Lambda^{-3} \right)$. The anomalous dimension matrix $\gamma^{(7,7)}$ has been calculated in Refs.~\cite{Liao:2016hru,Liao:2019tep}, and  the RGE of the operator $\Op^{}_{\ell H}$ including $\gamma^{(7,5)}$ and $\gamma^{(7,6)}$ is also acquired in Ref.~\cite{Chala:2021juk} with the same approximation that the down-type quark and lepton Yukawa couplings are neglected. In this work, we attempt to complete the RGEs for the dim-5 operator and also for all dim-7 operators up to $\mathcal{O} \left( \Lambda^{-3} \right)$ without any further approximation. However, we do not repeat the already achieved results for $\gamma^{(5,5)}$ and $\gamma^{(7,7)}$~\cite{Babu:1993qv,Chankowski:1993tx,Antusch:2001ck,Liao:2016hru,Liao:2019tep}. 
	
	\item $\hat{\gamma}^{(5,5)}$ and $\gamma^{(5,6)}$ have two types of possible contributions. One directly comes from 1PI diagrams involving triple insertions of the dim-5 operator (or one dim-5 operator together with one dim-6 operator) for $\hat{\gamma}^{(5,5)}$ (or $\gamma^{(5,6)}$). The other one results from 1PI diagrams containing double insertions of the dim-5 operator or single insertions of dim-6 operators by means of the SMEFT EoMs up to $\mathcal{O}\left( \Lambda^{-1} \right)$. The latter indicates that counterterms for dim-6 operators in the Green's basis are indispensable for calculating the RGEs of the dim-5 and dim-7 operators. Fortunately, not all dim-6 operators have a hand in the calculations, and only  those converted to physical operators with the help of EoMs of lepton and Higgs doublets need to be taken into account, as discussed at the end of Sec.~\ref{sec:framework}. The situation is the same for $\gamma^{(7,5)}$ and $\gamma^{(7,6)}$. All the dim-6 operators that may involve in calculations are listed in Table.~\ref{tab:dim-6}, and their contributions to Wilson coefficients of the dim-5 and dim-7 operators in the physical basis are given in Appendix~\ref{app:conversion}. However, as will be seen in the next subsection, $\hat{\gamma}^{(5,5)}$ for the dim-5 operator and $\gamma^{(7,5)}$ for dim-7 operators except $\Op^{(S)}_{\ell H}$ are vanishing, and a brief explanation for this is provided there.
	
	\item Since the dim-5 operator violates lepton number $L$ by 2 units  but preserves baryon number $B$, i.e., $\Delta L = \pm 2$ and $\Delta B = 0$, only the dim-6 operators preserving both lepton and baryon numbers ($\Delta L =\Delta B =0$) and the dim-7 operators violating lepton number while preserving baryon number ($\Delta L = \pm 2$, $\Delta B = 0$) can contribute to the RGE for the dim-5 operator. Similarly, $\gamma^{(7,5)}$ involving triple insertions of the dim-5 operator  and $\gamma^{(7,6)}$ induced by $\Delta L =\Delta B =0$ dim-6 operator can only exist for the dim-7 operators with $\Delta L = \pm 2$ and $\Delta B = 0$, and $\gamma^{(7,6)}$ resulting from $\Delta L = \Delta B = \pm 1$ dim-6 operators is only for the RGEs of $- \Delta L =\Delta B = \pm 1$ dim-7 operators.
	
	\item Up to $\mathcal{O} \left( \Lambda^{-3} \right)$, only dim-6 operators can make a contribution to wave-function renormalization constants of the SM fields. Since these corrections are already at $\mathcal{O} \left( \Lambda^{-2} \right)$, one does not need to take into consideration them for the RGEs of dim-7 operators, while they indeed contribute to $\gamma^{(5,6)}$ for the dim-5 operator. This indicates that the wave-function renormalization constants of lepton and Higgs doublets are adequate for our purpose.
\end{itemize}

According to the above discussions, besides counterterms for the dim-5 and dim-7 operators up to $\mathcal{O} \left( \Lambda^{-3} \right)$, counterterms for the dim-6 operators listed in Table~\ref{tab:dim-6} and the wave-function renormalization constants of lepton and Higgs doublets also need to be figured out for deriving the complete RGEs of the dim-5 and dim-7 operators. 

\subsection{Calculations for all counterterms} 

We adopt the background field method, dimension regularization in $d= 4 -2\varepsilon$ space-time dimensions, the modified minimal subtraction scheme, and the off-shell scheme to calculate all counterterms needed to cancel out all UV divergences in the SMEFT up to $\mathcal{O} \left( \Lambda^{-3} \right)$. With the off-shell scheme, we only need to calculate 1PI diagrams but the EoMs of relevant fields have to be applied in order to obtain final physical results. Due to the large amount of diagrams and intricate calculations, we only sketch out the main strategy we use to carry out all calculations in what follows. First, starting with the SMEFT Lagrangian in Eq.~\eqref{eq:LSMEFT}, we generate a set of 1PI diagrams in the light of field ingredients (not including gauge fields in the covariant derivative) of operators in the Green's basis so that all operators in the Green's basis can be covered by this set of 1PI diagrams. For example, to calculate counterterms for dim-7 operators in the $\psi^2 H^2 D^2$ class shown in Table~\ref{tab:green-basis} and also that for the dim-5 operator, one can generate all 1PI diagrams with external legs determined by two lepton-doublet fields and two Higgs-doublet fields. Then, we calculate all 1PI diagrams and contributions from the corresponding counterterms to work out explicit expressions of counterterms by requiring all UV divergences to be canceled out. Now, all results are in the Green's basis and we make use of the reduction relations in Appendix~\ref{app:conversion} to convert them to those in the physical basis. At last, from those results in the physical basis, the RGEs can be easily achieved by considering that all bare couplings are independent of the renormalization scale $\mu$.

For the calculations of amplitudes, we take advantage of Mathematica packages {\sf FeynRules}~\cite{Christensen:2008py,Alloul:2013bka}, {\sf FeynArts}~\cite{Hahn:2000kx} and {\sf FeynCalc}~\cite{Shtabovenko:2016sxi,Shtabovenko:2020gxv}, together with {\sf FeynHelper}~\cite{Shtabovenko:2016whf} connecting {\sf FeynCalc} to {\sf Package-X}~\cite{Patel:2015tea,Patel:2016fam}. However, {\sf FeynArts} can not deal with four-fermion vertices properly, and hence we calculate the 1PI diagrams involving four-fermion vertices by hand, where we adopt the Feynman rules in Refs.~\cite{Denner:1992me,Denner:1992vza} for the fermion-number-violating interactions. All results for counterterms and wave-function renormalization constants in the Green's basis and in the physical basis are collected in Appendix~\ref{app:counterterm-green} and Appendix~\ref{app:counterterm-phys}, respectively. Results in the Green's basis are partially crosschecked with the aid of package {\sf Matchmakereft}. Note that in Appendix~\ref{app:counterterm-green}, we only present the results for those contributing to the final results in the physical basis listed in Appendix~\ref{app:counterterm-phys}. Actually, during the calculations of counterterms in the Green's basis, some other operators are also involved and get a result for their counterterms, but they do not contribute to the desired results in Appendix~\ref{app:counterterm-phys}. For instance, to derive the counterterms for $\Opr^{\prime}_{HD}$ and $\Opr^{\prime\prime}_{HD}$ shown in Eq.~\eqref{eq:HDp}, one needs to calculate all 1PI diagrams with four Higgs-doublet legs and has to take into account the counterterms for $\Op^{}_{H\square}$ and $\Op^{}_{HD}$. In this case, not only are the counterterms for $\Opr^{\prime}_{HD}$ and $\Opr^{\prime\prime}_{HD}$ obtained, but also those for $\Op^{}_{H\square}$ and $\Op^{}_{HD}$ are achieved at the same time. The latter is only relevant to the RGEs of $C^{}_{H\square}$ and $C^{}_{HD}$ and hence not listed in Appendix~\ref{app:counterterm-green}. More discussions on those results can be found in Appendices~\ref{app:counterterm-green} and~\ref{app:counterterm-phys}. 

\subsection{Results for RGEs of the dim-5 and dim-7 operators}

Now we briefly show how to derive the RGEs from counterterms in general. The bare Wilson coefficient $C_0$ of an operator $\Op$ is related to its renormalized counterpart $C^{}_r$ via 
\begin{eqnarray}\label{eq:bare}
	C^{}_0 = \mu^{n^\prime \varepsilon} Z^{}_r C^{}_r \;,
\end{eqnarray}
where $n^\prime$ denotes the tree-level anomalous dimension of $C^{}_r$, and
\begin{eqnarray}\label{eq:rc}
	Z^{}_r = 1 - \sum^{}_\varphi \frac{1}{2} n^{}_\varphi \delta Z^{}_{\varphi} + \frac{\delta C^{}_r}{ C^{}_r}
\end{eqnarray}
with $\delta Z^{}_\varphi$ and $n^{}_\varphi$ being the wave-function renormalization constant and the number of the field $\varphi$ appearing in the corresponding operator, respectively, and $\delta C^{}_r$ being the counterterm of $C^{}_r$. Since the bare Wilson coefficient $C^{}_0$ is independent of the renormalization scale $\mu$, namely, $\mu {\rm d} C^{}_0 /{\rm d} \mu = 0 $, one can obtain the one-loop anomalous dimension of $C^{}_r$ with the help of Eq.~\eqref{eq:bare}, i.e.,
\begin{eqnarray}\label{eq:ad}
	\mu \frac{{\rm d} C^{}_r}{{\rm d} \mu} =  \varepsilon \left( \sum^{}_{i} n^\prime_i h^{}_i \frac{\partial Z^{}_r}{\partial h^{}_i}  \right) C^{}_r \;,
\end{eqnarray}
in which $h^{}_i$  and $n^\prime_i$ are the coupling or Wilson coefficient in the Lagrangian and the corresponding tree-level anomalous dimension, respectively. In Eq.~\eqref{eq:ad}, the tree-level relation $\mu {\rm d} h^{}_i/{\rm d}\mu = - \varepsilon\, n^\prime_i  h^{}_i $ has been exploited. Making use of Eqs.~\eqref{eq:rc} and \eqref{eq:ad} and taking into account the counterterms in Appendix~\ref{app:counterterm-phys}, we can achieve all RGEs for the Wilson coefficients of the dim-5 and dim-7 operators, which are presented in the following subsections. For simplicity, $\befun \equiv 16 \pi^2 \mu {\rm d} C/{\rm d} \mu$ is used in the expressions for all results.

\subsubsection{REGs for the dim-5 operator}

\begin{eqnarray} \label{eq:rge-wein5}
	\befun^{\alpha\beta}_5 &=& 2 m^2 \left\{ C^{\alpha\beta}_5 \left( 8C^{}_{H\square} - C^{}_{HD} \right) + 8C^{(S)\ast \alpha\beta}_{\ell H}  + \frac{3}{2} g^2_2 \left( 2 C^{(S)\ast \alpha\beta}_{\ell HD1} + C^{(S)\ast \alpha\beta}_{\ell HD2}  \right)  \right.
	\nonumber
	\\
	&& + \frac{1}{2} \left( Y^{}_l Y^\dagger_l  C^{(S)\dagger}_{\ell HD1} \right)^{\alpha\beta} + \frac{1}{2} \left( Y^{}_l Y^\dagger_l  C^{(S)\dagger}_{\ell HD1} \right)^{\beta\alpha}  - \frac{1}{4} \left( Y^{}_l Y^\dagger_l  C^{(S)\dagger}_{\ell HD2} \right)^{\alpha\beta}  - \frac{1}{4} \left( Y^{}_l Y^\dagger_l  C^{(S)\dagger}_{\ell HD2} \right)^{\beta\alpha} 
	\nonumber
	\\
	&& + \left( Y^{}_l C^\dagger_{\ell e HD} \right)^{\alpha\beta}  + \left( Y^{}_l C^\dagger_{\ell e HD}  \right)^{\beta\alpha}  - \left( Y^\dagger_l \right)^{}_{\gamma\lambda} \left( 3C^{(S)\ast\gamma\lambda\alpha\beta}_{\overline{e}\ell\ell\ell H}  + C^{(M)\ast\gamma\lambda\alpha\beta}_{\overline{e}\ell\ell\ell H} + C^{(M)\ast\gamma\lambda\beta\alpha}_{\overline{e}\ell\ell\ell H} \right) 
	\nonumber
	\\
	&& -  \left.  \frac{3}{2} \left( Y^{\dagger}_{\rm d} \right)^{}_{\gamma\lambda} \left( C^{\ast\gamma\alpha\lambda\beta}_{\overline{d}\ell q\ell H1}  + C^{\ast\gamma\beta\lambda\alpha}_{\overline{d}\ell q\ell H1} \right) 
	+  3 \left( Y^{}_{\rm u} \right)^{}_{\lambda\gamma} \left(  C^{\ast\lambda\gamma\alpha\beta}_{\overline{q}u\ell\ell H}  + C^{\ast\lambda\gamma\beta\alpha}_{\overline{q}u\ell\ell H}  \right)   \right\}  \;.
\end{eqnarray}

\subsubsection{REGs for dim-7 operators}

\noindent $\bullet~\bm{\psi^2H^4}$
\begin{eqnarray}\label{eq:rge-wein7}
	\befun^{(S)\alpha\beta}_{\ell H} &=&  2 C^{\ast \alpha\beta}_5 \left[ -3 C^{}_H - \frac{3}{4} \left( g^2_1 - g^2_2 + 4\lambda \right) C^{}_{HD} + \left( 16 \lambda - \frac{5}{3} g^2_2 \right) C^{}_{H \square} -3g^2_2 C^{}_{HW} \right. 
	\nonumber
	\\
	&& + \frac{3}{2} \rmi \left( g^2_1 C^{}_{H\widetilde{B}} + 3g^2_2 C^{}_{H\widetilde{W}} + g^{}_1g^{}_2 C^{}_{H\widetilde{W}B} \right) + \frac{1}{2} \tr{C^{}_5 C^\dagger_5} - \frac{2}{3} g^2_2 \tr{3 C^{(3)}_{Hq} +C^{(3)}_{H\ell} } 
	\nonumber
	\\
	&& - \tr{ C^{}_{eH} Y^\dagger_l + 3C^{}_{dH} Y^\dagger_{\rm d} + 3Y^{}_{\rm u} C^\dagger_{uH} } + 2 \tr{ Y^\dagger_l C^{(3)}_{H\ell} Y^{}_l + 3 Y^\dagger_{\rm d} C^{(3)}_{Hq} Y^{}_{\rm d} + 3 Y^\dagger_{\rm u} C^{(3)}_{Hq} Y^{}_{\rm u} } 
	\nonumber
	\\
	&& - \left. 3  \tr{ Y^{}_{\rm u} C^{}_{Hud} Y^\dagger_{\rm d} + Y^{}_{\rm d} C^\dagger_{Hud} Y^\dagger_{\rm u} } \right]  +  \frac{5}{2} \left( C^\dagger_5 C^{}_5 C^\dagger_5 \right)^{\alpha\beta} + \frac{3}{2} \left( g^2_1 + g^2_2 \right) \left[ \left( C^\dagger_5 C^{(3)}_{H\ell} \right)^{\alpha\beta} \right.
	\nonumber
	\\
	&& + \left. \left( C^\dagger_5 C^{(3)}_{H\ell} \right)^{\beta\alpha} - \left( C^\dagger_5 C^{(1)}_{H\ell} \right)^{\alpha\beta} - \left( C^\dagger_5 C^{(1)}_{H\ell} \right)^{\beta\alpha} \right] - 3 g^{}_2 \left[ \left( C^\dagger_5 Y^{}_l C^\dagger_{eW} \right)^{\alpha\beta} +  \left( C^\dagger_5 Y^{}_l C^\dagger_{eW} \right)^{\beta\alpha} \right] 
	\nonumber
	\\
	&& + \frac{1}{2} \left[ \left( C^\dagger_5 Y^{}_l C^\dagger_{eH} \right)^{\alpha\beta} +  \left( C^\dagger_5 Y^{}_l C^\dagger_{eH} \right)^{\beta\alpha} \right]  + \left[ \left( C^\dagger_5 C^{}_{eH} Y^\dagger_l  \right)^{\alpha\beta} +  \left( C^\dagger_5 C^{}_{eH} Y^\dagger_l  \right)^{\beta\alpha} \right] 
	\nonumber
	\\
	&& - 3 \left[ \left( C^\dagger_5 Y^{}_l Y^\dagger_l C^{(3)}_{H\ell} \right)^{\alpha\beta} + \left( C^\dagger_5 Y^{}_l Y^\dagger_l C^{(3)}_{H\ell} \right)^{\beta\alpha}  \right]  \;.
\end{eqnarray}

\noindent $\bullet~\bm{\psi^2H^3D}$
\begin{eqnarray}
	\befun^{\alpha\beta}_{\ell eHD} &=& \left( 2C^{}_{H\square} - \frac{3}{2} C^{}_{HD} \right) \left( C^\dagger_5 Y^{}_l \right)^{\alpha\beta} + 3g^{}_1 \left( C^\dagger_5 C^{}_{eB} \right)^{\alpha\beta} +  3 g^{}_2 \left( C^\dagger_5 C^{}_{eW} \right)^{\alpha\beta} + \left( C^\dagger_5 C^{}_{eH} \right)^{\alpha\beta} 
	\nonumber
	\\
	&& + \left[ \left( C^\dagger_5 C^{(1)}_{H\ell} \right)^{\alpha\gamma} - 2 \left( C^\dagger_5 C^{(1)}_{H\ell} \right)^{\gamma\alpha} -5 \left( C^\dagger_5 C^{(3)}_{H\ell} \right)^{\alpha\gamma} - 8 \left( C^\dagger_5 C^{(3)}_{H\ell} \right)^{\gamma\alpha} \right] \left(Y^{}_l \right)^{}_{\gamma\beta}  
	\nonumber
	\\
	&& - 3 \left( C^\dagger_5 Y^{}_l C^{}_{He} \right)^{\alpha\beta} - 4 \left( C^\dagger_5 Y^{}_l \right)^{\gamma\lambda} C^{\gamma\alpha\lambda\beta}_{\ell e} \;.
\end{eqnarray}

\noindent$\bullet~\bm{\psi^2H^2 D^2}$
\begin{eqnarray}
	\befun^{(S)\alpha\beta}_{\ell HD1} &=&  C^{\ast \alpha\beta}_5 \left( C^{}_{HD} + 2 C^{}_{H\square} \right) - 2\left( C^\dagger_5 C^{(1)}_{H\ell} \right)^{\alpha\beta} - 2\left( C^\dagger_5 C^{(1)}_{H\ell} \right)^{\beta\alpha} - 2 \left( C^\dagger_5 C^{(3)}_{H\ell} \right)^{\alpha\beta} 
	\nonumber
	\\
	&& - 2\left( C^\dagger_5 C^{(3)}_{H\ell} \right)^{\beta\alpha} + 8C^{\ast\gamma\lambda}_5 C^{\gamma\alpha\lambda\beta}_{\ell\ell} \;,
	\\
	\befun^{(S)\alpha\beta}_{\ell HD2} &=& -2C^{\ast \alpha\beta}_5 \left( C^{}_{HD} + 2 C^{}_{H\square} \right) + 4 \left( C^\dagger_5 C^{(1)}_{H\ell} \right)^{\alpha\beta} + 4 \left( C^\dagger_5 C^{(1)}_{H\ell} \right)^{\beta\alpha} - 8 \left( C^\dagger_5 C^{(3)}_{H\ell} \right)^{\alpha\beta} 
	\nonumber
	\\
	&& - 8 \left( C^\dagger_5 C^{(3)}_{H\ell} \right)^{\beta\alpha} - 16 C^{\ast\gamma\lambda}_5 C^{\gamma\alpha\lambda\beta}_{\ell\ell} \;.
\end{eqnarray}

\noindent$\bullet~\bm{\psi^2H^2X}$
\begin{eqnarray}
	\befun^{(A)\alpha\beta}_{\ell HB} &=& \frac{1}{2}  \left[  \left( C^\dagger_5 Y^{}_l C^\dagger_{eB} \right)^{\alpha\beta} -  \left( C^\dagger_5 Y^{}_l C^\dagger_{eB} \right)^{\beta\alpha} \right] \;,
	\\
	\befun^{\alpha\beta}_{\ell HW} &=& \frac{3}{2} g^{}_2 \left[  \left( C^\dagger_5 C^{(3)}_{H\ell } \right)^{\alpha\beta} + \left( C^\dagger_5 C^{(3)}_{H\ell } \right)^{\beta\alpha}  \right]  - \left( C^\dagger_5 Y^{}_l C^\dagger_{eW} \right)^{\alpha\beta} + 2 g^{}_2 \left( C^{}_{HW} - \rmi C^{}_{H\widetilde{W}} \right) C^{\ast\alpha\beta}_5
	\nonumber
	\\
	&&  - \frac{3}{2} g^2_2 \left( C^{}_{3W} - \rmi C^{}_{3\widetilde{W}} \right) C^{\ast\alpha\beta}_5 \;.
\end{eqnarray}

\noindent$\bullet~\bm{\psi^4H}$
\begin{eqnarray}
	\befun^{(S)\alpha\beta\gamma\lambda}_{\overline{e}\ell\ell\ell H} &=& -4 g^{}_2 \left[ \left( C^\dagger_{eW} \right)^{\alpha\beta} \left( C^\dagger_5 \right)^{\gamma\lambda} + \left( C^\dagger_{eW} \right)^{\alpha\lambda} \left( C^\dagger_5 \right)^{\beta\gamma} + \left( C^\dagger_{eW} \right)^{\alpha\gamma} \left( C^\dagger_5 \right)^{\lambda\beta} \right] 
	\nonumber
	\\
	&& + \frac{2}{3} C^{\alpha\rho}_{He} \left[ \left( Y^\dagger_l \right)^{}_{\rho\beta} \left( C^\dagger_5 \right)^{\gamma\lambda} + \left( Y^\dagger_l \right)^{}_{\rho\lambda} \left( C^\dagger_5 \right)^{\beta\gamma} + \left( Y^\dagger_l \right)^{}_{\rho\gamma} \left( C^\dagger_5 \right)^{\lambda\beta} \right] 
	\nonumber
	\\
	&&  - \frac{2}{3} \left( Y^\dagger_l \right)^{}_{\alpha\rho} \left[ \left( C^\dagger_5 \right)^{\beta\gamma} \left( C^{(1)}_{H\ell} +  C^{(3)}_{H\ell}  \right)^{\rho\lambda} +  \left( C^\dagger_5 \right)^{\gamma\lambda} \left( C^{(1)}_{H\ell} +  C^{(3)}_{H\ell}  \right)^{\rho\beta} \right. 
	\nonumber
	\\
	&& + \left. \left( C^\dagger_5 \right)^{\lambda\beta} \left( C^{(1)}_{H\ell} +  C^{(3)}_{H\ell}  \right)^{\rho\gamma} \right]  - \frac{4}{3} \left( Y^\dagger_l \right)^{}_{\alpha\sigma} \left[ \left( C^\dagger_5 \right)^{\beta\rho} \left( C^{\rho\gamma\sigma\lambda}_{\ell\ell} + C^{\sigma\gamma\rho\lambda}_{\ell\ell} \right)  \right.
	\nonumber
	\\
	&& + \left. \left( C^\dagger_5 \right)^{\lambda\rho} \left( C^{\rho\beta\sigma\gamma}_{\ell\ell} + C^{\sigma\beta\rho\gamma}_{\ell\ell} \right) + \left( C^\dagger_5 \right)^{\gamma\rho} \left( C^{\rho\lambda\sigma\beta}_{\ell\ell} + C^{\sigma\lambda\rho\beta}_{\ell\ell} \right) \right] + \frac{2}{3} C^{\sigma\gamma\alpha\rho}_{\ell e} 
	\nonumber
	\\
	&& \times \left[ \left( C^\dagger_5 \right)^{\beta\sigma} \left( Y^\dagger_l \right)^{}_{\rho\lambda} + \left( C^\dagger_5 \right)^{\lambda\sigma} \left( Y^\dagger_l \right)^{}_{\rho\beta} \right] + \frac{2}{3} C^{\sigma\beta\alpha\rho}_{\ell e} \left[ \left( C^\dagger_5 \right)^{\lambda\sigma} \left( Y^\dagger_l \right)^{}_{\rho\gamma} \right.
	\nonumber
	\\
	&& + \left. \left( C^\dagger_5 \right)^{\gamma\sigma} \left( Y^\dagger_l \right)^{}_{\rho\lambda} \right] + \frac{2}{3} C^{\sigma\lambda\alpha\rho}_{\ell e} \left[ \left( C^\dagger_5 \right)^{\gamma\sigma} \left( Y^\dagger_l \right)^{}_{\rho\beta} + \left( C^\dagger_5 \right)^{\beta\sigma} \left( Y^\dagger_l \right)^{}_{\rho\gamma} \right] 
	\nonumber
	\\
	&& + \frac{2}{3} \left[  \left(C^\dagger_{eH} \right)^{\alpha\beta} \left( C^\dagger_5 \right)^{\gamma\lambda} + \left(C^\dagger_{eH} \right)^{\alpha\gamma} \left( C^\dagger_5 \right)^{\lambda\beta} + \left(C^\dagger_{eH} \right)^{\alpha\lambda} \left( C^\dagger_5 \right)^{\beta\gamma} \right]  
	\nonumber
	\\
	&& - \frac{2}{3} \left( C^\dagger_5 \right)^{\gamma\lambda} \left[ 2 C^{\ast \beta \rho\sigma\alpha}_{\ell e} \left( Y^\dagger_l \right)^{}_{\sigma\rho} + 3 C^{(1)\ast\beta\alpha\rho\sigma}_{\ell equ} \left( Y^{}_{\rm u}  \right)^{}_{\rho\sigma} - 3 C^{\ast\beta\alpha\rho\sigma}_{\ell e d q} \left( Y^\dagger_{\rm d} \right)^{}_{\rho\sigma} \right]
	\nonumber
	\\
	&& - \frac{2}{3} \left( C^\dagger_5 \right)^{\beta\gamma} \left[ 2 C^{\ast \lambda\rho\sigma\alpha}_{\ell e} \left( Y^\dagger_l \right)^{}_{\sigma\rho} + 3 C^{(1)\ast\lambda\alpha\rho\sigma}_{\ell equ} \left( Y^{}_{\rm u}  \right)^{}_{\rho\sigma} - 3 C^{\ast\lambda\alpha\rho\sigma}_{\ell e d q} \left( Y^\dagger_{\rm d} \right)^{}_{\rho\sigma} \right]  
	\nonumber
	\\
	&& - \frac{2}{3} \left( C^\dagger_5 \right)^{\lambda\beta} \left[ 2 C^{\ast \gamma \rho\sigma\alpha}_{\ell e} \left( Y^\dagger_l \right)^{}_{\sigma\rho} + 3 C^{(1)\ast\gamma\alpha\rho\sigma}_{\ell equ} \left( Y^{}_{\rm u}  \right)^{}_{\rho\sigma} - 3 C^{\ast\gamma\alpha\rho\sigma}_{\ell e d q} \left( Y^\dagger_{\rm d} \right)^{}_{\rho\sigma} \right] 
	\nonumber
	\\
	&&  - \frac{2}{3} \left( Y^\dagger_l \right)^{}_{\alpha\beta} \left[ C^{\ast\gamma\lambda}_5 C^{}_{H\square} - \left( C^\dagger_5 C^{(1)}_{H\ell} \right)^{\gamma\lambda} - \left( C^\dagger_5 C^{(1)}_{H\ell} \right)^{\lambda\gamma} + 2  \left( C^\dagger_5 C^{(3)}_{H\ell} \right)^{\gamma\lambda} \right.
	\nonumber
	\\
	&& + \left. 2 \left( C^\dagger_5 C^{(3)}_{H\ell} \right)^{\lambda\gamma}  \right]  - \frac{2}{3} \left( Y^\dagger_l \right)^{}_{\alpha\lambda} \left[ C^{\ast\beta\gamma}_5 C^{}_{H\square} - \left( C^\dagger_5 C^{(1)}_{H\ell} \right)^{\beta\gamma} - \left( C^\dagger_5 C^{(1)}_{H\ell} \right)^{\gamma\beta} \right.
	\nonumber
	\\
	&& + \left. 2  \left( C^\dagger_5 C^{(3)}_{H\ell} \right)^{\beta\gamma} + 2 \left( C^\dagger_5 C^{(3)}_{H\ell} \right)^{\gamma\beta}  \right]  - \frac{2}{3} \left( Y^\dagger_l \right)^{}_{\alpha\gamma} \left[ C^{\ast\lambda\beta}_5 C^{}_{H\square} - \left( C^\dagger_5 C^{(1)}_{H\ell} \right)^{\lambda\beta} \right.
	\nonumber
	\\
	&& - \left. \left( C^\dagger_5 C^{(1)}_{H\ell} \right)^{\beta\lambda} + 2  \left( C^\dagger_5 C^{(3)}_{H\ell} \right)^{\lambda\beta} + 2 \left( C^\dagger_5 C^{(3)}_{H\ell} \right)^{\beta\lambda}  \right]   \;,\quad
	\\
	\befun^{(A)\alpha\beta\gamma\lambda}_{\overline{e}\ell\ell\ell H} &=&  4 \left( Y^\dagger_l \right)^{}_{\alpha\sigma} \left[ \left( C^\dagger_5 \right)^{\beta\rho} \left( C^{\rho\gamma\sigma\lambda}_{\ell\ell} - C^{\sigma\gamma\rho\lambda}_{\ell\ell} \right) + \left( C^\dagger_5 \right)^{\lambda\rho} \left( C^{\rho\beta\sigma\gamma}_{\ell\ell} - C^{\sigma\beta\rho\gamma}_{\ell\ell} \right) \right.
	\nonumber
	\\
	&& + \left.  \left( C^\dagger_5 \right)^{\gamma\rho} \left( C^{\rho\lambda\sigma\beta}_{\ell\ell} - C^{\sigma\lambda\rho\beta}_{\ell\ell} \right) \right] + 2C^{\sigma\gamma\alpha\rho}_{\ell e} \left[ \left( C^\dagger_5 \right)^{\beta\sigma} \left( Y^\dagger_l \right)^{}_{\rho\lambda} - \left( C^\dagger_5 \right)^{\lambda\sigma} \left( Y^\dagger_l \right)^{}_{\rho\beta} \right] 
	\nonumber
	\\
	&& + 2C^{\sigma\beta\alpha\rho}_{\ell e} \left[ \left( C^\dagger_5 \right)^{\lambda\sigma} \left( Y^\dagger_l \right)^{}_{\rho\gamma} - \left( C^\dagger_5 \right)^{\gamma\sigma} \left( Y^\dagger_l \right)^{}_{\rho\lambda} \right]  + 2C^{\sigma\lambda\alpha\rho}_{\ell e} \left[ \left( C^\dagger_5 \right)^{\gamma\sigma} \left( Y^\dagger_l \right)^{}_{\rho\beta} \right.
	\nonumber
	\\
	&& - \left. \left( C^\dagger_5 \right)^{\beta\sigma} \left( Y^\dagger_l \right)^{}_{\rho\gamma} \right] \;,
	\\
	\befun^{(M)\alpha\beta\gamma\lambda}_{\overline{e}\ell\ell\ell H} &=& 6 g^{}_2 \left[ \left( C^\dagger_{eW} \right)^{\alpha\beta} \left( C^\dagger_5 \right)^{\gamma\lambda} - \left( C^\dagger_{eW} \right)^{\alpha\lambda} \left( C^\dagger_5 \right)^{\beta\gamma} \right] + C^{\alpha\rho}_{He} \left[ \left( Y^\dagger_l \right)^{}_{\rho\beta} \left( C^\dagger_5 \right)^{\gamma\lambda} \right.
	\nonumber
	\\
	&& - \left. \left( Y^\dagger_l \right)^{}_{\rho\lambda} \left( C^\dagger_5 \right)^{\beta\gamma} \right] + \left( Y^\dagger_l \right)^{}_{\alpha\rho} \left[ \left( C^\dagger_5 \right)^{\beta\gamma} \left( C^{(1)}_{H\ell} +  C^{(3)}_{H\ell}  \right)^{\rho\lambda} -  \left( C^\dagger_5 \right)^{\gamma\lambda} \left( C^{(1)}_{H\ell} +  C^{(3)}_{H\ell}  \right)^{\rho\beta} \right]
	\nonumber
	\\
	&& - 4 \left( Y^\dagger_l \right)^{}_{\alpha\sigma} \left[ \left( C^\dagger_5 \right)^{\beta\rho}   C^{\sigma\gamma\rho\lambda}_{\ell\ell} - \left( C^\dagger_5 \right)^{\lambda\rho}  C^{\rho\beta\sigma\gamma}_{\ell\ell} + \left( C^\dagger_5 \right)^{\gamma\rho} \left( C^{\rho\lambda\sigma\beta}_{\ell\ell} - C^{\sigma\lambda\rho\beta}_{\ell\ell} \right) \right] 
	\nonumber
	\\
	&& - 2C^{\sigma\gamma\alpha\rho}_{\ell e} \left[ \left( C^\dagger_5 \right)^{\beta\sigma} \left( Y^\dagger_l \right)^{}_{\rho\lambda} - \left( C^\dagger_5 \right)^{\lambda\sigma} \left( Y^\dagger_l \right)^{}_{\rho\beta} \right] + C^{\sigma\beta\alpha\rho}_{\ell e} \left( C^\dagger_5 \right)^{\lambda\sigma} \left( Y^\dagger_l \right)^{}_{\rho\gamma} 
	\nonumber
	\\
	&& - 2C^{\sigma\lambda\alpha\rho}_{\ell e} \left( C^\dagger_5 \right)^{\beta\sigma} \left( Y^\dagger_l \right)^{}_{\rho\gamma} + \frac{1}{2} \left[  \left(C^\dagger_{eH} \right)^{\alpha\beta} \left( C^\dagger_5 \right)^{\gamma\lambda} - \left(C^\dagger_{eH} \right)^{\alpha\lambda} \left( C^\dagger_5 \right)^{\beta\gamma} \right]  
	\nonumber
	\\
	&& - \left( C^\dagger_5 \right)^{\gamma\lambda} \left[ 2 C^{\ast \beta \rho\sigma\alpha}_{\ell e} \left( Y^\dagger_l \right)^{}_{\sigma\rho} + 3 C^{(1)\ast\beta\alpha\rho\sigma}_{\ell equ} \left( Y^{}_{\rm u}  \right)^{}_{\rho\sigma} - 3 C^{\ast\beta\alpha\rho\sigma}_{\ell e d q} \left( Y^\dagger_{\rm d} \right)^{}_{\rho\sigma} \right]
	\nonumber
	\\
	&& + \left( C^\dagger_5 \right)^{\beta\gamma} \left[ 2 C^{\ast \lambda\rho\sigma\alpha}_{\ell e} \left( Y^\dagger_l \right)^{}_{\sigma\rho} + 3 C^{(1)\ast\lambda\alpha\rho\sigma}_{\ell equ} \left( Y^{}_{\rm u}  \right)^{}_{\rho\sigma} - 3 C^{\ast\lambda\alpha\rho\sigma}_{\ell e d q} \left( Y^\dagger_{\rm d} \right)^{}_{\rho\sigma} \right]  
	\nonumber
	\\
	&&  - \left( Y^\dagger_l \right)^{}_{\alpha\beta} \left[ C^{\ast\gamma\lambda}_5 C^{}_{H\square} - \left( C^\dagger_5 C^{(1)}_{H\ell} \right)^{\gamma\lambda} - \left( C^\dagger_5 C^{(1)}_{H\ell} \right)^{\lambda\gamma} + 2  \left( C^\dagger_5 C^{(3)}_{H\ell} \right)^{\gamma\lambda} \right.
	\nonumber
	\\
	&& + \left. 2 \left( C^\dagger_5 C^{(3)}_{H\ell} \right)^{\lambda\gamma}  \right]  + \left( Y^\dagger_l \right)^{}_{\alpha\lambda} \left[ C^{\ast\beta\gamma}_5 C^{}_{H\square} - \left( C^\dagger_5 C^{(1)}_{H\ell} \right)^{\beta\gamma} - \left( C^\dagger_5 C^{(1)}_{H\ell} \right)^{\gamma\beta} \right.
	\nonumber
	\\
	&& + \left. 2  \left( C^\dagger_5 C^{(3)}_{H\ell} \right)^{\beta\gamma} + 2 \left( C^\dagger_5 C^{(3)}_{H\ell} \right)^{\gamma\beta}  \right] \;,
	\\
	\befun^{\alpha\beta\gamma\lambda}_{\overline{d}\ell q \ell H 1} &=&   - 4 \left( Y^\dagger_{\rm d} \right)^{}_{\alpha\rho} \left( C^\dagger_5 \right)^{\beta\lambda} \left( C^{(1)}_{Hq} + C^{(3)}_{Hq} \right)^{\rho\gamma}  - 8 \left( Y^\dagger_{\rm d} \right)^{}_{\alpha\sigma}  \left( C^\dagger_5 \right)^{\beta\rho} \left( C^{(1)}_{\ell q} - 2 C^{(3)}_{\ell q} \right)^{\rho\lambda\sigma\gamma} 
	\nonumber
	\\
	&& + 4 \left( Y^\dagger_{\rm d} \right)^{}_{\rho\gamma}  \left( C^\dagger_5 \right)^{\beta\lambda} C^{\alpha\rho}_{Hd}   - 2C^{\rho\sigma\alpha\gamma}_{\ell e d q} \left[ \left( Y^\dagger_l \right)^{}_{\sigma\beta}  \left( C^\dagger_5 \right)^{\rho\lambda} + \left( Y^\dagger_l \right)^{}_{\sigma\lambda}  \left( C^\dagger_5 \right)^{\rho\beta} \right]
	\nonumber
	\\
	&& + 8 \left( Y^\dagger_{\rm d} \right)^{}_{\sigma\gamma}  \left( C^\dagger_5 \right)^{\rho\lambda} C^{\rho\beta\alpha\sigma}_{\ell d} + 4 \left( C^\dagger_{dH} \right)^{\alpha\gamma} \left( C^\dagger_5\right)^{\beta\lambda} + 2C^{\ast\beta\lambda}_5 \left[ 2 \left( Y^\dagger_l \right)^{}_{\rho\sigma} C^{\sigma\rho\alpha\gamma}_{\ell edq} \right.
	\nonumber
	\\
	&& - 4 \left( Y^\dagger_{\rm d} \right)^{}_{\rho \sigma} \left(C^{(1)\ast \gamma\sigma\rho\alpha}_{qd} + \frac{4}{3} C^{(8)\ast\gamma\sigma\rho\alpha}_{qd} \right)  + \left( Y^{}_{\rm u} \right)^{}_{\rho\sigma} \left( 6 C^{(1)\ast\rho\sigma\gamma\alpha}_{quqd} + C^{(1)\ast\gamma\sigma\rho\alpha}_{quqd} \right.
	\nonumber
	\\
	&& + \left.\left. \frac{4}{3} C^{(8)\ast\gamma\sigma\rho\alpha}_{quqd} \right)  \right] - 4 \left( Y^\dagger_{\rm d} \right)^{}_{\alpha\gamma} \left[ C^{\ast\beta\lambda}_5 C^{}_{H\square} - \left( C^\dagger_5 C^{(1)}_{H\ell} \right)^{\beta\lambda} - \left( C^\dagger_5 C^{(1)}_{H\ell} \right)^{\lambda\beta} \right.
	\nonumber
	\\
	&& +  \left. 2 \left( C^\dagger_5 C^{(3)}_{H\ell} \right)^{\beta\lambda} + 2\left( C^\dagger_5 C^{(3)}_{H\ell} \right)^{\lambda\beta} \right]  \;,
	\\
	\befun^{\alpha\beta\gamma\lambda}_{\overline{d}\ell q \ell H 2} &=&  -12g^{}_2 \left( C^\dagger_{dW} \right)^{\alpha\gamma} \left( C^\dagger_5 \right)^{\beta\lambda} + 2\left( Y^\dagger_{\rm d} \right)^{}_{\alpha\rho} \left( C^\dagger_5 \right)^{\beta\lambda} \left( C^{(1)}_{Hq} + C^{(3)}_{Hq} \right)^{\rho\gamma} 
	\nonumber
	\\
	&& - 2 \left( Y^\dagger_{\rm d} \right)^{}_{\rho\gamma}  \left( C^\dagger_5 \right)^{\beta\lambda} C^{\alpha\rho}_{Hd}  + 4 \left( Y^\dagger_{\rm d} \right)^{}_{\alpha\sigma}  \left( C^\dagger_5 \right)^{\beta\rho} \left( C^{(1)}_{\ell q} - 5 C^{(3)}_{\ell q} \right)^{\rho\lambda\sigma\gamma}  
	\nonumber
	\\
	&& - 4 \left( Y^\dagger_{\rm d} \right)^{}_{\sigma\gamma}  \left( C^\dagger_5 \right)^{\rho\lambda} C^{\rho\beta\alpha\sigma}_{\ell d}  - 2C^{\rho\sigma\alpha\gamma}_{\ell e d q} \left[ \left( Y^\dagger_l \right)^{}_{\sigma\beta}  \left( C^\dagger_5 \right)^{\rho\lambda} - 2 \left( Y^\dagger_l \right)^{}_{\sigma\lambda}  \left( C^\dagger_5 \right)^{\rho\beta} \right] 
	\nonumber
	\\
	&& - 2 \left( C^\dagger_{dH} \right)^{\alpha\gamma} \left( C^\dagger_5\right)^{\beta\lambda} -  C^{\ast\beta\lambda}_5 \left[ 2 \left( Y^\dagger_l \right)^{}_{\rho\sigma} C^{\sigma\rho\alpha\gamma}_{\ell edq} - 4 \left( Y^\dagger_{\rm d} \right)^{}_{\rho \sigma} \left(C^{(1)\ast \gamma\sigma\rho\alpha}_{qd} + \frac{4}{3} C^{(8)\ast\gamma\sigma\rho\alpha}_{qd} \right) \right.
	\nonumber
	\\
	&&  + \left. \left( Y^{}_{\rm u} \right)^{}_{\rho\sigma} \left( 6 C^{(1)\ast\rho\sigma\gamma\alpha}_{quqd} + C^{(1)\ast\gamma\sigma\rho\alpha}_{quqd} +  \frac{4}{3} C^{(8)\ast\gamma\sigma\rho\alpha}_{quqd} \right)  \right] + 2 \left( Y^\dagger_{\rm d} \right)^{}_{\alpha\gamma} \left[ C^{\ast\beta\lambda}_5 C^{}_{H\square} \right.
	\nonumber
	\\
	&& - \left. \left( C^\dagger_5 C^{(1)}_{H\ell} \right)^{\beta\lambda} - \left( C^\dagger_5 C^{(1)}_{H\ell} \right)^{\lambda\beta} + 2 \left( C^\dagger_5 C^{(3)}_{H\ell} \right)^{\beta\lambda} + 2\left( C^\dagger_5 C^{(3)}_{H\ell} \right)^{\lambda\beta} \right]  \;,
	\\
	\befun^{\alpha\beta\gamma\lambda}_{\overline{d}\ell u e H} &=&  -6 \left( C^\dagger_{Hud} \right)^{\alpha\gamma} \left( C^\dagger_5 \right)^{\beta\rho} \left( Y^{}_l \right)^{}_{\rho\lambda} - 3 \left( C^\dagger_5 \right)^{\beta\sigma} \left( Y^{}_{\rm u} \right)^{}_{\rho\gamma} C^{\sigma\lambda\alpha\rho}_{\ell e d q} + 3 \left( Y^\dagger_{\rm d} \right)^{}_{\alpha\rho} \left( C^\dagger_5 \right)^{\beta\sigma}
	\nonumber
	\\
	&& \times  \left( C^{(1)}_{\ell e qu} - 12 C^{(3)}_{\ell e q u} \right)^{\sigma\lambda\rho\gamma} \;,
	\\
	\befun^{\alpha\beta\gamma\lambda}_{\overline{q}u\ell\ell H} &=&  - 2 \left( C^{(1)}_{Hq} - C^{(3)}_{Hq} \right)^{\alpha\rho} \left( Y^{}_{\rm u} \right)^{}_{\rho \beta} \left( C^\dagger_5 \right)^{\gamma\lambda} + 2\left( Y^{}_{\rm u} \right)^{}_{\alpha\rho} C^{\rho\beta}_{Hu} \left( C^\dagger_5 \right)^{\gamma\lambda} 
	\nonumber
	\\
	&& - 2\left( Y^{}_{\rm u} \right)^{}_{\rho \beta} \left[ \left( C^{(1)}_{\ell q} + 5 C^{(3)}_{\ell q} \right)^{\sigma\gamma\alpha\rho} \left( C^\dagger_5 \right)^{\sigma\lambda} + \left( C^{(1)}_{\ell q} - C^{(3)}_{\ell q} \right)^{\sigma\lambda\alpha\rho} \left( C^\dagger_5 \right)^{\sigma\gamma}  \right]
	\nonumber
	\\
	&& + 2 \left( Y^{}_{\rm u} \right)^{}_{\alpha\rho} \left[ \left( C^\dagger_5 \right)^{\lambda\sigma} C^{\sigma\gamma\rho\beta}_{\ell u} +   \left( C^\dagger_5 \right)^{\gamma\sigma} C^{\sigma\lambda\rho\beta}_{\ell u}  \right] + 2 C^{(1)\sigma\rho\alpha\beta}_{\ell e q u} \left[ \left( C^\dagger_5 \right)^{\sigma\gamma} \left( Y^\dagger_l \right)^{}_{\rho\lambda} \right.
	\nonumber
	\\
	&& - \left. 2 \left( C^\dagger_5 \right)^{\sigma\lambda} \left( Y^\dagger_l \right)^{}_{\rho\gamma} \right]  - 2C^{\alpha\beta}_{uH} \left( C^\dagger_5 \right)^{\gamma\lambda}   + C^{\ast\gamma\lambda}_5 \left[ 2 C^{(1)\rho\sigma\alpha\beta}_{\ell e qu} \left( Y^{\dagger}_l \right)^{}_{\sigma\rho}  \right.
	\nonumber
	\\
	&& +  4 \left( C^{(1)\alpha\rho\sigma\beta}_{qu} + \frac{4}{3} C^{(8)\alpha\rho\sigma\beta}_{qu} \right) \left( Y^{}_{\rm u} \right)^{}_{\rho\sigma}  - \left( Y^{\dagger}_{\rm d} \right)^{}_{\rho\sigma} \left( 6C^{(1)\alpha\beta\sigma\rho}_{quqd} + C^{(1)\sigma\beta\alpha\rho}_{quqd} \right.
	\nonumber
	\\
	&& + \left.\left. \frac{4}{3} C^{(8)\sigma\beta\alpha\rho}_{quqd} \right) \right] + 2\left( Y^{}_{\rm u} \right)^{}_{\alpha\beta} \left[ C^{\ast\gamma\lambda}_5 C^{}_{H\square} - \left( C^\dagger_5 C^{(1)}_{H\ell} \right)^{\gamma\lambda} - \left( C^\dagger_5 C^{(1)}_{H\ell} \right)^{\lambda\gamma} \right.
	\nonumber
	\\
	&&  + \left. 2 \left( C^\dagger_5 C^{(3)}_{H\ell} \right)^{\gamma\lambda} + 2 \left( C^\dagger_5 C^{(3)}_{H\ell} \right)^{\lambda\gamma}  \right] \;,
	\\
	\befun^{\alpha\beta\gamma\lambda}_{\overline{\ell}dud\widetilde{H}} &=& -  6 C^{\alpha\rho}_5 \left( Y^{}_{\rm d} \right)^{}_{\sigma\beta} C^{\lambda\gamma\sigma\rho}_{duq} \;,
	\\
	\befun^{\alpha\beta\gamma\lambda}_{\overline{\ell}dqq\widetilde{H}} &=& - 2C^{\alpha\rho}_5 \left( Y^\dagger_{\rm u} \right)^{}_{\sigma\gamma} C^{\beta\sigma\lambda\rho}_{duq} -  C^{\alpha\rho}_5 \left( Y^\dagger_{\rm u} \right)^{}_{\sigma\lambda} C^{\beta\sigma\gamma\rho}_{duq} + C^{\alpha\rho}_5 \left( Y^{}_{\rm d} \right)^{}_{\sigma\beta} 
	\nonumber
	\\
	&& \times \left( 9 C^{(S)\gamma\lambda\sigma\rho}_{qqq} - 3 C^{(A)\gamma\lambda\sigma\rho}_{qqq} + 8 C^{(M)\gamma\lambda\sigma\rho}_{qqq} + 10 C^{(M)\sigma\gamma\lambda\rho}_{qqq}  \right) \vphantom{\frac{1}{2}} \;.\label{eq:rge-ldqqH}
\end{eqnarray}
In Eq.~\eqref{eq:rge-ldqqH}, $C^{(S),(A),(M)}_{qqq}$ are the Wilson coefficients of the dim-6 operators $\Op^{(S),(A),(M)}_{qqq}$, whose definitions together with an explanation can be found in Eq.~\eqref{eq:Oqqq} and the text around. Finally, some comments on the above results and possible applications are briefly given in order.
\begin{itemize}
	\item For the RGE of the dim-5 operator in Eq.~\eqref{eq:rge-wein5}, all contributions are proportional to $m^2$ as expected, and no $C^{}_5$ cubic contribution exists, i.e., $\hat{\gamma}^{(5,5)} = 0$. The reason for the latter is that there is no one-loop diagram with two lepton-doublet and two Higgs-doublet legs involving triple insertions of the dim-5 operator. At the one-loop level, triple insertions of the dim-5 operator lead to at least six (Higgs- and lepton-doublet) external lines. This is also the reason why only $C^{(S)}_{\ell H}$ can acquire such a $C^{}_5$ cubic contribution, as shown in Eq.~\eqref{eq:rge-wein7}. Other dim-7 operators consist of five fields at most, where gauge fields in the covariant derivative are not counted. The $C^{}_5$ cubic contributions to $C^{(S)}_{\ell H}$ are not made directly by the 1PI diagrams with triple insertions of the dim-5 operator, but by those with double insertions of the dim-5 operator, which contribute to $\Opr^{\prime}_{HD}$, $\Opr^{\prime(1)}_{H\ell}$ and $\Opr^{\prime(3)}_{H\ell}$, and then indirectly to $C^{(S)}_{\ell H}$ via the reduction relation in Eq.~\eqref{eq:convertion-clh} (equivalently by reducible diagrams with triple insertions of the dim-5 operator).
	
	\item The above results may shed light on the non-linear non-renormalization theorem, which can predict zero entries in the anomalous dimension matrices or tenors induced by the mixing among different dimensional operators~\footnote{The non-renormalization theorem for the mixing between the same dimensional operators have been studied and established in Refs.~\cite{Elias-Miro:2014eia,Cheung:2015aba,Bern:2019wie}.}. Very recently, an attempt on such a non-linear non-renormalization theorem has been made in Ref.~\cite{Cao:2023adc}, but the results are still preliminary. More efforts need to be made to improve the present results and to get a more promising and general theorem. In this case, the above results may give some hints and also work as a specific and helpful example to test the theorem.
	
	\item If both $\Op^{}_5$ and $\Op^{(S)}_{\ell H}$ which can generate neutrino masses after spontaneous gauge symmetry breaking are absent at the tree level, they may be generated by other dim-7 operators, e.g., $\Op^{}_{\overline{d}\ell q\ell H1}$, and hence neutrinos can acquire non-zero masses radiatively. Some similar but slightly different studies on radiative neutrino masses can be found in Refs.~\cite{Babu:2001ex,deGouvea:2007qla} (see e.g., Ref~\cite{Cai:2017jrq} for a comprehensive review).  On the other hand, one can make use of the results in Eqs.~\eqref{eq:rge-wein5} and \eqref{eq:rge-wein7} to discuss RG-running corrections to neutrino masses in the SMEFT~\cite{Chala:2021juk}, namely
	\begin{eqnarray}\label{eq:neutrino-mass}
		\delta M^{\alpha\beta}_\nu 	&=& - \frac{v^2}{32\pi^2} \ln\left( \frac{v}{\Lambda} \right) \left\{ C^{\alpha\beta}_5 \left[ - 3 g^2_2 + 4\lambda + 6 \tr{Y^{}_{\rm u} Y^\dagger_{\rm u}} \right]  \right.
		\nonumber
		\\
		&& + 2 v^2 C^{\alpha\beta}_5 \left[ -3 C^{}_H + \left( 8\lambda - \frac{5}{3} g^2_2 \right) C^{}_{H\square} - \left( 2\lambda + \frac{3}{4} g^2_1 - \frac{3}{4} g^2_2 \right) C^{}_{HD} -3g^2_2 C^{}_{HW} \right.
		\nonumber
		\\
		&& - \frac{3}{2} \rmi \left( g^2_1 C^{}_{H\widetilde{B}} + 3g^2_2 C^{}_{H\widetilde{W}} + g^{}_1g^{}_2 C^{}_{H\widetilde{W}B} \right) + \frac{1}{2} \tr{C^{}_5 C^\dagger_5} - \frac{2}{3} g^2_2 \tr{3 C^{(3)\dagger}_{Hq} +C^{(3)\dagger}_{H\ell} } 
		\nonumber
		\\
		&& - \left. 3 \tr{ C^{}_{uH}Y^\dagger_{\rm u} } + 6 \tr{  Y^\dagger_{\rm u} C^{(3)\dagger}_{Hq} Y^{}_{\rm u} } \right] +  \frac{5}{2} v^2 \left( C^{}_5 C^\dagger_5 C^{}_5 \right)^{\alpha\beta}  + \frac{3}{2} \left( g^2_1 + g^2_2 \right) v^2
		\nonumber
		\\
		&& \times \left.  \left[ \left( C^{(3)\dagger}_{H\ell} C^{}_5  \right)^{\alpha\beta} + \left( C^{(3)\dagger}_{H\ell} C^{}_5  \right)^{\beta\alpha}  - \left( C^{(1)\dagger}_{H\ell} C^{}_5  \right)^{\alpha\beta}  -  \left( C^{(1)\dagger}_{H\ell} C^{}_5  \right)^{\beta\alpha} \right]  \right\}
	\end{eqnarray}
	to the leading-logarithmic approximation from the cut-off scale $\Lambda$ down to the electroweak scale $\Lambda^{}_{\rm EW} \sim v \simeq 246$ GeV, where the small down-type quark and lepton Yukawa couplings, and contributions from dim-7 operators are ignored. The terms in the first line of Eq.~\eqref{eq:neutrino-mass} result from single insertions of the dim-5 operator, which have been derived a long time ago~\cite{Babu:1993qv,Chankowski:1993tx,Antusch:2001ck} and are not repeated in this work. Apart from the $C^{}_5$ cubic term in the fourth line of Eq.~\eqref{eq:neutrino-mass}, the result is consistent with that obtained in Ref.~\cite{Chala:2021juk} after different conventions are taken into account. This $C^{}_5$ cubic term comes from two redundant dim-6 operators, i.e., $\Opr^{\prime(1)}_{H\ell}$ and $\Opr^{\prime (3)}_{H \ell}$, after applying field redefinitions or the SMEFT EoMs. However, this contribution seems to be overlooked in Ref.~\cite{Chala:2021juk} somehow. Ref.~\cite{Chala:2021juk} has carried out a numerical analysis of the above corrections, which shows that contributions from insertions of one dim-5 operator and one dim-6 operator can reach $50\%$ of those from single insertions of the dim-5 operator for $\Lambda = 1$ TeV, and the relative size of the total corrections is about $4\%$-$8\%$ compared to the tree-level neutrino masses for $\Lambda \in [1~{\rm TeV},3~{\rm TeV}]$. More details and discussions can be found in Ref.~\cite{Chala:2021juk}.
	
	\item Dim-7 operators can lead to some appealing lepton-number-violating (LNV) processes and the above results can result in new corrections to those LNV processes. Though usually those new effects are suppressed by neutrino masses, it is still attractive to study how they can affect the present results and to see whether there are some unexpected consequences. One highly concerned LNV process is neutrinoless double beta decay ($0\nu\beta\beta$) related to Majorana neutrinos. In the SMEFT, besides the dim-5 operator, nine dim-7 operators, i.e., $\{ \Op^{(S)}_{\ell H}, \Op^{(S)}_{\ell HD1}, \Op^{}_{\ell HW}, \Op^{(S)}_{\overline{d}u\ell\ell D}, \Op^{}_{\ell eHD}, \Op^{}_{\overline{d}\ell ue H}, \Op^{}_{\overline{d}\ell q\ell H1}, \Op^{}_{\overline{d}\ell q\ell H2}, \Op^{}_{\overline{q}u\ell\ell H} \}$ can contribute to $0\nu\beta\beta$ as well. Different from contributions from $\Op^{}_5$ and $\Op^{(S)}_{\ell H}$, those from the other eight dim-7 operators are not proportional to the effective neutrino mass~\cite{Cirigliano:2017djv,Cirigliano:2018yza,Liao:2019tep}. The RGEs for most of these dim-7 operators receive contributions from insertions of one dim-5 operator and one dim-6 operator, which have not been taken into consideration in the previous analyses. Similarly, one may apply the above results to study meson decays (e.g., $K^\pm \to \pi^\mp l^\pm l^\pm$~\cite{Liao:2019gex,Liao:2020roy}), nucleon decays (e.g., $p^+ \to \pi^+ \nu$~\cite{Liao:2016hru}) and the neutrino transition moment as well.
\end{itemize}

\section{Conclusions}\label{sec:conclusions}

In the past decade right after the discovery of the SM Higgs boson at the Large Hadron Collider, great attention has been paid to the SMEFT and plenty of progress has been achieved in various aspects of the SMEFT. Among them, the operator basis and RGEs of the SMEFT have attracted a lot of interest and been extensively discussed. In this work, we are concerned with the operator basis for dim-7 operators and the RGEs both for the dim-5 operator and for dim-7 operators up to $\mathcal{O}\left( \Lambda^{-3} \right)$. A Green's basis for dim-7 operators is proposed for the first time, where eight redundant operators (barring hermitian conjugates and flavor structures) are added. The operators with some flavor relations thanks to symmetries among flavor indices are decomposed into several combinations according to tensor decompositions of ${\rm SU(}n{\rm )}$ group. In this case, some combinations are vanishing due to the flavor relations and then redundant degrees of freedom are automatically removed. Following this strategy, we construct a new physical basis for dim-7 operators. Compared with the one put forward in Ref.~\cite{Liao:2019tep}, this new basis is more convenient to deal with matching and running in the SMEFT since flavor indices of operators are not restricted and can run over all flavors. Moreover, the reduction relations to convert operators in the Green's basis to those in the physic basis have been established, to which some redundant dim-6 operators in the Green's basis also make contributions due to field redefinitions (or EOMs) including the dim-5 operator. As a result, besides counterterms for dim-7 operators in the Green's basis, counterterms for some dim-6 operators are also needed when deriving the RGEs for dim-7 operators. Thus, we calculate all relevant counterterms for the dim-5, dim-6, and dim-7 operators in the Green's basis, and convert them to those for the dim-5 and dim-7 operators in the physics basis by means of the reduction relations. Then, the RGEs for the dim-5 and dim-7 operators up to $\mathcal{O}\left( \Lambda^{-3} \right)$ resulting from the mixing between operators of different dimensions are derived with the help of those counterterms in the physical basis. Together with previous results~\cite{Babu:1993qv,Chankowski:1993tx,Antusch:2001ck,Liao:2016hru,Liao:2019tep,Chala:2021juk}, they constitute the complete RGEs for the dim-5 and dim-7 operators in the SMEFT up to $\mathcal{O}\left( \Lambda^{-3} \right)$, and can be exploited for discussions about running effects on some appealing observables or processes, such as neutrino masses, neutrinoless double beta decay, meson and nucleon decays. 

The Green's and physical bases together with the reduction relations for dim-7 operators are not only used for the deviation of RGEs in the SMEFT but also essential and indispensable for the (one-loop) matching of UV models onto the SMEFT at $\mathcal{O}\left( \Lambda^{-3} \right)$, especially with the Feynman diagrammatic approach~\footnote{However, different from the calculations of one-loop RGEs, some evanescent operators may also be involved in one-loop matching procedure~\cite{Fuentes-Martin:2022vvu} (see also references therein) and hence they and reduction relations for them have to be taken into account and added to those in the Green's basis and the corresponding reduction relations~\cite{Carmona:2021xtq}.}. One may embed them in the package {\sf Matchmakereft} to automatically carry out the one-loop matching and derive RGEs in the SMEFT. Additionally, the achieved results for RGEs in this work may shed light on the non-linear non-renormalization theorem, which predicts zero entries in the anomalous dimension matrices or tensors resulting from the mixing among different dimensional operators. Such a theorem is pretty intriguing and worthy of more attention and effort.

{\bf Note added:} After one and a half months our work appeared on arXiv, a preprint~\cite{Li:2023cwy} aiming to establish a tree-level correspondence between all possible UV resonances and the SMEFT dim-5, -6, and -7 operators came out. Ref.~\cite{Li:2023cwy} also took into account a Green's basis and the reduction of redundancies for dim-7 operators. The Green's basis in Ref.~\cite{Li:2023cwy} is slightly different from ours, but they are equivalent. However, Ref.~\cite{Li:2023cwy} did not give any explicit reduction relations between the Green's and physical bases.

\section*{Acknowledgements}

This work was supported by the Alexander von Humboldt Foundation.

\begin{appendices}

\section{Conversion between bases} \label{app:conversion}

In this appendix, we present the reduction relations between redundant and non-redundant operators, where $C^{\dots}_{\dots}$ and $G^{\dots}_{\dots}$ denote the Wilson coefficients of operators in the Green's and physics bases, respectively. This includes the reductions of redundant dim-6 operators in Table~\ref{tab:dim-6} and dim-7 operators in Table~\ref{tab:green-basis} with the help of the SMEFT EoMs up to $\mathcal{O} \left( \Lambda^{-1} \right)$ given in Eq.~\eqref{eq:eom}. The reductions of all dim-6 operators in the Green's basis to those in the Warsaw basis by means of the SM EoMs can be found in Ref.~\cite{Gherardi:2020det}. In the following expressions, we adopt the physical basis in Table~\ref{tab:physical-basis} for independent dim-7 operators, and we do not decompose the unique dim-5 operator into symmetric and antisymmetric combinations since it is trivial to figure out the symmetry among two flavor indices. Meanwhile, contributions from redundant dim-6 and dim-7 operators are highlighted in blue and in red, respectively.

\subsection{Reduction relations for the dim-5 operator}

\begin{eqnarray}\label{eq:convertion-c5}
	C^{\alpha\beta}_5 &=& G^{\alpha\beta}_5  \textcolor{blue}{ - 2 m^2 G^{}_{DH} G^{\alpha\beta}_5}  \textcolor{red}{ + 2m^2 \left( G^{(S)\dagger}_{\ell HD3} \right)^{\alpha\beta}} \;.
\end{eqnarray}

\subsection{Reduction relations for dim-7 operators}

\subsubsection*{$\bullet~ \bm{\psi^2 H^4}$}

\begin{eqnarray}\label{eq:convertion-clh}
	C^{(S)\alpha\beta}_{\ell H} &=& G^{(S)\alpha\beta}_{\ell H} \textcolor{blue}{ + \left( G^\dagger_5 \right)^{\alpha\beta}  \left( \frac{1}{4} g^2_2 G^{}_{2W} - g^{}_2 G^{}_{WDH} - 2\lambda G^{}_{DH} - \frac{1}{2} G^\prime_{HD} - \rmi G^{\prime\prime}_{HD} \right)  }
	\nonumber
	\\
	&& \textcolor{blue}{ + \frac{1}{2} \left( G^\dagger_5 \right)^{\alpha\gamma} \left( G^{\prime(3)\gamma\beta}_{H\ell} - \rmi G^{\prime\prime(3)\gamma\beta}_{H\ell} - G^{\prime(1)\gamma\beta}_{H\ell} + \rmi G^{\prime\prime(1)\gamma\beta}_{H\ell} \right) + \frac{1}{2}\left( G^\dagger_5 \right)^{\beta\gamma}  }
	\nonumber
	\\
	&& \textcolor{blue}{ \times \left( G^{\prime(3)\gamma\alpha}_{H\ell} - \rmi G^{\prime\prime(3)\gamma\alpha}_{H\ell}  - G^{\prime(1)\gamma\alpha}_{H\ell} + \rmi G^{\prime\prime(1)\gamma\alpha}_{H\ell} \right) }  \textcolor{red}{ + 2\lambda G^{(S)\alpha\beta}_{\ell HD3}} \;.
\end{eqnarray}

\subsubsection*{$\bullet~ \bm{\psi^2 H^3 D}$}
\begin{eqnarray}
	C^{\alpha\beta}_{\ell eHD} &=& G^{\alpha\beta}_{\ell eHD} \textcolor{blue}{ +  \frac{1}{2} \left( G^\dagger_5 \right)^{\alpha\gamma} \left[ G^{\gamma\beta}_{eHD2} - G^{\gamma\beta}_{eHD4} - 2 G^{\gamma\lambda}_{\ell D} \left( Y^{}_l \right)^{}_{\lambda\beta} \right] }
	\nonumber
	\\
	&& \textcolor{red}{ + \left( \rmi G^{\alpha\gamma}_{\ell HD5} + \rmi G^{(S)\alpha\gamma}_{\ell HD6} - G^{(S)\alpha\gamma}_{\ell H D 4} \right) \left( Y^{}_l \right)^{}_{\gamma\beta} - \frac{1}{4}  \left( G^{\alpha\gamma}_{\ell HD2} - G^{\gamma\alpha}_{\ell HD2} \right)\left( Y^{}_l \right)^{}_{\gamma \beta} } \;.
\end{eqnarray}

\subsubsection*{$\bullet~ \bm{\psi^2 H^2 X}$}

\begin{eqnarray}
	C^{\alpha\beta}_{\ell H W} &=& G^{\alpha\beta}_{\ell H W} \textcolor{blue}{ + \frac{1}{8} \left( G^\dagger_5 \right)^{\alpha\gamma} \left( 2\rmi G^{\prime \gamma\beta}_{W\ell} + 2 G^{\prime \gamma\beta}_{\widetilde{W}\ell} + g^{}_2 G^{\gamma\beta}_{\ell D} \right)  }  \textcolor{red}{ + \frac{1}{4} g^{}_2 \left( G^{(S)\alpha\beta}_{\ell HD4} - \rmi G^{(S)\alpha\beta}_{\ell HD6} \right) }  
	\nonumber
	\\
	&& \textcolor{red}{ + \frac{1}{16} g^{}_2 \left( G^{\alpha\beta}_{\ell HD2} - G^{\beta\alpha}_{\ell HD2} \right)  } \;,
	\\
	C^{(A)\alpha\beta}_{\ell H B} &=& G^{(A)\alpha\beta}_{\ell H B} \textcolor{blue}{ - \frac{1}{16} \left( G^\dagger_5 \right)^{\alpha\gamma} \left( 2\rmi G^{\prime \gamma\beta}_{B\ell} + 2 G^{\prime \gamma\beta}_{\widetilde{B}\ell} - g^{}_1 G^{\gamma\beta}_{\ell D} \right) }
	\nonumber
	\\
	&& \textcolor{blue}{ + \frac{1}{16} \left( G^\dagger_5 \right)^{\beta\gamma} \left( 2\rmi G^{\prime \gamma\alpha}_{B\ell} + 2 G^{\prime \gamma\alpha}_{\widetilde{B}\ell} - g^{}_1 G^{\gamma\alpha}_{\ell D} \right) } \textcolor{red}{ + \frac{1}{16} g^{}_1 \left( G^{\alpha\beta}_{\ell HD2} - G^{\beta\alpha}_{\ell HD2} \right)  } \;.
\end{eqnarray}	

\subsubsection*{$\bullet~ \bm{\psi^2 H^2 D^2}$}

\begin{eqnarray}
	C^{(S)\alpha\beta}_{\ell HD1} &=& \frac{1}{2} \left( G^{\alpha\beta}_{\ell HD1} + G^{\beta\alpha}_{\ell HD1} \right) \textcolor{red}{ + G^{(S)\alpha\beta}_{\ell HD3} + G^{(S)\alpha\beta}_{\ell HD4} + \frac{1}{2} \rmi \left( G^{\alpha\beta}_{\ell HD5} + G^{\beta\alpha}_{\ell HD5} \right) - \rmi G^{(S)\alpha\beta}_{\ell HD6} } \;,
	\\
	C^{(S)\alpha\beta}_{\ell HD2} &=&  \frac{1}{2} \left( G^{\alpha\beta}_{\ell HD2} + G^{\beta\alpha}_{\ell HD2} \right)  \textcolor{red}{ - 2 \left( G^{(S)\alpha\beta}_{\ell HD3} + G^{(S)\alpha\beta}_{\ell HD4} - \rmi G^{(S)\alpha\beta}_{\ell HD6} \right) } \;.
\end{eqnarray}

\subsubsection*{$\bullet~ \bm{\psi^4 H}$}

\begin{eqnarray}
	C^{(S)\alpha\beta\gamma\lambda}_{\overline{e}\ell\ell\ell H} &=& 
	G^{(S)\alpha\beta\gamma\lambda}_{\overline{e}\ell\ell\ell H}  
	\textcolor{blue}{ + \frac{1}{3}  \left[ G^{}_{DH} \left( Y^\dagger_l \right)^{}_{\alpha\beta} - \left( G^\dagger_{eHD1} \right)^{\alpha\beta} - \frac{1}{2}  \left( G^\dagger_{eHD2} \right)^{\alpha\beta} + \frac{1}{2}  \left( G^\dagger_{eHD4} \right)^{\alpha\beta} \right] \left( G^\dagger_5 \right)^{\gamma\lambda} } 
	\nonumber
	\\
	&& \textcolor{blue}{ + \frac{1}{3} \left[ G^{}_{DH} \left( Y^\dagger_l \right)^{}_{\alpha\lambda} - \left( G^\dagger_{eHD1} \right)^{\alpha\lambda} - \frac{1}{2}  \left( G^\dagger_{eHD2} \right)^{\alpha\lambda} + \frac{1}{2}  \left( G^\dagger_{eHD4} \right)^{\alpha\lambda} \right] \left( G^\dagger_5 \right)^{\beta\gamma} }
	\nonumber
	\\
	&& \textcolor{blue}{ + \frac{1}{3} \left[ G^{}_{DH} \left( Y^\dagger_l \right)^{}_{\alpha\gamma} - \left( G^\dagger_{eHD1} \right)^{\alpha\gamma} - \frac{1}{2}  \left( G^\dagger_{eHD2} \right)^{\alpha\gamma} + \frac{1}{2}  \left( G^\dagger_{eHD4} \right)^{\alpha\gamma} \right] \left( G^\dagger_5 \right)^{\lambda\beta} }
	\nonumber
	\\
	&& \textcolor{red}{ - \frac{1}{3} \left[ \left( Y^\dagger_l \right)^{}_{\alpha\beta} G^{(S)\gamma\lambda}_{\ell HD3} + \left( Y^\dagger_l \right)^{}_{\alpha\lambda} G^{(S)\beta\gamma}_{\ell HD3} + \left( Y^\dagger_l \right)^{}_{\alpha\gamma} G^{(S)\lambda\beta}_{\ell HD3} \right] } \;,
	\\
	C^{(A)\alpha\beta\gamma\lambda}_{\overline{e}\ell\ell\ell H} &=& 
	G^{(A)\alpha\beta\gamma\lambda}_{\overline{e}\ell\ell\ell H}
	\textcolor{red}{ + \frac{1}{6} \left( Y^\dagger_l \right)^{}_{\alpha\beta} \left[ G^{\gamma\lambda}_{\ell HD1} - G^{\lambda\gamma}_{\ell HD1} + \frac{1}{2} \left(G^{\gamma\lambda}_{\ell HD2} - G^{\lambda\gamma}_{\ell HD2}  \right) + \rmi \left(G^{\gamma\lambda}_{\ell HD5} - G^{\lambda\gamma}_{\ell HD5}  \right)  \right] } 
	\nonumber
	\\
	&& \textcolor{red}{ + \frac{1}{6} \left( Y^\dagger_l \right)^{}_{\alpha\lambda} \left[ G^{\beta\gamma}_{\ell HD1} - G^{\gamma\beta}_{\ell HD1} + \frac{1}{2} \left(G^{\beta\gamma}_{\ell HD2} - G^{\gamma\beta}_{\ell HD2}  \right) + \rmi \left(G^{\beta\gamma}_{\ell HD5} - G^{\gamma\beta}_{\ell HD5}  \right)  \right] } 
	\nonumber
	\\
	&& \textcolor{red}{ + \frac{1}{6} \left( Y^\dagger_l \right)^{}_{\alpha\gamma} \left[ G^{\lambda\beta}_{\ell HD1} - G^{\beta\lambda}_{\ell HD1} + \frac{1}{2} \left(G^{\lambda\beta}_{\ell HD2} - G^{\beta\lambda}_{\ell HD2}  \right) + \rmi \left(G^{\lambda\beta}_{\ell HD5} - G^{\beta\lambda}_{\ell HD5}  \right)  \right] } \;,
	\\
	C^{(M)\alpha\beta\gamma\lambda}_{\overline{e}\ell\ell\ell H} &=& 
	G^{(M)\alpha\beta\gamma\lambda}_{\overline{e}\ell\ell\ell H}
	\textcolor{blue}{ + \frac{1}{2} \left[ G^{}_{DH} \left( Y^\dagger_l \right)^{}_{\alpha\beta} - \left( G^\dagger_{eHD1} \right)^{\alpha\beta} - \frac{1}{2}  \left( G^\dagger_{eHD2} \right)^{\alpha\beta} + \frac{1}{2}  \left( G^\dagger_{eHD4} \right)^{\alpha\beta} \right] \left( G^\dagger_5 \right)^{\gamma\lambda} }
	\nonumber
	\\
	&& \textcolor{blue}{ - \frac{1}{2} \left[ G^{}_{DH} \left( Y^\dagger_l \right)^{}_{\alpha\lambda} - \left( G^\dagger_{eHD1} \right)^{\alpha\lambda} - \frac{1}{2}  \left( G^\dagger_{eHD2} \right)^{\alpha\lambda} + \frac{1}{2}  \left( G^\dagger_{eHD4} \right)^{\alpha\lambda} \right] \left( G^\dagger_5 \right)^{\beta\gamma} }
	\nonumber
	\\
	&& \textcolor{red}{ - \frac{1}{2} \left[ \left( Y^\dagger_l \right)^{}_{\alpha\beta} G^{(S)\gamma\lambda}_{\ell HD3} - \left( Y^\dagger_l \right)^{}_{\alpha\lambda} G^{(S)\beta\gamma}_{\ell HD3} \right] } 
	\nonumber
	\\
	&& \textcolor{red}{ + \frac{1}{12} \left( Y^\dagger_l \right)^{}_{\alpha\beta} \left[ G^{\gamma\lambda}_{\ell HD1} - G^{\lambda\gamma}_{\ell HD1} + \frac{1}{2} \left(G^{\gamma\lambda}_{\ell HD2} - G^{\lambda\gamma}_{\ell HD2}  \right) + \rmi \left(G^{\gamma\lambda}_{\ell HD5} - G^{\lambda\gamma}_{\ell HD5}  \right)  \right] } 
	\nonumber
	\\
	&& \textcolor{red}{ + \frac{1}{12} \left( Y^\dagger_l \right)^{}_{\alpha\lambda} \left[ G^{\beta\gamma}_{\ell HD1} - G^{\gamma\beta}_{\ell HD1} + \frac{1}{2} \left(G^{\beta\gamma}_{\ell HD2} - G^{\gamma\beta}_{\ell HD2}  \right) + \rmi \left(G^{\beta\gamma}_{\ell HD5} - G^{\gamma\beta}_{\ell HD5}  \right)  \right] } 
	\nonumber
	\\
	&& \textcolor{red}{ - \frac{1}{6} \left( Y^\dagger_l \right)^{}_{\alpha\gamma} \left[ G^{\lambda\beta}_{\ell HD1} - G^{\beta\lambda}_{\ell HD1} + \frac{1}{2} \left(G^{\lambda\beta}_{\ell HD2} - G^{\beta\lambda}_{\ell HD2}  \right) + \rmi \left(G^{\lambda\beta}_{\ell HD5} - G^{\beta\lambda}_{\ell HD5}  \right)  \right] } \;,
	\\
	C^{\alpha\beta\gamma\lambda}_{\overline{d}\ell q \ell H1} &=& G^{\alpha\beta\gamma\lambda}_{\overline{d}\ell q \ell H1} \textcolor{blue}{ + \left[ 2G^{}_{DH} \left( Y^\dagger_{\rm d} \right)^{}_{\alpha\gamma} - 2\left( G^\dagger_{dHD1} \right)^{\alpha\gamma} - \left( G^\dagger_{dHD2} \right)^{\alpha\gamma} + \left( G^\dagger_{dHD4} \right)^{\alpha\gamma} \right] \left( G^\dagger_5 \right)^{\beta\lambda} }
	\nonumber
	\\
	&&  \textcolor{red}{ -2 \left( Y^\dagger_{\rm d} \right)^{}_{\alpha\gamma} G^{(S)\beta\lambda}_{\ell HD3} } \;,
	\\
	C^{\alpha\beta\gamma\lambda}_{\overline{d}\ell q \ell H2} &=& G^{\alpha\beta\gamma\lambda}_{\overline{d}\ell q \ell H2} \textcolor{blue}{  -\frac{1}{2} \left[ 2G^{}_{DH} \left( Y^\dagger_{\rm d} \right)^{}_{\alpha\gamma} - 2\left( G^\dagger_{dHD1} \right)^{\alpha\gamma} - \left( G^\dagger_{dHD2} \right)^{\alpha\gamma} + \left( G^\dagger_{dHD4} \right)^{\alpha\gamma} \right] \left( G^\dagger_5 \right)^{\beta\lambda} }
	\nonumber
	\\
	&& \textcolor{red}{ + \left( Y^\dagger_{\rm d} \right)^{}_{\alpha\gamma} \left[ \frac{1}{2} \left(G^{\beta\lambda}_{\ell HD1} - G^{\lambda\beta}_{\ell HD1}  \right) + \frac{1}{4} \left(G^{\beta\lambda}_{\ell HD2} - G^{\lambda\beta}_{\ell HD2}  \right) + G^{(S)\beta\lambda}_{\ell HD3} + \frac{1}{2} \rmi \left(G^{\beta\lambda}_{\ell HD5} \right.\right. }  
	\nonumber
	\\
	&& \textcolor{red}{ - \left.\left. G^{\lambda\beta}_{\ell HD5}  \right)  \right]  + G^{\alpha\beta\lambda\rho}_{\overline{d}\ell\ell D u} \left( Y^\dagger_{\rm u} \right)^{}_{\rho \gamma} + \frac{1}{2} \left( G^{\alpha\rho\beta\lambda}_{\overline{d}u\ell\ell D} - G^{\alpha\rho\lambda\beta}_{\overline{d}u\ell\ell D} + G^{\alpha\lambda\beta\rho}_{\overline{d}D\ell\ell u} - G^{\alpha\beta\lambda\rho}_{\overline{d}D\ell\ell u} \right) \left( Y^\dagger_{\rm u} \right)^{}_{\rho\gamma} } \;,
	\\
	C^{\alpha\beta\gamma\lambda}_{\overline{d}\ell ueH} &=& G^{\alpha\beta\gamma\lambda}_{\overline{d}\ell ueH}  \textcolor{red}{ - G^{\alpha\rho\beta\gamma}_{\overline{d}D\ell\ell u} \left( Y^{}_l \right)^{}_{\rho\lambda} } \;,
	\\
	C^{\alpha\beta\gamma\lambda}_{\overline{q} u \ell\ell H} &=& G^{\alpha\beta\gamma\lambda}_{\overline{q} u \ell\ell H} \textcolor{blue}{ - \left[ G^{}_{DH} \left( Y^{}_{\rm u} \right)^{}_{\alpha\beta} -  \left( G^{}_{uHD1} \right)^{\alpha\beta} - \frac{1}{2}  \left( G^{}_{uHD2} \right)^{\alpha\beta} +  \frac{1}{2}  \left( G^{}_{uHD4} \right)^{\alpha\beta}  \right] \left( G^\dagger_5 \right)^{\gamma\lambda} } 
	\nonumber
	\\
	&& \textcolor{red}{ + \left( Y^{}_{\rm u} \right)^{}_{\alpha\beta} \left[ - \frac{1}{2} \left(G^{\gamma\lambda}_{\ell HD1} - G^{\lambda\gamma}_{\ell HD1}  \right) - \frac{1}{4} \left(G^{\gamma\lambda}_{\ell HD2} - G^{\lambda\gamma}_{\ell HD2}  \right) + G^{(S)\gamma\lambda}_{\ell HD3}  -  \frac{1}{2} \rmi \left(G^{\gamma\lambda}_{\ell HD5}  \right.\right. } 
	\nonumber
	\\
	&& \textcolor{red}{ - \left.\left. G^{\lambda\gamma}_{\ell HD5}  \right)  \right] + \frac{1}{2} \left( Y^{}_{\rm d} \right)^{}_{\alpha\rho} \left( G^{\rho\beta\gamma\lambda}_{\overline{d}u\ell\ell D} - G^{\rho\beta\lambda\gamma}_{\overline{d}u\ell\ell D} + G^{\rho\lambda\gamma\beta}_{\overline{d}D\ell\ell u} - G^{\rho\gamma\lambda\beta}_{\overline{d}D\ell\ell u} \right) } \;,
	\\
	C^{\alpha\beta\gamma\lambda}_{\overline{\ell}dud\widetilde{H}} &=& G^{\alpha\beta\gamma\lambda}_{\overline{\ell}dud\widetilde{H}}  \textcolor{red}{ - \frac{1}{2} \left( 2 G^{\alpha\beta\rho\lambda}_{\overline{\ell}dDqd} + G^{\alpha\rho\beta\lambda}_{\overline{\ell}qddD} - G^{\alpha\rho\lambda\beta}_{\overline{\ell}qddD} \right) \left( Y^{}_{\rm u} \right)^{}_{\rho\gamma} } \;,
	\\
	C^{(M)\alpha\beta\gamma\lambda}_{\overline{\ell}dddH} &=&   
	G^{(M)\alpha\beta\gamma\lambda}_{\overline{\ell}dddH}
	\textcolor{red}{ + \frac{1}{6} \left[ 2G^{\alpha\beta\rho\gamma}_{\overline{\ell}dDqd} \left( Y^{}_{\rm d} \right)^{}_{\rho\lambda} - G^{\alpha\lambda\rho\beta}_{\overline{\ell}dDqd} \left( Y^{}_{\rm d} \right)^{}_{\rho\gamma} - G^{\alpha\gamma\rho\lambda}_{\overline{\ell}dDqd} \left( Y^{}_{\rm d} \right)^{}_{\rho\beta} - 2 G^{\alpha\beta\rho\lambda}_{\overline{\ell}dDqd} \left( Y^{}_{\rm d} \right)^{}_{\rho\gamma} \right.}
	\nonumber
	\\
	&& \textcolor{red}{ + \left. G^{\alpha\gamma\rho\beta}_{\overline{\ell}dDqd} \left( Y^{}_{\rm d} \right)^{}_{\rho\lambda} + G^{\alpha\lambda\rho\gamma}_{\overline{\ell}dDqd} \left( Y^{}_{\rm d} \right)^{}_{\rho\beta}  \right] +  \frac{1}{6} \left( Y^{}_l \right)^{}_{\alpha\rho} \left( G^{\rho\beta\gamma\lambda}_{\overline{e}dddD} + G^{\rho\gamma\beta\lambda}_{\overline{e}dddD} - G^{\rho\beta\lambda\gamma}_{\overline{e}dddD} - G^{\rho\lambda\beta\gamma}_{\overline{e}dddD} \right)  } 
	\nonumber
	\\
	&& \textcolor{red}{ - \frac{1}{6} \left( G^{\alpha\rho\gamma\lambda}_{\overline{\ell}qddD} - G^{\alpha\rho\lambda\gamma}_{\overline{\ell}qddD} \right) \left( Y^{}_{\rm d} \right)^{}_{\rho\beta} - \frac{1}{12} \left( G^{\alpha\rho\beta\lambda}_{\overline{\ell}qddD} - G^{\alpha\rho\lambda\beta}_{\overline{\ell}qddD} \right) \left( Y^{}_{\rm d} \right)^{}_{\rho\gamma}  }
	\nonumber
	\\
	&& \textcolor{red}{ - \frac{1}{12} \left( G^{\alpha\rho\gamma\beta}_{\overline{\ell}qddD} - G^{\alpha\rho\beta\gamma}_{\overline{\ell}qddD} \right) \left( Y^{}_{\rm d} \right)^{}_{\rho\lambda} } \;,
	\\
	C^{(A)\alpha\beta\gamma\lambda}_{\overline{e}qdd\widetilde{H}} &=&  G^{(A)\alpha\beta\gamma\lambda}_{\overline{e}qdd\widetilde{H}} \textcolor{red}{ + \frac{1}{4} \left( Y^\dagger_{\rm d} \right)^{}_{\rho\beta} \left( G^{\alpha\rho\gamma\lambda}_{\overline{e}dddD} - G^{\alpha\lambda\rho\gamma}_{\overline{e}dddD} + G^{\alpha\gamma\lambda\rho}_{\overline{e}dddD} - G^{\alpha\rho\lambda\gamma}_{\overline{e}dddD} + G^{\alpha\gamma\rho\lambda}_{\overline{e}dddD} - G^{\alpha\lambda\gamma\rho}_{\overline{e}dddD} \right) }
	\nonumber
	\\
	&& \textcolor{red}{ - \frac{1}{4} \left( Y^\dagger_l \right)^{}_{\alpha\rho} \left( G^{\rho\beta\gamma\lambda}_{\overline{\ell}qddD} - G^{\rho\beta\lambda\gamma}_{\overline{\ell}qddD} \right) } \;,
	\\
	C^{\alpha\beta\gamma\lambda}_{\overline{\ell}dqq\widetilde{H}} &=& G^{\alpha\beta\gamma\lambda}_{\overline{\ell}dqq\widetilde{H}}  \textcolor{red}{ - G^{\alpha\beta\gamma\rho}_{\overline{\ell}dqDd} \left( Y^\dagger_{\rm d} \right)^{}_{\rho\lambda} } \;.
\end{eqnarray}

\subsubsection*{$\bullet~ \bm{\psi^4D}$}

\begin{eqnarray}
	C^{(S)\alpha\beta\gamma\lambda}_{\overline{e} dddD} &=& \frac{1}{6} \left( G^{\alpha\beta\gamma\lambda}_{\overline{e}dddD} + G^{\alpha\lambda\beta\gamma}_{\overline{e}dddD} +  G^{\alpha\gamma\lambda\beta}_{\overline{e}dddD} + G^{\alpha\beta\lambda\gamma}_{\overline{e}dddD} + G^{\alpha\gamma\beta\lambda}_{\overline{e}dddD} + G^{\alpha\lambda\gamma\beta}_{\overline{e}dddD} \right) \;,
	\\
	C^{(S)\alpha\beta\gamma\lambda}_{\overline{d} u \ell\ell D} &=& \frac{1}{2} \left( G^{\alpha\beta\gamma\lambda}_{\overline{d}u\ell\ell D} + G^{\alpha\beta\lambda\gamma}_{\overline{d}u\ell\ell D} \right) \textcolor{red}{ + \frac{1}{2} \left( G^{\alpha\lambda\gamma\beta}_{\overline{d}D\ell\ell u} + G^{\alpha\gamma\lambda\beta}_{\overline{d}D\ell\ell u} \right) }  \;,
	\\
	C^{(S)\alpha\beta\gamma\lambda}_{\overline{\ell}qddD} &=& \frac{1}{2} \left( G^{\alpha\beta\gamma\lambda}_{\overline{\ell} qdd D} + G^{\alpha\beta\lambda\gamma}_{\overline{\ell} qdd D} \right) \;.
\end{eqnarray}

\section{Counterterms in the Green's basis}\label{app:counterterm-green}

As discussed in Sec.~\ref{subsec:gsr}, we need to calculate the counterterms not only for the dim-5 and dim-7 operators but also for the dim-6 operators listed in Table~\ref{tab:dim-6} to achieve RGEs of the dim-5 and dim-7 operators. Though we do not consider contributions from single insertions of dim-7 operators and wave-function renormalization constants to RGEs of dim-7 operators, we still have to take into account such contributions to the RGE of the dim-5 operator up to $\mathcal{O} \left( \Lambda^{-3} \right)$. For counterterms of the dim-6 operators in Table~\ref{tab:dim-6}, they may receive contributions from both single insertions of dim-6 operators and double insertions of the dim-5 operator. All results for those counterterms in the Green's basis are listed in the following subsections, where $\delta Z^{G}_{H}$ and $\delta G^{\dots}_{\dots}$ on the left-hand side of each equality represent the wave-function renormalization constant and counterterms in the Green's basis, and $C^{\dots}_{\dots}$ on the right-hand side are Wilson coefficients of the non-redundant operators in the physical basis. Notice that other operators in the same operator class as those appearing on the left-hand side of the following equalities are also involved in calculations and the counterterms for them in the Green's basis are obtained simultaneously, but they eventually make no contribution to the RGEs under consideration and are not shown here.

\subsection{Wave-function renormalization constants}\label{subs:wave-function}

\begin{eqnarray}
	\delta Z^{G}_H = \loopf m^2 \left( C^{}_{HD} - 2C^{}_{H \square} \right) \;,
\end{eqnarray}
in which only new corrections from non-renormalization operators up to $\mathcal{O}\left( \Lambda^{-3}\right)$ are given, and the SM part is omitted. There are no such new corrections to the wave-function renormalization constant of the lepton doublet $\delta Z^G_\ell$ up to $\mathcal{O}\left( \Lambda^{-3}\right)$.

\subsection{Counterterm for the dim-5 operator}

The counterterm for the dim-5 operator has two types of contributions, as separately listed below.

\subsubsection{Contributions from insertions of one dim-5 operator and one dim-6 operator}
\begin{eqnarray}
	\delta G^{}_5 &=& \loopf 2m^2 \left[ 2 C^{}_5 C^{}_{H\square} + C^{(1)\dagger}_{H\ell}C^{}_5  + \left( C^{(1)\dagger}_{H\ell}C^{}_5 \right)^{\rm T} - 2C^{(3)\dagger}_{H\ell}C^{}_5 - 2\left( C^{(3)\dagger}_{H\ell}C^{}_5 \right)^{\rm T} \right] \;.
\end{eqnarray}

\subsubsection{Contributions from insertions of one dim-7 operator}
\begin{eqnarray}
	\delta G^{}_5 &=& \loopf \left\{ 8m^2C^{(S)\dagger}_{\ell H} + m^2 Y^{}_l C^\dagger_{\ell e HD}  + m^2 \left( Y^{}_l C^\dagger_{\ell e HD}  \right)^{\rm T} - m^2 Y^{}_l Y^\dagger_l  \left( C^{(S)\dagger}_{\ell HD1} + C^{(S)\dagger}_{\ell HD2} \right) \right.
	\nonumber
	\\
	&& - \left. m^2 \left[ Y^{}_l Y^\dagger_l  \left( C^{(S)\dagger}_{\ell HD1} + C^{(S)\dagger}_{\ell HD2} \right)  \right]^{\rm T} \right\} \;.
\end{eqnarray}

\subsection{Counterterms for dim-6 operators}

The counterterms for the redundant dim-6 operators in Table~\ref{tab:dim-6} are presented as below:

\noindent$\bullet~\bm{X^2D^2}$

\begin{eqnarray}
	\delta G^{}_{2W} = \loopf 12g^{}_2 C^{}_{3W} \;.
\end{eqnarray}

\noindent$\bullet~\bm{H^2XD^2}$

\begin{eqnarray}
	\delta G^{}_{WDH} &=& \loopf \left[ \frac{1}{6}g^{}_2 C^{}_{H\square} + 3 g^2_2 C^{}_{3W} + \frac{2}{3} g^{}_2 \tr{C^{(3)}_{H\ell}} + 2g^{}_2 \tr{C^{(3)}_{Hq}} \right] \;.
\end{eqnarray}

\noindent$\bullet~\bm{H^4D^2}$

\begin{eqnarray}
	\delta G^\prime_{HD} &=& \loopf \left\{ \left( 3g^2_2 - 4\lambda \right) C^{}_{H\square} + \left[ \frac{3}{2}\left( g^2_1 - g^2_2 \right) + 2\lambda \right] C^{}_{HD} + 3 g^2_1 C^{}_{HB} + 9 g^2_2 C^{}_{HW}   \right.
	\nonumber
	\\
	&& + 3g^{}_1 g^{}_2 C^{}_{HWB} + \tr{ C^{}_{eH} Y^\dagger_l + Y^{}_l C^\dagger_{eH} + 3 C^{}_{dH} Y^\dagger_{\rm d} + 3 Y^{}_{\rm d} C^\dagger_{dH} + 3 C^{}_{uH} Y^\dagger_{\rm u} + 3 Y^{}_{\rm u} C^\dagger_{uH} }
	\nonumber
	\\
	&& - 4\tr{ Y^\dagger_l C^{(3)}_{H\ell} Y^{}_l + 3 Y^\dagger_{\rm d} C^{(3)}_{Hq} Y^{}_{\rm d} + 3 Y^\dagger_{\rm u} C^{(3)}_{Hq} Y^{}_{\rm u} } + 6 \tr{ Y^{}_{\rm u} C^{}_{Hud} Y^\dagger_{\rm d} + Y^{}_{\rm d} C^\dagger_{Hud} Y^\dagger_{\rm u} }
	\nonumber
	\\
	&&  - \left. \tr{C^{}_5 C^\dagger_5}  \right\} \;,\label{eq:HDp}
	\nonumber
	\\
	\delta G^{\prime\prime}_{HD} &=& - \loopf \frac{\rmi}{2} \tr{ C^{}_{eH} Y^\dagger_l - Y^{}_l C^\dagger_{eH} + 3 C^{}_{dH} Y^\dagger_{\rm d} - 3Y^{}_{\rm d} C^\dagger_{dH} - 3 C^{}_{uH} Y^\dagger_{\rm u} + 3 Y^{}_{\rm u} C^\dagger_{uH} } \;.\label{eq:HDpp}
\end{eqnarray}

\noindent$\bullet~\bm{\psi^2 H D^2}$

\begin{eqnarray}
	\delta G^{\alpha\beta}_{eHD1} &=& \loopf \left[ - \frac{3}{2} g^{}_1 C^{\alpha\beta}_{eB} + \frac{9}{2} g^{}_2 C^{\alpha\beta}_{eW} - \left( Y^{}_l \right)^{}_{\alpha\gamma} C^{\gamma\beta}_{He}  + 2 C^{\alpha\gamma\lambda\beta}_{\ell e} \left( Y^{}_l \right)^{}_{\gamma\lambda} + 3 C^{(1)\alpha\beta\gamma\lambda}_{\ell equ} \left( Y^\dagger_{\rm u} \right)^{}_{\lambda \gamma} \right.
	\nonumber
	\\
	&& - \left. 3 C^{\alpha\beta\gamma\lambda}_{\ell edq} \left( Y^{}_{\rm d} \right)^{}_{\lambda\gamma} \right] \;,
	\nonumber
	\\
	\delta G^{}_{eHD2} &=& \loopf \frac{1}{2} \left( Y^{}_l C^{}_{He} - C^{(1)}_{H\ell} Y^{}_l - 3 C^{(3)}_{H\ell} Y^{}_l \right) \;,
	\nonumber
	\\
	\delta G^{}_{eHD4} &=& \loopf \left( - 3g^{}_1 C^{}_{eB} + 9 g^{}_2 C^{}_{eW} - \frac{3}{2} Y^{}_l C^{}_{He} - \frac{1}{2} C^{(1)}_{H\ell} Y^{}_l - \frac{3}{2} C^{(3)}_{H\ell} Y^{}_l \right) \;,
	\nonumber
	\\
	\delta G^{\alpha\beta}_{uHD1} &=& \loopf \left[ \frac{1}{2} g^{}_1 C^{\alpha\beta}_{uB} + \frac{9}{2} g^{}_2 C^{\alpha\beta}_{uW} + 4 g^{}_s C^{\alpha\beta}_{uG} + \left( Y^{}_{\rm u} \right)^{}_{\alpha\gamma} C^{\gamma\beta}_{Hu} - \left( Y^{}_{\rm d} \right)^{}_{\alpha\gamma} \left(C^\dagger_{Hud} \right)^{\gamma\beta} \right.
	\nonumber
	\\
	&& + C^{(1)\gamma\lambda\alpha\beta}_{\ell equ} \left( Y^\dagger_l \right)^{}_{\lambda\gamma } + 2 \left( C^{(1)\alpha\gamma\lambda\beta}_{qu} + \frac{4}{3} C^{(8)\alpha\gamma\lambda\beta}_{qu}  \right) \left( Y^{}_{\rm u} \right)^{}_{\gamma\lambda} - 3 C^{(1)\alpha\beta\gamma\lambda}_{quqd} \left( Y^\dagger_{\rm d} \right)^{}_{\lambda\gamma} 
	\nonumber
	\\
	&&-  \left.  \frac{1}{2} \left( C^{(1)\lambda\beta\alpha\gamma}_{quqd} + \frac{4}{3} C^{(8)\lambda\beta\alpha\gamma}_{quqd} \right) \left( Y^\dagger_{\rm d} \right)^{}_{\gamma\lambda}  \right] \;,
	\nonumber
	\\
	\delta G^{}_{uHD2} &=& \loopf \frac{1}{2} \left( C^{(1)}_{Hq} Y^{}_{\rm u} - 3 C^{(3)}_{Hq} Y^{}_{\rm u} - Y^{}_{\rm u} C^{}_{Hu} + Y^{}_{\rm d} C^\dagger_{Hud} \right) \;,
	\nonumber
	\\
	\delta G^{}_{uHD4} &=& \loopf \left( g^{}_1 C^{}_{uB} + 9g^{}_2 C^{}_{uW} + 8g^{}_s C^{}_{uG} + \frac{1}{2} C^{(1)}_{Hq} Y^{}_{\rm u} - \frac{3}{2} C^{(3)}_{Hq} Y^{}_{\rm u} + \frac{3}{2} Y^{}_{\rm u} C^{}_{Hu} - \frac{3}{2} Y^{}_{\rm d} C^\dagger_{Hud} \right) \;,
	\nonumber
	\\
	\delta G^{\alpha\beta}_{dHD1} &=& \loopf \left[ \frac{1}{2} g^{}_1 C^{\alpha\beta}_{dB} + \frac{9}{2} g^{}_2 C^{\alpha\beta}_{dW} + 4 g^{}_s C^{\alpha\beta}_{dG} - \left( Y^{}_{\rm d} \right)^{}_{\alpha\gamma} C^{\gamma\beta}_{Hd} - \left( Y^{}_{\rm u} \right)^{}_{\alpha\gamma} C^{\gamma\beta}_{Hud} -  C^{\ast\gamma\lambda\beta\alpha}_{\ell edq} \left( Y^{}_l \right)^{}_{\gamma\lambda}   \right.
	\nonumber
	\\
	&& + 2 \left( C^{(1)\alpha\gamma\lambda\beta}_{qd} + \frac{4}{3} C^{(8)\alpha\gamma\lambda\beta}_{qd}  \right) \left( Y^{}_{\rm d} \right)^{}_{\gamma\lambda} - 3 C^{(1)\gamma\lambda\alpha\beta}_{quqd} \left( Y^\dagger_{\rm u} \right)^{}_{\lambda\gamma} 
	\nonumber
	\\
	&&-  \left. \frac{1}{2} \left( C^{(1)\alpha\gamma\lambda\beta}_{quqd} + \frac{4}{3} C^{(8)\alpha\gamma\lambda\beta}_{quqd} \right) \left( Y^\dagger_{\rm u} \right)^{}_{\gamma\lambda} \right] \;,
	\nonumber
	\\
	\delta G^{}_{dHD2} &=& \loopf \frac{1}{2} \left( - C^{(1)}_{Hq} Y^{}_{\rm d} - 3 C^{(3)}_{Hq} Y^{}_{\rm d} + Y^{}_{\rm d} C^{}_{Hd} + Y^{}_{\rm u} C^{}_{Hud} \right) \;,
	\nonumber
	\\
	\delta G^{}_{dHD4} &=& \loopf \left( g^{}_1 C^{}_{dB} + 9g^{}_2 C^{}_{dW} + 8g^{}_s C^{}_{dG} - \frac{1}{2} C^{(1)}_{Hq} Y^{}_{\rm d} - \frac{3}{2} C^{(3)}_{Hq} Y^{}_{\rm d} - \frac{3}{2} Y^{}_{\rm d} C^{}_{Hd} - \frac{3}{2} Y^{}_{\rm u} C^{}_{Hud} \right) \;.\;\;\quad
\end{eqnarray}

\noindent$\bullet~\bm{\psi^2XD}$

\begin{eqnarray}
	\delta G^\prime_{B\ell} &=& \loopf \rmi  \left( C^{}_{eB} Y^\dagger_l - Y^{}_l C^\dagger_{eB} \right) \;,
	\nonumber
	\\
	\delta G^\prime_{\widetilde{B}\ell} &=& \loopf \left( C^{}_{eB} Y^\dagger_l + Y^{}_l C^\dagger_{eB} \right) \;,
	\nonumber
	\\
	\delta G^\prime_{W\ell} &=& \loopf \rmi \left( C^{}_{eW} Y^\dagger_l - Y^{}_l C^\dagger_{eW} \right) \;,
	\nonumber
	\\
	\delta G^\prime_{\widetilde{W}\ell} &=& \loopf \left( C^{}_{eW} Y^\dagger_l + Y^{}_l C^\dagger_{eW} \right) \;.
\end{eqnarray}

\noindent$\bullet~\bm{\psi^2 D H^2}$
\begin{eqnarray}\label{eq:coun-dim6-hl}
	\delta G^{\prime(1)}_{H\ell} &=& \loopf \frac{1}{8} \left[ \left( C^{}_{HD} - 2C^{}_{H\square} \right) Y^{}_l Y^\dagger_l - 3 g^{}_1 \left( C^{}_{eB} Y^\dagger_l + Y^{}_l C^\dagger_{eB} \right) + 9 g^{}_2 \left( C^{}_{eW} Y^\dagger_l + Y^{}_l C^\dagger_{eW} \right) \right.
	\nonumber
	\\
	&& + \left. 3 \left( C^{}_{eH} Y^\dagger_l + Y^{}_l C^\dagger_{eH} \right) + Y^{}_l Y^\dagger_l C^{(1)}_{H\ell} + C^{(1)}_{H\ell} Y^{}_l Y^\dagger_l + 3Y^{}_l Y^\dagger_l C^{(3)}_{H\ell} + 3C^{(3)}_{H\ell} Y^{}_l Y^\dagger_l  - 6 C^{}_5 C^\dagger_5  \right] \;,
	\nonumber
	\\
	\delta G^{\prime(3)}_{H\ell} &=& \loopf \frac{1}{8} \left[ \left( C^{}_{HD} - 2C^{}_{H\square} \right) Y^{}_l Y^\dagger_l - 3 g^{}_1 \left( C^{}_{eB} Y^\dagger_l + Y^{}_l C^\dagger_{eB} \right) - 3 g^{}_2 \left( C^{}_{eW} Y^\dagger_l + Y^{}_l C^\dagger_{eW} \right) \right.
	\nonumber
	\\
	&& + \left. C^{}_{eH} Y^\dagger_l + Y^{}_l C^\dagger_{eH} + Y^{}_l Y^\dagger_l C^{(1)}_{H\ell} + C^{(1)}_{H\ell} Y^{}_l Y^\dagger_l - Y^{}_l Y^\dagger_l C^{(3)}_{H\ell} - C^{(3)}_{H\ell} Y^{}_l Y^\dagger_l  + 4 C^{}_5 C^\dagger_5 \right] \;,
	\nonumber
	\\
	\delta G^{\prime\prime(1)}_{H\ell} &=& - \loopf \frac{\rmi}{8} \left[ 3g^{}_1 \left( C^{}_{eB} Y^\dagger_l - Y^{}_l C^\dagger_{eB} \right) + 9 g^{}_2 \left( C^{}_{eW} Y^\dagger_l - Y^{}_l C^\dagger_{eW}  \right) + 12 \rmi \left( g^2_1 C^{}_{H\widetilde{B}} + 3 g^2_2 C^{}_{H\widetilde{W}} \right) \mathbbm{1} \right.
	\nonumber
	\\
	&&+ \left. 3 \left( C^{}_{eH} Y^\dagger_l - Y^{}_l C^\dagger_{eH} \right) + \left( C^{(1)}_{H\ell} Y^{}_l Y^\dagger_l - Y^{}_l Y^\dagger_l C^{(1)}_{H\ell} \right) +3  \left( C^{(3)}_{H\ell} Y^{}_l Y^\dagger_l - Y^{}_l Y^\dagger_l C^{(3)}_{H\ell} \right)  \right] \;,
	\nonumber
	\\
	\delta G^{\prime\prime(3)}_{H\ell} &=& - \loopf \frac{\rmi}{8} \left[ 3g^{}_1 \left( C^{}_{eB} Y^\dagger_l - Y^{}_l C^\dagger_{eB} \right) -3 g^{}_2 \left( C^{}_{eW} Y^\dagger_l - Y^{}_l C^\dagger_{eW}  \right) - 12\rmi g^{}_1 g^{}_2 C^{}_{H\widetilde{W}B} \mathbbm{1} \right.
	\nonumber
	\\
	&&+ \left.  \left( C^{}_{eH} Y^\dagger_l - Y^{}_l C^\dagger_{eH} \right) + \left( C^{(1)}_{H\ell} Y^{}_l Y^\dagger_l - Y^{}_l Y^\dagger_l C^{(1)}_{H\ell} \right) -  \left( C^{(3)}_{H\ell} Y^{}_l Y^\dagger_l - Y^{}_l Y^\dagger_l C^{(3)}_{H\ell} \right)  \right] \;.
\end{eqnarray}

As can be seen from the above results, $\Opr^{}_{DH}$ and $\Opr^{}_{\ell D}$ do not have the corresponding counterterms in the SMEFT up to $\mathcal{O} \left( \Lambda^{-3} \right)$, and the counterterms for $\Opr^\prime_{HD}$, $\Opr^{\prime(1)}_{H\ell}$ and $\Opr^{\prime(3)}_{H\ell}$ have a contribution from double insertions of the dim-5 operator, besides that from single insertions of dim-6 operators. Therefore, they will contribute both to $\gamma^{(7,5)}$ and to $\gamma^{(7,6)}$ after the SMEFT EoMs up to $\mathcal{O} \left( \Lambda^{-1} \right)$ are exploited.

\subsection{Counterterms for dim-7 operators}

There is no contribution from 1PI diagrams with triple insertions of the dim-5 operator to the counterterms for dim-7 operators. Those from insertions of one dim-5 operator and one dim-6 operator and single insertions of dim-7 operators are given in the following two subsections, respectively. For the latter, only the result for $\Opr^{(S)}_{\ell HD3}$ is presented, which makes a contribution to the RGE of the dim-5 operator via the reduction relation in Eq.~\eqref{eq:convertion-c5}.

\subsubsection{Contributions from insertions of one dim-5 operator and one dim-6 operator}

\noindent$\bullet~\bm{\psi^2H^4}$
\begin{eqnarray}
	\delta G^{(S)}_{\ell H} &=& \loopf \left\{ - C^\dagger_5 \left[ 3C^{}_H + 2\lambda \left( C^{}_{HD} - 6 C^{}_{H\square} \right) \right] + \frac{3}{2} C^\dagger_5 \left( g^2_1 C^{}_{HB} +g^2_2 C^{}_{HW} + g^{}_1 g^{}_2 C^{}_{HWB} \right) \right.
	\nonumber
	\\
	&& + \frac{1}{4} \left[ 8\lambda - 3\left( g^2_1 + g^2_2 \right) \right] \left[ C^\dagger_5 C^{(1)}_{H\ell} +  \left( C^\dagger_5 C^{(1)}_{H\ell} \right)^{\rm T} \right] - \frac{1}{4} \left[ 16\lambda - 3\left( g^2_1 + g^2_2 \right) \right]
	\nonumber
	\\
	&& \times \left[ C^\dagger_5 C^{(3)}_{H\ell} + \left( C^\dagger_5 C^{(3)}_{H\ell} \right)^{\rm T}  \right] - C^\dagger_5 Y^{}_l Y^\dagger_l C^{(3)}_{H\ell} - \left( C^\dagger_5 Y^{}_l Y^\dagger_l C^{(3)}_{H\ell} \right)^{\rm T} + \frac{1}{2} C^\dagger_5 C^{}_{eH} Y^\dagger_l
	\nonumber
	\\
	&& + \left. \frac{1}{2} \left( C^\dagger_5 C^{}_{eH} Y^\dagger_l \right)^{\rm T}  + \frac{1}{2} C^\dagger_5 Y^{}_l C^\dagger_{eH} + \frac{1}{2} \left( C^\dagger_5 Y^{}_l C^\dagger_{eH} \right)^{\rm T} \right\}  \;.
\end{eqnarray}

\noindent$\bullet~\bm{\psi^2H^3 D}$
\begin{eqnarray}
	\delta G^{\alpha\beta}_{\ell eHD} &=& \loopf \left[ \left( C^{}_{H\square} - \frac{3}{4} C^{}_{HD} \right) \left( C^\dagger_5 Y^{}_l \right)^{\alpha\beta} + 6g^{}_2 \left( C^\dagger_5 C^{}_{eW} \right)^{\alpha\beta} +  \frac{1}{2} \left( C^\dagger_5 C^{}_{eH} \right)^{\alpha\beta} \right.
	\nonumber
	\\
	&& - \left. \frac{5}{2} \left( C^\dagger_5 Y^{}_l C^{}_{He} \right)^{\alpha\beta} - 2 \left( C^{(1)\gamma\alpha}_{H\ell} + C^{(3)\gamma\alpha}_{H\ell} \right) \left( C^\dagger_5 Y^{}_l \right)^{\gamma\beta}  - 2 \left( C^\dagger_5 Y^{}_l \right)^{\gamma\lambda} C^{\gamma\alpha\lambda\beta}_{\ell e } \right] \;.
\end{eqnarray}

\noindent$\bullet~\bm{\psi^2H^2 D^2}$
\begin{eqnarray}
	\delta G^{\alpha\beta}_{\ell HD1} &=& \loopf \left[ \frac{1}{2} \left( C^\dagger_5 \right)^{\alpha\beta} C^{}_{HD} + \frac{1}{2} \left( C^\dagger_5 C^{(1)}_{H\ell} \right)^{\alpha\beta} + \left( C^\dagger_5 C^{(1)}_{H\ell} \right)^{\beta\alpha} - \frac{5}{2} \left( C^\dagger_5 C^{(3)}_{H\ell} \right)^{\alpha\beta}   \right. 
	\nonumber
	\\
	&& - \left. 5 \left( C^\dagger_5 C^{(3)}_{H\ell} \right)^{\beta\alpha} + 4  \left( C^\dagger_5 \right)^{\gamma\lambda} C^{\gamma\alpha\lambda\beta}_{\ell\ell} \right] \;,
	\nonumber
	\\
	\delta G^{\alpha\beta}_{\ell HD2} &=& \loopf \left[ - \left( C^\dagger_5 \right)^{\alpha\beta} C^{}_{HD} - \left( C^\dagger_5 C^{(1)}_{H\ell} \right)^{\alpha\beta} - 2 \left( C^\dagger_5 C^{(1)}_{H\ell} \right)^{\beta\alpha} + 2 \left( C^\dagger_5 C^{(3)}_{H\ell} \right)^{\alpha\beta}  \right.
	\nonumber
	\\
	&& + \left. 4 \left( C^\dagger_5 C^{(3)}_{H\ell} \right)^{\beta\alpha}  - 8 \left( C^\dagger_5 \right)^{\gamma\lambda} C^{\gamma\alpha\lambda\beta}_{\ell\ell} \right] \;,
	\nonumber
	\\
	\delta G^{(S)\alpha\beta}_{\ell HD3} &=& \loopf \left[ \left( C^\dagger_5 \right)^{\alpha\beta} C^{}_{H\square} -  \left( C^\dagger_5 C^{(1)}_{H\ell} \right)^{\alpha\beta} - \left( C^\dagger_5 C^{(1)}_{H\ell} \right)^{\beta\alpha} + 2 \left( C^\dagger_5 C^{(3)}_{H\ell} \right)^{\alpha\beta} + 2 \left( C^\dagger_5 C^{(3)}_{H\ell} \right)^{\beta\alpha} \right] \;,
	\nonumber
	\\
	\delta G^{(S)\alpha\beta}_{\ell HD4} &=& \loopf \left[ - \frac{1}{2}  \left( C^\dagger_5 C^{(1)}_{H\ell} \right)^{\alpha\beta} - \frac{1}{2} \left( C^\dagger_5 C^{(1)}_{H\ell} \right)^{\beta\alpha} +  \left( C^\dagger_5 C^{(3)}_{H\ell} \right)^{\alpha\beta} +  \left( C^\dagger_5 C^{(3)}_{H\ell} \right)^{\beta\alpha} \right] \;,
	\nonumber
	\\
	\delta G^{\alpha\beta}_{\ell HD5} &=&\loopf  \frac{3}{2} \rmi \left( C^\dagger_5 C^{(3)}_{H\ell} \right)^{\alpha\beta}  \;,
	\nonumber
	\\
	\delta G^{(S)\alpha\beta}_{\ell HD6} &=& \loopf \left[  - \frac{1}{4} \rmi  \left( C^\dagger_5 C^{(1)}_{H\ell} \right)^{\alpha\beta} - \frac{1}{4} \rmi \left( C^\dagger_5 C^{(1)}_{H\ell} \right)^{\beta\alpha} + \frac{1}{2} \rmi \left( C^\dagger_5 C^{(3)}_{H\ell} \right)^{\alpha\beta} + \frac{1}{2} \rmi \left( C^\dagger_5 C^{(3)}_{H\ell} \right)^{\beta\alpha} \right] \;.
\end{eqnarray}

\noindent$\bullet~\bm{\psi^2H^2X}$
\begin{eqnarray}
	\delta G^{(A)}_{\ell HB} &=& \loopf \left\{ - \frac{1}{16} g^{}_1 \left[ C^\dagger_5 C^{(1)}_{H\ell} - \left( C^\dagger_5 C^{(1)}_{H\ell}  \right)^{\rm T} \right] + \frac{1}{8} g^{}_1 \left[ C^\dagger_5 C^{(3)}_{H\ell} - \left( C^\dagger_5 C^{(3)}_{H\ell}  \right)^{\rm T} \right]  \right.
	\nonumber
	\\
	&& + \left. \frac{1}{2} \left[  C^\dagger_5 Y^{}_l C^\dagger_{eB} - \left( C^\dagger_5 Y^{}_l C^\dagger_{eB} \right)^{\rm T} \right] \right\}\;,
	\nonumber
	\\
	\delta G^{}_{\ell HW} &=& \loopf \left\{ \frac{1}{8} g^{}_2 \left[ C^\dagger_5 C^{(1)}_{H\ell} + 2 \left( C^\dagger_5 C^{(1)}_{H\ell}  \right)^{\rm T} +  4C^\dagger_5 C^{(3)}_{H\ell} + 2 \left( C^\dagger_5 C^{(3)}_{H\ell}  \right)^{\rm T} \right]  - C^\dagger_5 Y^{}_l C^\dagger_{eW}  \right.
	\nonumber
	\\
	&& + \left. g^{}_2 \left( C^{}_{HW} -\rmi C^{}_{H\widetilde{W}} \right) C^\dagger_5  - \frac{3}{4} g^2_2 \left( C^{}_{3W} - \rmi C^{}_{3\widetilde{W} } \right) C^\dagger_5  \right\}  \;.
\end{eqnarray}

\noindent$\bullet~\bm{\psi^4H}$
\begin{eqnarray}\label{eq:ct-psi4h}
	\delta G^{(S)\alpha\beta\gamma\lambda}_{\overline{e}\ell\ell\ell H} &=& \loopf \left\{  -2 g^{}_2 \left[ \left( C^\dagger_{eW} \right)^{\alpha\beta} \left( C^\dagger_5 \right)^{\gamma\lambda} + \left( C^\dagger_{eW} \right)^{\alpha\lambda} \left( C^\dagger_5 \right)^{\beta\gamma} + \left( C^\dagger_{eW} \right)^{\alpha\gamma} \left( C^\dagger_5 \right)^{\lambda\beta} \right] \right.
	\nonumber
	\\
	&& + \frac{1}{3} C^{\alpha\rho}_{He} \left[ \left( Y^\dagger_l \right)^{}_{\rho\beta} \left( C^\dagger_5 \right)^{\gamma\lambda} + \left( Y^\dagger_l \right)^{}_{\rho\lambda} \left( C^\dagger_5 \right)^{\beta\gamma} + \left( Y^\dagger_l \right)^{}_{\rho\gamma} \left( C^\dagger_5 \right)^{\lambda\beta} \right]
	\nonumber
	\\
	&& - \frac{1}{3} \left( Y^\dagger_l \right)^{}_{\alpha\rho} \left[ \left( C^\dagger_5 \right)^{\beta\gamma} \left( C^{(1)}_{H\ell} +  C^{(3)}_{H\ell}  \right)^{\rho\lambda} +  \left( C^\dagger_5 \right)^{\gamma\lambda} \left( C^{(1)}_{H\ell} +  C^{(3)}_{H\ell}  \right)^{\rho\beta} \right.
	\nonumber
	\\
	&& + \left. \left( C^\dagger_5 \right)^{\lambda\beta} \left( C^{(1)}_{H\ell} +  C^{(3)}_{H\ell}  \right)^{\rho\gamma} \right] - \frac{2}{3} \left( Y^\dagger_l \right)^{}_{\alpha\sigma} \left[ \left( C^\dagger_5 \right)^{\beta\rho} \left( C^{\rho\gamma\sigma\lambda}_{\ell\ell} + C^{\sigma\gamma\rho\lambda}_{\ell\ell} \right) \right.
	\nonumber
	\\
	&& + \left.  \left( C^\dagger_5 \right)^{\lambda\rho} \left( C^{\rho\beta\sigma\gamma}_{\ell\ell} + C^{\sigma\beta\rho\gamma}_{\ell\ell} \right) + \left( C^\dagger_5 \right)^{\gamma\rho} \left( C^{\rho\lambda\sigma\beta}_{\ell\ell} + C^{\sigma\lambda\rho\beta}_{\ell\ell} \right) \right] 
	\nonumber
	\\
	&& + \frac{1}{3} C^{\sigma\gamma\alpha\rho}_{\ell e} \left[ \left( C^\dagger_5 \right)^{\beta\sigma} \left( Y^\dagger_l \right)^{}_{\rho\lambda} + \left( C^\dagger_5 \right)^{\lambda\sigma} \left( Y^\dagger_l \right)^{}_{\rho\beta} \right] + \frac{1}{3} C^{\sigma\beta\alpha\rho}_{\ell e} \left[ \left( C^\dagger_5 \right)^{\lambda\sigma} \left( Y^\dagger_l \right)^{}_{\rho\gamma} \right.
	\nonumber
	\\
	&& + \left. \left( C^\dagger_5 \right)^{\gamma\sigma} \left( Y^\dagger_l \right)^{}_{\rho\lambda} \right] + \frac{1}{3} C^{\sigma\lambda\alpha\rho}_{\ell e} \left[ \left( C^\dagger_5 \right)^{\gamma\sigma} \left( Y^\dagger_l \right)^{}_{\rho\beta} + \left( C^\dagger_5 \right)^{\beta\sigma} \left( Y^\dagger_l \right)^{}_{\rho\gamma} \right]
	\nonumber
	\\
	&& +\left. \frac{1}{3} \left[  \left(C^\dagger_{eH} \right)^{\alpha\beta} \left( C^\dagger_5 \right)^{\gamma\lambda} + \left(C^\dagger_{eH} \right)^{\alpha\gamma} \left( C^\dagger_5 \right)^{\lambda\beta} + \left(C^\dagger_{eH} \right)^{\alpha\lambda} \left( C^\dagger_5 \right)^{\beta\gamma} \right] \right\}  \;,
	\nonumber
	\\
	\delta G^{(A)\alpha\beta\gamma\lambda}_{\overline{e}\ell\ell\ell H} &=& \loopf \left\{ 2 \left( Y^\dagger_l \right)^{}_{\alpha\sigma} \left[ \left( C^\dagger_5 \right)^{\beta\rho} \left( C^{\rho\gamma\sigma\lambda}_{\ell\ell} - C^{\sigma\gamma\rho\lambda}_{\ell\ell} \right) + \left( C^\dagger_5 \right)^{\lambda\rho} \left( C^{\rho\beta\sigma\gamma}_{\ell\ell} - C^{\sigma\beta\rho\gamma}_{\ell\ell} \right) \right.\right.
	\nonumber
	\\
	&& + \left.  \left( C^\dagger_5 \right)^{\gamma\rho} \left( C^{\rho\lambda\sigma\beta}_{\ell\ell} - C^{\sigma\lambda\rho\beta}_{\ell\ell} \right) \right] + C^{\sigma\gamma\alpha\rho}_{\ell e} \left[ \left( C^\dagger_5 \right)^{\beta\sigma} \left( Y^\dagger_l \right)^{}_{\rho\lambda} - \left( C^\dagger_5 \right)^{\lambda\sigma} \left( Y^\dagger_l \right)^{}_{\rho\beta} \right] 
	\nonumber
	\\
	&& + C^{\sigma\beta\alpha\rho}_{\ell e} \left[ \left( C^\dagger_5 \right)^{\lambda\sigma} \left( Y^\dagger_l \right)^{}_{\rho\gamma} - \left( C^\dagger_5 \right)^{\gamma\sigma} \left( Y^\dagger_l \right)^{}_{\rho\lambda} \right]  + C^{\sigma\lambda\alpha\rho}_{\ell e} \left[ \left( C^\dagger_5 \right)^{\gamma\sigma} \left( Y^\dagger_l \right)^{}_{\rho\beta} \right.
	\nonumber
	\\
	&& - \left.\left.  \left( C^\dagger_5 \right)^{\beta\sigma} \left( Y^\dagger_l \right)^{}_{\rho\gamma} \right] \right\} \;,
	\nonumber
	\\
	\delta G^{(M)\alpha\beta\gamma\lambda}_{\overline{e}\ell\ell\ell H} &=& \loopf \left\{ 3 g^{}_2 \left[ \left( C^\dagger_{eW} \right)^{\alpha\beta} \left( C^\dagger_5 \right)^{\gamma\lambda} - \left( C^\dagger_{eW} \right)^{\alpha\lambda} \left( C^\dagger_5 \right)^{\beta\gamma} \right] \right.
	\nonumber
	\\
	&& + \frac{1}{2} C^{\alpha\rho}_{He} \left[ \left( Y^\dagger_l \right)^{}_{\rho\beta} \left( C^\dagger_5 \right)^{\gamma\lambda} - \left( Y^\dagger_l \right)^{}_{\rho\lambda} \left( C^\dagger_5 \right)^{\beta\gamma} \right]
	\nonumber
	\\
	&& + \frac{1}{2} \left( Y^\dagger_l \right)^{}_{\alpha\rho} \left[ \left( C^\dagger_5 \right)^{\beta\gamma} \left( C^{(1)}_{H\ell} +  C^{(3)}_{H\ell}  \right)^{\rho\lambda} -  \left( C^\dagger_5 \right)^{\gamma\lambda} \left( C^{(1)}_{H\ell} +  C^{(3)}_{H\ell}  \right)^{\rho\beta} \right]
	\nonumber
	\\
	&& - 2 \left( Y^\dagger_l \right)^{}_{\alpha\sigma} \left[ \left( C^\dagger_5 \right)^{\beta\rho}   C^{\sigma\gamma\rho\lambda}_{\ell\ell} - \left( C^\dagger_5 \right)^{\lambda\rho}  C^{\rho\beta\sigma\gamma}_{\ell\ell} + \left( C^\dagger_5 \right)^{\gamma\rho} \left( C^{\rho\lambda\sigma\beta}_{\ell\ell} - C^{\sigma\lambda\rho\beta}_{\ell\ell} \right) \right] 
	\nonumber
	\\
	&& - C^{\sigma\gamma\alpha\rho}_{\ell e} \left[ \left( C^\dagger_5 \right)^{\beta\sigma} \left( Y^\dagger_l \right)^{}_{\rho\lambda} - \left( C^\dagger_5 \right)^{\lambda\sigma} \left( Y^\dagger_l \right)^{}_{\rho\beta} \right] + C^{\sigma\beta\alpha\rho}_{\ell e} \left( C^\dagger_5 \right)^{\lambda\sigma} \left( Y^\dagger_l \right)^{}_{\rho\gamma} 
	\nonumber
	\\
	&& - \left. C^{\sigma\lambda\alpha\rho}_{\ell e} \left( C^\dagger_5 \right)^{\beta\sigma} \left( Y^\dagger_l \right)^{}_{\rho\gamma} + \frac{1}{2} \left[  \left(C^\dagger_{eH} \right)^{\alpha\beta} \left( C^\dagger_5 \right)^{\gamma\lambda} - \left(C^\dagger_{eH} \right)^{\alpha\lambda} \left( C^\dagger_5 \right)^{\beta\gamma} \right]  \right\} \;,
	\nonumber
	\\
	\delta G^{\alpha\beta\gamma\lambda}_{\overline{d}\ell q \ell H 1} &=&  \loopf \left\{ - 2 \left( Y^\dagger_{\rm d} \right)^{}_{\alpha\rho} \left( C^\dagger_5 \right)^{\beta\lambda} \left( C^{(1)}_{Hq} + C^{(3)}_{Hq} \right)^{\rho\gamma}  - 4 \left( Y^\dagger_{\rm d} \right)^{}_{\alpha\sigma}  \left( C^\dagger_5 \right)^{\beta\rho} \left( C^{(1)}_{\ell q} - 2 C^{(3)}_{\ell q} \right)^{\rho\lambda\sigma\gamma} \right.
	\nonumber
	\\
	&& + 2 \left( Y^\dagger_{\rm d} \right)^{}_{\rho\gamma}  \left( C^\dagger_5 \right)^{\beta\lambda} C^{\alpha\rho}_{Hd}   - C^{\rho\sigma\alpha\gamma}_{\ell e d q} \left[ \left( Y^\dagger_l \right)^{}_{\sigma\beta}  \left( C^\dagger_5 \right)^{\rho\lambda} + \left( Y^\dagger_l \right)^{}_{\sigma\lambda}  \left( C^\dagger_5 \right)^{\rho\beta} \right]
	\nonumber
	\\
	&& + \left. 4  \left( Y^\dagger_{\rm d} \right)^{}_{\sigma\gamma}  \left( C^\dagger_5 \right)^{\rho\lambda} C^{\rho\beta\alpha\sigma}_{\ell d} + 2 \left( C^\dagger_{dH} \right)^{\alpha\gamma} \left( C^\dagger_5\right)^{\beta\lambda} \right\} \;,
	\nonumber
	\\
	\delta G^{\alpha\beta\gamma\lambda}_{\overline{d}\ell q \ell H 2} &=&  \loopf \left\{ -6g^{}_2 \left( C^\dagger_{dW} \right)^{\alpha\gamma} \left( C^\dagger_5 \right)^{\beta\lambda} + \left( Y^\dagger_{\rm d} \right)^{}_{\alpha\rho} \left( C^\dagger_5 \right)^{\beta\lambda} \left( C^{(1)}_{Hq} + C^{(3)}_{Hq} \right)^{\rho\gamma}   \right.
	\nonumber
	\\
	&& -  \left( Y^\dagger_{\rm d} \right)^{}_{\rho\gamma}  \left( C^\dagger_5 \right)^{\beta\lambda} C^{\alpha\rho}_{Hd} + 2 \left( Y^\dagger_{\rm d} \right)^{}_{\alpha\sigma}  \left( C^\dagger_5 \right)^{\beta\rho} \left( C^{(1)}_{\ell q} - 5 C^{(3)}_{\ell q} \right)^{\rho\lambda\sigma\gamma}  
	\nonumber
	\\
	&& - 2 \left( Y^\dagger_{\rm d} \right)^{}_{\sigma\gamma}  \left( C^\dagger_5 \right)^{\rho\lambda} C^{\rho\beta\alpha\sigma}_{\ell d} - C^{\rho\sigma\alpha\gamma}_{\ell e d q} \left[ \left( Y^\dagger_l \right)^{}_{\sigma\beta}  \left( C^\dagger_5 \right)^{\rho\lambda} - 2 \left( Y^\dagger_l \right)^{}_{\sigma\lambda}  \left( C^\dagger_5 \right)^{\rho\beta} \right] 
	\nonumber
	\\
	&& - \left.  \left( C^\dagger_{dH} \right)^{\alpha\gamma} \left( C^\dagger_5\right)^{\beta\lambda} \right\} \;,
	\nonumber
	\\
	\delta G^{\alpha\beta\gamma\lambda}_{\overline{d}\ell u e H} &=&  \loopf \left[ -3 \left( C^\dagger_{Hud} \right)^{\alpha\gamma} \left( C^\dagger_5 \right)^{\beta\rho} \left( Y^{}_l \right)^{}_{\rho\lambda} - \frac{3}{2} \left( C^\dagger_5 \right)^{\beta\sigma} \left( Y^{}_{\rm u} \right)^{}_{\rho\gamma} C^{\sigma\lambda\alpha\rho}_{\ell e d q} \right.
	\nonumber
	\\
	&& + \left.  \frac{3}{2} \left( Y^\dagger_{\rm d} \right)^{}_{\alpha\rho} \left( C^\dagger_5 \right)^{\beta\sigma} \left( C^{(1)}_{\ell e qu} - 12 C^{(3)}_{\ell e q u} \right)^{\sigma\lambda\rho\gamma} \right] \;,
	\nonumber
	\\
	\delta G^{\alpha\beta\gamma\lambda}_{\overline{q}u\ell\ell H} &=&  \loopf \left\{ - \left( C^{(1)}_{Hq} - C^{(3)}_{Hq} \right)^{\alpha\rho} \left( Y^{}_{\rm u} \right)^{}_{\rho \beta} \left( C^\dagger_5 \right)^{\gamma\lambda} + \left( Y^{}_{\rm u} \right)^{}_{\alpha\rho} C^{\rho\beta}_{Hu} \left( C^\dagger_5 \right)^{\gamma\lambda} \right.
	\nonumber
	\\
	&& - \left( Y^{}_{\rm u} \right)^{}_{\rho \beta} \left[ \left( C^{(1)}_{\ell q} + 5 C^{(3)}_{\ell q} \right)^{\sigma\gamma\alpha\rho} \left( C^\dagger_5 \right)^{\sigma\lambda} + \left( C^{(1)}_{\ell q} - C^{(3)}_{\ell q} \right)^{\sigma\lambda\alpha\rho} \left( C^\dagger_5 \right)^{\sigma\gamma}  \right]
	\nonumber
	\\
	&& + \left( Y^{}_{\rm u} \right)^{}_{\alpha\rho} \left[ \left( C^\dagger_5 \right)^{\lambda\sigma} C^{\sigma\gamma\rho\beta}_{\ell u} +   \left( C^\dagger_5 \right)^{\gamma\sigma} C^{\sigma\lambda\rho\beta}_{\ell u}  \right] + C^{(1)\sigma\rho\alpha\beta}_{\ell e q u} \left[ \left( C^\dagger_5 \right)^{\sigma\gamma} \left( Y^\dagger_l \right)^{}_{\rho\lambda} \right.
	\nonumber
	\\
	&& - \left.\left. 2 \left( C^\dagger_5 \right)^{\sigma\lambda} \left( Y^\dagger_l \right)^{}_{\rho\gamma} \right]  - C^{\alpha\beta}_{uH} \left( C^\dagger_5 \right)^{\gamma\lambda} \right\}  \;,
	\nonumber
	\\
	\delta G^{\alpha\beta\gamma\lambda}_{\overline{\ell}dud\widetilde{H}} &=& - \loopf 3 C^{\alpha\rho}_5 \left( Y^{}_{\rm d} \right)^{}_{\sigma\beta} C^{\lambda\gamma\sigma\rho}_{duq} \;,
	\nonumber
	\\
	\delta G^{\alpha\beta\gamma\lambda}_{\overline{\ell}dqq\widetilde{H}} &=&  \loopf \left[ - C^{\alpha\rho}_5 \left( Y^\dagger_{\rm u} \right)^{}_{\sigma\gamma} C^{\beta\sigma\lambda\rho}_{duq} - \frac{1}{2} C^{\alpha\rho}_5 \left( Y^\dagger_{\rm u} \right)^{}_{\sigma\lambda} C^{\beta\sigma\gamma\rho}_{duq} + \frac{1}{2} C^{\alpha\rho}_5 \left( Y^{}_{\rm d} \right)^{}_{\sigma\beta} \right.
	\nonumber
	\\
	&& \times \left. \left( 9 C^{(S)\gamma\lambda\sigma\rho}_{qqq} - 3 C^{(A)\gamma\lambda\sigma\rho}_{qqq} + 8 C^{(M)\gamma\lambda\sigma\rho}_{qqq} +10 C^{(M)\sigma\gamma\lambda\rho}_{qqq}  \right) \vphantom{\frac{1}{2}}\right]  \;.
\end{eqnarray}
In the last equality of Eq.~\eqref{eq:ct-psi4h}, $C^{(S),(A),(M)}_{qqq}$ are the Wilson coefficients of the dim-6 operators $\Op^{(S),(A),(M)}_{qqq}$, which are the totally symmetric, totally antisymmetric and mixed symmetric combinations of $\Op^{\alpha\beta\gamma\lambda}_{qqq} = \epsilon^{ijk} \epsilon^{ad} \epsilon^{be} \left( Q^{ia}_{\alpha\rm L} C Q^{jb}_{\beta\rm L} \right) \left( Q^{ke}_{\gamma\rm L} C \ell^d_{\lambda \rm L} \right)$, namely,
\begin{eqnarray}\label{eq:Oqqq}
	\Op^{(S)\alpha\beta\gamma\lambda}_{qqq} &=& \frac{1}{6} \left( \Op^{\alpha\beta\gamma\lambda}_{qqq} +  \Op^{\gamma\alpha\beta\lambda}_{qqq} + \Op^{\beta\gamma\alpha\lambda}_{qqq} + \Op^{\alpha\gamma\beta\lambda}_{qqq} + \Op^{\beta\alpha\gamma\lambda}_{qqq} + \Op^{\gamma\beta\alpha\lambda}_{qqq} \right) \;,
	\nonumber
	\\
	\Op^{(A)\alpha\beta\gamma\lambda}_{qqq} &=& \frac{1}{6} \left( \Op^{\alpha\beta\gamma\lambda}_{qqq} +  \Op^{\gamma\alpha\beta\lambda}_{qqq} + \Op^{\beta\gamma\alpha\lambda}_{qqq} - \Op^{\alpha\gamma\beta\lambda}_{qqq} - \Op^{\beta\alpha\gamma\lambda}_{qqq} - \Op^{\gamma\beta\alpha\lambda}_{qqq} \right) \;,
	\nonumber
	\\
	\Op^{(M)\alpha\beta\gamma\lambda}_{qqq} &=& \frac{1}{3} \left( \Op^{\alpha\beta\gamma\lambda}_{qqq} +  \Op^{\gamma\beta\alpha\lambda}_{qqq} - \Op^{\beta\alpha\gamma\lambda}_{qqq} - \Op^{\beta\gamma\alpha\lambda}_{qqq} \right) \;,
\end{eqnarray}
with 
\begin{eqnarray}
	C^{(S)\alpha\beta\gamma\lambda}_{qqq} &=& \frac{1}{6} \left( C^{\alpha\beta\gamma\lambda}_{qqq} +  C^{\gamma\alpha\beta\lambda}_{qqq} + C^{\beta\gamma\alpha\lambda}_{qqq} + C^{\alpha\gamma\beta\lambda}_{qqq} +C^{\beta\alpha\gamma\lambda}_{qqq} + C^{\gamma\beta\alpha\lambda}_{qqq} \right) \;,
	\nonumber
	\\
	C^{(A)\alpha\beta\gamma\lambda}_{qqq} &=& \frac{1}{6} \left( C^{\alpha\beta\gamma\lambda}_{qqq} +  C^{\gamma\alpha\beta\lambda}_{qqq} + C^{\beta\gamma\alpha\lambda}_{qqq} - C^{\alpha\gamma\beta\lambda}_{qqq} - C^{\beta\alpha\gamma\lambda}_{qqq} - C^{\gamma\beta\alpha\lambda}_{qqq} \right) \;,
	\nonumber
	\\
	C^{(M)\alpha\beta\gamma\lambda}_{qqq} &=& \frac{1}{6} \left( 2C^{\alpha\beta\gamma\lambda}_{qqq} - C^{\gamma\alpha\beta\lambda}_{qqq} - C^{\beta\gamma\alpha\lambda}_{qqq} - C^{\alpha\gamma\beta\lambda}_{qqq} - C^{\beta\alpha\gamma\lambda}_{qqq} + 2 C^{\gamma\beta\alpha\lambda}_{qqq} \right).
\end{eqnarray}
The other mixed symmetric combination $\Op^{(M^\prime)}_{qqq}$ is vanishing thanks to $\Op^{\alpha\beta\gamma\lambda}_{qqq} + \Op^{\beta\alpha\gamma\lambda}_{qqq} - \Op^{\gamma\alpha\beta\lambda}_{qqq} - \Op^{\gamma\beta\alpha\lambda}_{qqq} = 0 $.

\subsubsection{Contributions from insertions of one dim-7 operator}

\begin{eqnarray}
	\delta G^{(S)\alpha\beta}_{\ell HD3} &=& \loopf \left[ \frac{3}{2} g^2_2 C^{(S)\alpha\beta}_{\ell HD1} + \frac{3}{4} g^2_2 C^{(S)\alpha\beta}_{\ell HD2} + \frac{3}{4} \left( C^{(S)}_{\ell HD1} Y^{}_l Y^\dagger_l \right)^{\alpha\beta} + \frac{3}{4} \left( C^{(S)}_{\ell HD1} Y^{}_l Y^\dagger_l \right)^{\beta\alpha} \right.
	\nonumber
	\\
	&& - \frac{1}{2} \left( Y^{}_l \right)^{}_{\lambda\gamma} \left( 3C^{(S)\gamma\lambda\alpha\beta}_{\overline{e}\ell\ell\ell H} + C^{(M)\gamma\lambda\alpha\beta}_{\overline{e}\ell\ell\ell H} + C^{(M)\gamma\lambda\beta\alpha}_{\overline{e}\ell\ell\ell H}   \right) + \frac{3}{8} \left( C^{(S)}_{\ell HD2} Y^{}_l Y^\dagger_l \right)^{\alpha\beta}
	\nonumber
	\\
	&& + \frac{3}{8} \left( C^{(S)}_{\ell HD2} Y^{}_l Y^\dagger_l \right)^{\beta\alpha}  - \frac{3}{4} \left( Y^{}_{\rm d} \right)^{}_{\lambda\gamma} \left( C^{\gamma\alpha\lambda\beta}_{\overline{d}\ell q\ell H1}  + C^{\gamma\beta\lambda\alpha}_{\overline{d}\ell q\ell H1} \right) 
	\nonumber
	\\
	&& + \left. \frac{3}{2} \left( Y^\dagger_{\rm u} \right)^{}_{\gamma\lambda} \left(  C^{\lambda\gamma\alpha\beta}_{\overline{q}u\ell\ell H}  + C^{\lambda\gamma\beta\alpha}_{\overline{q}u\ell\ell H}  \right) \right] \;.
\end{eqnarray}

\section{Counterterms in the physical basis} \label{app:counterterm-phys}

With the help of the reduction relations in Appendix~\ref{app:conversion}, one can convert all counterterms in the Green's basis presented in Appendix~\ref{app:counterterm-green} to those in the physical basis. All results are arranged in the following subsections.

\subsection{Wave-function renormalization}\label{subs:wave-function-1}

\begin{eqnarray}
	\delta Z^{}_H = \loopf m^2 \left( C^{}_{HD} - 2C^{}_{H \square} \right) \;.
\end{eqnarray}

\subsection{Counterterms for the dim-5 operator}

\begin{eqnarray}
	\delta C^{\alpha\beta}_5 &=& \loopf m^2 \left\{ 6 C^{\alpha\beta}_5 C^{}_{H\square} + 8C^{(S)\ast \alpha\beta}_{\ell H}  + \frac{3}{2} g^2_2 \left( 2 C^{(S)\ast \alpha\beta}_{\ell HD1} + C^{(S)\ast \alpha\beta}_{\ell HD2}  \right) + \frac{1}{2} \left( Y^{}_l Y^\dagger_l  C^{(S)\dagger}_{\ell HD1} \right)^{\alpha\beta} \right.
	\nonumber
	\\
	&& + \frac{1}{2} \left( Y^{}_l Y^\dagger_l  C^{(S)\dagger}_{\ell HD1} \right)^{\beta\alpha}  - \frac{1}{4} \left( Y^{}_l Y^\dagger_l  C^{(S)\dagger}_{\ell HD2} \right)^{\alpha\beta}  - \frac{1}{4} \left( Y^{}_l Y^\dagger_l  C^{(S)\dagger}_{\ell HD2} \right)^{\beta\alpha} + \left( Y^{}_l C^\dagger_{\ell e HD} \right)^{\alpha\beta}  
	\nonumber
	\\
	&& + \left( Y^{}_l C^\dagger_{\ell e HD}  \right)^{\beta\alpha}  - \left( Y^\dagger_l \right)^{}_{\gamma\lambda} \left( 3C^{(S)\ast\gamma\lambda\alpha\beta}_{\overline{e}\ell\ell\ell H}  + C^{(M)\ast\gamma\lambda\alpha\beta}_{\overline{e}\ell\ell\ell H} + C^{(M)\ast\gamma\lambda\beta\alpha}_{\overline{e}\ell\ell\ell H} \right) 
	\nonumber
	\\
	&& -  \left.  \frac{3}{2} \left( Y^{\dagger}_{\rm d} \right)^{}_{\gamma\lambda} \left( C^{\ast\gamma\alpha\lambda\beta}_{\overline{d}\ell q\ell H1}  + C^{\ast\gamma\beta\lambda\alpha}_{\overline{d}\ell q\ell H1} \right) 
	+  3 \left( Y^{}_{\rm u} \right)^{}_{\lambda\gamma} \left(  C^{\ast\lambda\gamma\alpha\beta}_{\overline{q}u\ell\ell H}  + C^{\ast\lambda\gamma\beta\alpha}_{\overline{q}u\ell\ell H}  \right)   \right\}  \;.
\end{eqnarray}

\subsection{Counterterms for dim-7 operators}

\noindent $\bullet~\bm{\psi^2H^4}$
\begin{eqnarray}\label{eq:coun-dim7-lh}
	\delta C^{(S)\alpha\beta}_{\ell H} &=& \loopf \left\{ C^{\ast \alpha\beta}_5 \left[ -3 C^{}_H - \frac{3}{4} \left( g^2_1 - g^2_2 + 4\lambda \right) C^{}_{HD} + \left( 16 \lambda - \frac{5}{3} g^2_2 \right) C^{}_{H \square} -3g^2_2 C^{}_{HW} \right. \right.
	\nonumber
	\\
	&& + \frac{3}{2} \rmi \left( g^2_1 C^{}_{H\widetilde{B}} + 3g^2_2 C^{}_{H\widetilde{W}} + g^{}_1g^{}_2 C^{}_{H\widetilde{W}B} \right) + \frac{1}{2} \tr{C^{}_5 C^\dagger_5} - \frac{2}{3} g^2_2 \tr{3 C^{(3)}_{Hq} +C^{(3)}_{H\ell} } 
	\nonumber
	\\
	&& - \tr{ C^{}_{eH} Y^\dagger_l + 3C^{}_{dH} Y^\dagger_{\rm d} + 3Y^{}_{\rm u} C^\dagger_{uH} } + 2 \tr{ Y^\dagger_l C^{(3)}_{H\ell} Y^{}_l + 3 Y^\dagger_{\rm d} C^{(3)}_{Hq} Y^{}_{\rm d} + 3 Y^\dagger_{\rm u} C^{(3)}_{Hq} Y^{}_{\rm u} } 
	\nonumber
	\\
	&& - \left. 3  \tr{ Y^{}_{\rm u} C^{}_{Hud} Y^\dagger_{\rm d} + Y^{}_{\rm d} C^\dagger_{Hud} Y^\dagger_{\rm u} } \right]  +  \frac{5}{4} \left( C^\dagger_5 C^{}_5 C^\dagger_5 \right)^{\alpha\beta} + \frac{3}{4} \left( g^2_1 + g^2_2 \right) \left[ \left( C^\dagger_5 C^{(3)}_{H\ell} \right)^{\alpha\beta} \right.
	\nonumber
	\\
	&& + \left. \left( C^\dagger_5 C^{(3)}_{H\ell} \right)^{\beta\alpha} - \left( C^\dagger_5 C^{(1)}_{H\ell} \right)^{\alpha\beta} - \left( C^\dagger_5 C^{(1)}_{H\ell} \right)^{\beta\alpha} \right] - \frac{3}{2} g^{}_2 \left[ \left( C^\dagger_5 Y^{}_l C^\dagger_{eW} \right)^{\alpha\beta} +  \left( C^\dagger_5 Y^{}_l C^\dagger_{eW} \right)^{\beta\alpha} \right] 
	\nonumber
	\\
	&& + \frac{1}{4} \left[ \left( C^\dagger_5 Y^{}_l C^\dagger_{eH} \right)^{\alpha\beta} +  \left( C^\dagger_5 Y^{}_l C^\dagger_{eH} \right)^{\beta\alpha} \right]  + \frac{1}{2} \left[ \left( C^\dagger_5 C^{}_{eH} Y^\dagger_l  \right)^{\alpha\beta} +  \left( C^\dagger_5 C^{}_{eH} Y^\dagger_l  \right)^{\beta\alpha} \right] 
	\nonumber
	\\
	&& -  \left.  \frac{3}{2} \left[ \left( C^\dagger_5 Y^{}_l Y^\dagger_l C^{(3)}_{H\ell} \right)^{\alpha\beta} + \left( C^\dagger_5 Y^{}_l Y^\dagger_l C^{(3)}_{H\ell} \right)^{\beta\alpha}  \right]  \right\}  \;.
\end{eqnarray}

\noindent $\bullet~\bm{\psi^2H^3D}$
\begin{eqnarray}
	\delta C^{\alpha\beta}_{\ell eHD} &=& \loopf \left\{ \left( C^{}_{H\square} - \frac{3}{4} C^{}_{HD} \right) \left( C^\dagger_5 Y^{}_l \right)^{\alpha\beta} + \frac{3}{2} g^{}_1 \left( C^\dagger_5 C^{}_{eB} \right)^{\alpha\beta} +  \frac{3}{2} g^{}_2 \left( C^\dagger_5 C^{}_{eW} \right)^{\alpha\beta} \right.
	\nonumber
	\\
	&& + \frac{1}{2} \left( C^\dagger_5 C^{}_{eH} \right)^{\alpha\beta} - \frac{3}{2} \left( C^\dagger_5 Y^{}_l C^{}_{He} \right)^{\alpha\beta} + \left[ \frac{1}{2} \left( C^\dagger_5 C^{(1)}_{H\ell} \right)^{\alpha\gamma} -  \left( C^\dagger_5 C^{(1)}_{H\ell} \right)^{\gamma\alpha} \right.
	\nonumber
	\\
	&& - \left.\left. \frac{5}{2} \left( C^\dagger_5 C^{(3)}_{H\ell} \right)^{\alpha\gamma} - 4 \left( C^\dagger_5 C^{(3)}_{H\ell} \right)^{\gamma\alpha} \right] \left(Y^{}_l \right)^{}_{\gamma\beta} - 2 \left( C^\dagger_5 Y^{}_l \right)^{\gamma\lambda} C^{\gamma\alpha\lambda\beta}_{\ell e}  \right\} \;.
\end{eqnarray}

\noindent $\bullet~\bm{\psi^2H^2D^2}$
\begin{eqnarray}
	\delta C^{(S)\alpha\beta}_{\ell HD1} &=& \loopf \left[ \frac{1}{2} C^{\ast \alpha\beta}_5 \left( C^{}_{HD} + 2 C^{}_{H\square} \right) - \left( C^\dagger_5 C^{(1)}_{H\ell} \right)^{\alpha\beta} - \left( C^\dagger_5 C^{(1)}_{H\ell} \right)^{\beta\alpha} -  \left( C^\dagger_5 C^{(3)}_{H\ell} \right)^{\alpha\beta} \right.
	\nonumber
	\\
	&& - \left. \left( C^\dagger_5 C^{(3)}_{H\ell} \right)^{\beta\alpha} + 4C^{\ast\gamma\lambda}_5 C^{\gamma\alpha\lambda\beta}_{\ell\ell} \right] \;,
	\nonumber
	\\
	\delta C^{(S)\alpha\beta}_{\ell HD2} &=& \loopf \left[ -C^{\ast \alpha\beta}_5 \left( C^{}_{HD} + 2 C^{}_{H\square} \right) + 2 \left( C^\dagger_5 C^{(1)}_{H\ell} \right)^{\alpha\beta} + 2 \left( C^\dagger_5 C^{(1)}_{H\ell} \right)^{\beta\alpha} - 4 \left( C^\dagger_5 C^{(3)}_{H\ell} \right)^{\alpha\beta} \right.
	\nonumber
	\\
	&& - \left. 4 \left( C^\dagger_5 C^{(3)}_{H\ell} \right)^{\beta\alpha} - 8 C^{\ast\gamma\lambda}_5 C^{\gamma\alpha\lambda\beta}_{\ell\ell} \right] \;.
\end{eqnarray}

\noindent $\bullet~\bm{\psi^2H^2X}$
\begin{eqnarray}
	\delta C^{(A)\alpha\beta}_{\ell HB} &=& \loopf \frac{1}{4}  \left[  \left( C^\dagger_5 Y^{}_l C^\dagger_{eB} \right)^{\alpha\beta} -  \left( C^\dagger_5 Y^{}_l C^\dagger_{eB} \right)^{\beta\alpha} \right] \;,
	\nonumber
	\\
	\delta C^{\alpha\beta}_{\ell HW} &=& \loopf \left\{ \frac{3}{4} g^{}_2 \left[  \left( C^\dagger_5 C^{(3)}_{H\ell } \right)^{\alpha\beta} + \left( C^\dagger_5 C^{(3)}_{H\ell } \right)^{\beta\alpha}  \right]  - \frac{1}{2} \left( C^\dagger_5 Y^{}_l C^\dagger_{eW} \right)^{\alpha\beta} \right.
	\nonumber
	\\
	&&  + \left. g^{}_2 \left( C^{}_{HW} - \rmi C^{}_{H\widetilde{W}} \right)C^{\ast\alpha\beta}_5 - \frac{3}{4} g^2_2 \left( C^{}_{3W} - \rmi C^{}_{3\widetilde{W}} \right)C^{\ast\alpha\beta}_5 \right\} \;.
\end{eqnarray}

\noindent $\bullet~\bm{\psi^4H}$
\begin{eqnarray}
	\delta C^{(S)\alpha\beta\gamma\lambda}_{\overline{e}\ell\ell\ell H} &=& \loopf \left\{  -2 g^{}_2 \left[ \left( C^\dagger_{eW} \right)^{\alpha\beta} \left( C^\dagger_5 \right)^{\gamma\lambda} + \left( C^\dagger_{eW} \right)^{\alpha\lambda} \left( C^\dagger_5 \right)^{\beta\gamma} + \left( C^\dagger_{eW} \right)^{\alpha\gamma} \left( C^\dagger_5 \right)^{\lambda\beta} \right] \right.
	\nonumber
	\\
	&& + \frac{1}{3} C^{\alpha\rho}_{He} \left[ \left( Y^\dagger_l \right)^{}_{\rho\beta} \left( C^\dagger_5 \right)^{\gamma\lambda} + \left( Y^\dagger_l \right)^{}_{\rho\lambda} \left( C^\dagger_5 \right)^{\beta\gamma} + \left( Y^\dagger_l \right)^{}_{\rho\gamma} \left( C^\dagger_5 \right)^{\lambda\beta} \right]
	\nonumber
	\\
	&& - \frac{1}{3} \left( Y^\dagger_l \right)^{}_{\alpha\rho} \left[ \left( C^\dagger_5 \right)^{\beta\gamma} \left( C^{(1)}_{H\ell} +  C^{(3)}_{H\ell}  \right)^{\rho\lambda} +  \left( C^\dagger_5 \right)^{\gamma\lambda} \left( C^{(1)}_{H\ell} +  C^{(3)}_{H\ell}  \right)^{\rho\beta} \right.
	\nonumber
	\\
	&& + \left. \left( C^\dagger_5 \right)^{\lambda\beta} \left( C^{(1)}_{H\ell} +  C^{(3)}_{H\ell}  \right)^{\rho\gamma} \right] - \frac{2}{3} \left( Y^\dagger_l \right)^{}_{\alpha\sigma} \left[ \left( C^\dagger_5 \right)^{\beta\rho} \left( C^{\rho\gamma\sigma\lambda}_{\ell\ell} + C^{\sigma\gamma\rho\lambda}_{\ell\ell} \right) \right.
	\nonumber
	\\
	&& + \left.  \left( C^\dagger_5 \right)^{\lambda\rho} \left( C^{\rho\beta\sigma\gamma}_{\ell\ell} + C^{\sigma\beta\rho\gamma}_{\ell\ell} \right) + \left( C^\dagger_5 \right)^{\gamma\rho} \left( C^{\rho\lambda\sigma\beta}_{\ell\ell} + C^{\sigma\lambda\rho\beta}_{\ell\ell} \right) \right] 
	\nonumber
	\\
	&& + \frac{1}{3} C^{\sigma\gamma\alpha\rho}_{\ell e} \left[ \left( C^\dagger_5 \right)^{\beta\sigma} \left( Y^\dagger_l \right)^{}_{\rho\lambda} + \left( C^\dagger_5 \right)^{\lambda\sigma} \left( Y^\dagger_l \right)^{}_{\rho\beta} \right] + \frac{1}{3} C^{\sigma\beta\alpha\rho}_{\ell e} \left[ \left( C^\dagger_5 \right)^{\lambda\sigma} \left( Y^\dagger_l \right)^{}_{\rho\gamma} \right.
	\nonumber
	\\
	&& + \left. \left( C^\dagger_5 \right)^{\gamma\sigma} \left( Y^\dagger_l \right)^{}_{\rho\lambda} \right] + \frac{1}{3} C^{\sigma\lambda\alpha\rho}_{\ell e} \left[ \left( C^\dagger_5 \right)^{\gamma\sigma} \left( Y^\dagger_l \right)^{}_{\rho\beta} + \left( C^\dagger_5 \right)^{\beta\sigma} \left( Y^\dagger_l \right)^{}_{\rho\gamma} \right]
	\nonumber
	\\
	&& + \frac{1}{3} \left[  \left(C^\dagger_{eH} \right)^{\alpha\beta} \left( C^\dagger_5 \right)^{\gamma\lambda} + \left(C^\dagger_{eH} \right)^{\alpha\gamma} \left( C^\dagger_5 \right)^{\lambda\beta} + \left(C^\dagger_{eH} \right)^{\alpha\lambda} \left( C^\dagger_5 \right)^{\beta\gamma} \right]  
	\nonumber
	\\
	&& - \frac{1}{3} \left( C^\dagger_5 \right)^{\gamma\lambda} \left[ 2 C^{\ast \beta \rho\sigma\alpha}_{\ell e} \left( Y^\dagger_l \right)^{}_{\sigma\rho} + 3 C^{(1)\ast\beta\alpha\rho\sigma}_{\ell equ} \left( Y^{}_{\rm u}  \right)^{}_{\rho\sigma} - 3 C^{\ast\beta\alpha\rho\sigma}_{\ell e d q} \left( Y^\dagger_{\rm d} \right)^{}_{\rho\sigma} \right]
	\nonumber
	\\
	&& - \frac{1}{3} \left( C^\dagger_5 \right)^{\beta\gamma} \left[ 2 C^{\ast \lambda\rho\sigma\alpha}_{\ell e} \left( Y^\dagger_l \right)^{}_{\sigma\rho} + 3 C^{(1)\ast\lambda\alpha\rho\sigma}_{\ell equ} \left( Y^{}_{\rm u}  \right)^{}_{\rho\sigma} - 3 C^{\ast\lambda\alpha\rho\sigma}_{\ell e d q} \left( Y^\dagger_{\rm d} \right)^{}_{\rho\sigma} \right]  
	\nonumber
	\\
	&& - \frac{1}{3} \left( C^\dagger_5 \right)^{\lambda\beta} \left[ 2 C^{\ast \gamma \rho\sigma\alpha}_{\ell e} \left( Y^\dagger_l \right)^{}_{\sigma\rho} + 3 C^{(1)\ast\gamma\alpha\rho\sigma}_{\ell equ} \left( Y^{}_{\rm u}  \right)^{}_{\rho\sigma} - 3 C^{\ast\gamma\alpha\rho\sigma}_{\ell e d q} \left( Y^\dagger_{\rm d} \right)^{}_{\rho\sigma} \right] 
	\nonumber
	\\
	&&  - \frac{1}{3} \left( Y^\dagger_l \right)^{}_{\alpha\beta} \left[ C^{\ast\gamma\lambda}_5 C^{}_{H\square} - \left( C^\dagger_5 C^{(1)}_{H\ell} \right)^{\gamma\lambda} - \left( C^\dagger_5 C^{(1)}_{H\ell} \right)^{\lambda\gamma} + 2  \left( C^\dagger_5 C^{(3)}_{H\ell} \right)^{\gamma\lambda} \right.
	\nonumber
	\\
	&& + \left. 2 \left( C^\dagger_5 C^{(3)}_{H\ell} \right)^{\lambda\gamma}  \right] - \frac{1}{3} \left( Y^\dagger_l \right)^{}_{\alpha\lambda} \left[ C^{\ast\beta\gamma}_5 C^{}_{H\square} - \left( C^\dagger_5 C^{(1)}_{H\ell} \right)^{\beta\gamma} - \left( C^\dagger_5 C^{(1)}_{H\ell} \right)^{\gamma\beta} \right.
	\nonumber
	\\
	&& + \left. 2  \left( C^\dagger_5 C^{(3)}_{H\ell} \right)^{\beta\gamma} + 2 \left( C^\dagger_5 C^{(3)}_{H\ell} \right)^{\gamma\beta}  \right]  - \frac{1}{3} \left( Y^\dagger_l \right)^{}_{\alpha\gamma} \left[ C^{\ast\lambda\beta}_5 C^{}_{H\square} - \left( C^\dagger_5 C^{(1)}_{H\ell} \right)^{\lambda\beta} \right.
	\nonumber
	\\
	&& - \left.\left. \left( C^\dagger_5 C^{(1)}_{H\ell} \right)^{\beta\lambda} + 2  \left( C^\dagger_5 C^{(3)}_{H\ell} \right)^{\lambda\beta} + 2 \left( C^\dagger_5 C^{(3)}_{H\ell} \right)^{\beta\lambda}  \right]  \right\} \;,
	\nonumber
	\\
	\delta C^{(A)\alpha\beta\gamma\lambda}_{\overline{e}\ell\ell\ell H} &=& \loopf \left\{ 2 \left( Y^\dagger_l \right)^{}_{\alpha\sigma} \left[ \left( C^\dagger_5 \right)^{\beta\rho} \left( C^{\rho\gamma\sigma\lambda}_{\ell\ell} - C^{\sigma\gamma\rho\lambda}_{\ell\ell} \right) + \left( C^\dagger_5 \right)^{\lambda\rho} \left( C^{\rho\beta\sigma\gamma}_{\ell\ell} - C^{\sigma\beta\rho\gamma}_{\ell\ell} \right) \right.\right.
	\nonumber
	\\
	&& + \left.  \left( C^\dagger_5 \right)^{\gamma\rho} \left( C^{\rho\lambda\sigma\beta}_{\ell\ell} - C^{\sigma\lambda\rho\beta}_{\ell\ell} \right) \right] + C^{\sigma\gamma\alpha\rho}_{\ell e} \left[ \left( C^\dagger_5 \right)^{\beta\sigma} \left( Y^\dagger_l \right)^{}_{\rho\lambda} - \left( C^\dagger_5 \right)^{\lambda\sigma} \left( Y^\dagger_l \right)^{}_{\rho\beta} \right] 
	\nonumber
	\\
	&& + C^{\sigma\beta\alpha\rho}_{\ell e} \left[ \left( C^\dagger_5 \right)^{\lambda\sigma} \left( Y^\dagger_l \right)^{}_{\rho\gamma} - \left( C^\dagger_5 \right)^{\gamma\sigma} \left( Y^\dagger_l \right)^{}_{\rho\lambda} \right]  + C^{\sigma\lambda\alpha\rho}_{\ell e} \left[ \left( C^\dagger_5 \right)^{\gamma\sigma} \left( Y^\dagger_l \right)^{}_{\rho\beta} \right.
	\nonumber
	\\
	&& - \left.\left.  \left( C^\dagger_5 \right)^{\beta\sigma} \left( Y^\dagger_l \right)^{}_{\rho\gamma} \right] \right\} \;,
	\nonumber
	\\
	\delta C^{(M)\alpha\beta\gamma\lambda}_{\overline{e}\ell\ell\ell H} &=& \loopf \left\{ 3 g^{}_2 \left[ \left( C^\dagger_{eW} \right)^{\alpha\beta} \left( C^\dagger_5 \right)^{\gamma\lambda} - \left( C^\dagger_{eW} \right)^{\alpha\lambda} \left( C^\dagger_5 \right)^{\beta\gamma} \right] \right.
	\nonumber
	\\
	&& + \frac{1}{2} C^{\alpha\rho}_{He} \left[ \left( Y^\dagger_l \right)^{}_{\rho\beta} \left( C^\dagger_5 \right)^{\gamma\lambda} - \left( Y^\dagger_l \right)^{}_{\rho\lambda} \left( C^\dagger_5 \right)^{\beta\gamma} \right]
	\nonumber
	\\
	&& + \frac{1}{2} \left( Y^\dagger_l \right)^{}_{\alpha\rho} \left[ \left( C^\dagger_5 \right)^{\beta\gamma} \left( C^{(1)}_{H\ell} +  C^{(3)}_{H\ell}  \right)^{\rho\lambda} -  \left( C^\dagger_5 \right)^{\gamma\lambda} \left( C^{(1)}_{H\ell} +  C^{(3)}_{H\ell}  \right)^{\rho\beta} \right]
	\nonumber
	\\
	&& - 2 \left( Y^\dagger_l \right)^{}_{\alpha\sigma} \left[ \left( C^\dagger_5 \right)^{\beta\rho}   C^{\sigma\gamma\rho\lambda}_{\ell\ell} - \left( C^\dagger_5 \right)^{\lambda\rho}  C^{\rho\beta\sigma\gamma}_{\ell\ell} + \left( C^\dagger_5 \right)^{\gamma\rho} \left( C^{\rho\lambda\sigma\beta}_{\ell\ell} - C^{\sigma\lambda\rho\beta}_{\ell\ell} \right) \right] 
	\nonumber
	\\
	&& - C^{\sigma\gamma\alpha\rho}_{\ell e} \left[ \left( C^\dagger_5 \right)^{\beta\sigma} \left( Y^\dagger_l \right)^{}_{\rho\lambda} - \left( C^\dagger_5 \right)^{\lambda\sigma} \left( Y^\dagger_l \right)^{}_{\rho\beta} \right] + C^{\sigma\beta\alpha\rho}_{\ell e} \left( C^\dagger_5 \right)^{\lambda\sigma} \left( Y^\dagger_l \right)^{}_{\rho\gamma} 
	\nonumber
	\\
	&& - C^{\sigma\lambda\alpha\rho}_{\ell e} \left( C^\dagger_5 \right)^{\beta\sigma} \left( Y^\dagger_l \right)^{}_{\rho\gamma} + \frac{1}{2} \left[  \left(C^\dagger_{eH} \right)^{\alpha\beta} \left( C^\dagger_5 \right)^{\gamma\lambda} - \left(C^\dagger_{eH} \right)^{\alpha\lambda} \left( C^\dagger_5 \right)^{\beta\gamma} \right]  
	\nonumber
	\\
	&& - \frac{1}{2} \left( C^\dagger_5 \right)^{\gamma\lambda} \left[ 2 C^{\ast \beta \rho\sigma\alpha}_{\ell e} \left( Y^\dagger_l \right)^{}_{\sigma\rho} + 3 C^{(1)\ast\beta\alpha\rho\sigma}_{\ell equ} \left( Y^{}_{\rm u}  \right)^{}_{\rho\sigma} - 3 C^{\ast\beta\alpha\rho\sigma}_{\ell e d q} \left( Y^\dagger_{\rm d} \right)^{}_{\rho\sigma} \right]
	\nonumber
	\\
	&& + \frac{1}{2} \left( C^\dagger_5 \right)^{\beta\gamma} \left[ 2 C^{\ast \lambda\rho\sigma\alpha}_{\ell e} \left( Y^\dagger_l \right)^{}_{\sigma\rho} + 3 C^{(1)\ast\lambda\alpha\rho\sigma}_{\ell equ} \left( Y^{}_{\rm u}  \right)^{}_{\rho\sigma} - 3 C^{\ast\lambda\alpha\rho\sigma}_{\ell e d q} \left( Y^\dagger_{\rm d} \right)^{}_{\rho\sigma} \right]  
	\nonumber
	\\
	&&  - \frac{1}{2} \left( Y^\dagger_l \right)^{}_{\alpha\beta} \left[ C^{\ast\gamma\lambda}_5 C^{}_{H\square} - \left( C^\dagger_5 C^{(1)}_{H\ell} \right)^{\gamma\lambda} - \left( C^\dagger_5 C^{(1)}_{H\ell} \right)^{\lambda\gamma} + 2  \left( C^\dagger_5 C^{(3)}_{H\ell} \right)^{\gamma\lambda} \right.
	\nonumber
	\\
	&& + \left. 2 \left( C^\dagger_5 C^{(3)}_{H\ell} \right)^{\lambda\gamma}  \right] + \frac{1}{2} \left( Y^\dagger_l \right)^{}_{\alpha\lambda} \left[ C^{\ast\beta\gamma}_5 C^{}_{H\square} - \left( C^\dagger_5 C^{(1)}_{H\ell} \right)^{\beta\gamma} - \left( C^\dagger_5 C^{(1)}_{H\ell} \right)^{\gamma\beta} \right.
	\nonumber
	\\
	&& + \left.\left. 2  \left( C^\dagger_5 C^{(3)}_{H\ell} \right)^{\beta\gamma} + 2 \left( C^\dagger_5 C^{(3)}_{H\ell} \right)^{\gamma\beta}  \right]   \right\} \;,
	\nonumber
	\\
	\delta C^{\alpha\beta\gamma\lambda}_{\overline{d}\ell q \ell H 1} &=&  \loopf \left\{ - 2 \left( Y^\dagger_{\rm d} \right)^{}_{\alpha\rho} \left( C^\dagger_5 \right)^{\beta\lambda} \left( C^{(1)}_{Hq} + C^{(3)}_{Hq} \right)^{\rho\gamma} + 2 \left( Y^\dagger_{\rm d} \right)^{}_{\rho\gamma}  \left( C^\dagger_5 \right)^{\beta\lambda} C^{\alpha\rho}_{Hd}  \right.
	\nonumber
	\\
	&&  - 4 \left( Y^\dagger_{\rm d} \right)^{}_{\alpha\sigma}  \left( C^\dagger_5 \right)^{\beta\rho} \left( C^{(1)}_{\ell q} - 2 C^{(3)}_{\ell q} \right)^{\rho\lambda\sigma\gamma}  + 2 \left( C^\dagger_{dH} \right)^{\alpha\gamma} \left( C^\dagger_5\right)^{\beta\lambda} 
	\nonumber
	\\
	&& + 4  \left( Y^\dagger_{\rm d} \right)^{}_{\sigma\gamma}  \left( C^\dagger_5 \right)^{\rho\lambda} C^{\rho\beta\alpha\sigma}_{\ell d} - C^{\rho\sigma\alpha\gamma}_{\ell e d q} \left[ \left( Y^\dagger_l \right)^{}_{\sigma\beta}  \left( C^\dagger_5 \right)^{\rho\lambda} + \left( Y^\dagger_l \right)^{}_{\sigma\lambda}  \left( C^\dagger_5 \right)^{\rho\beta} \right] 
	\nonumber
	\\
	&& + C^{\ast\beta\lambda}_5 \left[ 2 \left( Y^\dagger_l \right)^{}_{\rho\sigma} C^{\sigma\rho\alpha\gamma}_{\ell edq}  - 4 \left( Y^\dagger_{\rm d} \right)^{}_{\rho \sigma} \left(C^{(1)\ast \gamma\sigma\rho\alpha}_{qd} + \frac{4}{3} C^{(8)\ast\gamma\sigma\rho\alpha}_{qd} \right)  \right.
	\nonumber
	\\
	&& + \left. \left( Y^{}_{\rm u} \right)^{}_{\rho\sigma} \left( 6 C^{(1)\ast\rho\sigma\gamma\alpha}_{quqd} + C^{(1)\ast\gamma\sigma\rho\alpha}_{quqd}  + \frac{4}{3} C^{(8)\ast\gamma\sigma\rho\alpha}_{quqd} \right)  \right] - 2 \left( Y^\dagger_{\rm d} \right)^{}_{\alpha\gamma} \left[ C^{\ast\beta\lambda}_5 C^{}_{H\square} \right.
	\nonumber
	\\
	&& 
	\nonumber
	\\
	&& -  \left.\left. \left( C^\dagger_5 C^{(1)}_{H\ell} \right)^{\beta\lambda} - \left( C^\dagger_5 C^{(1)}_{H\ell} \right)^{\lambda\beta} + 2 \left( C^\dagger_5 C^{(3)}_{H\ell} \right)^{\beta\lambda} + 2\left( C^\dagger_5 C^{(3)}_{H\ell} \right)^{\lambda\beta} \right] \right\} \;,
	\nonumber
	\\
	\delta C^{\alpha\beta\gamma\lambda}_{\overline{d}\ell q \ell H 2} &=&  \loopf \left\{ -6g^{}_2 \left( C^\dagger_{dW} \right)^{\alpha\gamma} \left( C^\dagger_5 \right)^{\beta\lambda} + \left( Y^\dagger_{\rm d} \right)^{}_{\alpha\rho} \left( C^\dagger_5 \right)^{\beta\lambda} \left( C^{(1)}_{Hq} + C^{(3)}_{Hq} \right)^{\rho\gamma}  \right.
	\nonumber
	\\
	&& -  \left( Y^\dagger_{\rm d} \right)^{}_{\rho\gamma}  \left( C^\dagger_5 \right)^{\beta\lambda} C^{\alpha\rho}_{Hd}  + 2 \left( Y^\dagger_{\rm d} \right)^{}_{\alpha\sigma}  \left( C^\dagger_5 \right)^{\beta\rho} \left( C^{(1)}_{\ell q} - 5 C^{(3)}_{\ell q} \right)^{\rho\lambda\sigma\gamma}  
	\nonumber
	\\
	&& - 2 \left( Y^\dagger_{\rm d} \right)^{}_{\sigma\gamma}  \left( C^\dagger_5 \right)^{\rho\lambda} C^{\rho\beta\alpha\sigma}_{\ell d}  - C^{\rho\sigma\alpha\gamma}_{\ell e d q} \left[ \left( Y^\dagger_l \right)^{}_{\sigma\beta}  \left( C^\dagger_5 \right)^{\rho\lambda} - 2 \left( Y^\dagger_l \right)^{}_{\sigma\lambda}  \left( C^\dagger_5 \right)^{\rho\beta} \right] 
	\nonumber
	\\
	&& -  \left( C^\dagger_{dH} \right)^{\alpha\gamma} \left( C^\dagger_5\right)^{\beta\lambda} - \frac{1}{2} C^{\ast\beta\lambda}_5 \left[- 4 \left( Y^\dagger_{\rm d} \right)^{}_{\rho \sigma} \left(C^{(1)\ast \gamma\sigma\rho\alpha}_{qd} + \frac{4}{3} C^{(8)\ast\gamma\sigma\rho\alpha}_{qd} \right) \right.
	\nonumber
	\\
	&&  + \left. 2 \left( Y^\dagger_l \right)^{}_{\rho\sigma} C^{\sigma\rho\alpha\gamma}_{\ell edq} + \left( Y^{}_{\rm u} \right)^{}_{\rho\sigma} \left( 6 C^{(1)\ast\rho\sigma\gamma\alpha}_{quqd} + C^{(1)\ast\gamma\sigma\rho\alpha}_{quqd} +  \frac{4}{3} C^{(8)\ast\gamma\sigma\rho\alpha}_{quqd} \right)  \right] 
	\nonumber
	\\
	&& + \left( Y^\dagger_{\rm d} \right)^{}_{\alpha\gamma} \left[ C^{\ast\beta\lambda}_5 C^{}_{H\square} - \left( C^\dagger_5 C^{(1)}_{H\ell} \right)^{\beta\lambda} -  \left( C^\dagger_5 C^{(1)}_{H\ell} \right)^{\lambda\beta} + 2 \left( C^\dagger_5 C^{(3)}_{H\ell} \right)^{\beta\lambda} \right.
	\nonumber
	\\
	&& + \left.\left. 2\left( C^\dagger_5 C^{(3)}_{H\ell} \right)^{\lambda\beta} \right] \right\} \;,
	\nonumber
	\\
	\delta C^{\alpha\beta\gamma\lambda}_{\overline{d}\ell u e H} &=&  \loopf \left[ -3 \left( C^\dagger_{Hud} \right)^{\alpha\gamma} \left( C^\dagger_5 \right)^{\beta\rho} \left( Y^{}_l \right)^{}_{\rho\lambda} - \frac{3}{2} \left( C^\dagger_5 \right)^{\beta\sigma} \left( Y^{}_{\rm u} \right)^{}_{\rho\gamma} C^{\sigma\lambda\alpha\rho}_{\ell e d q} \right.
	\nonumber
	\\
	&& + \left.  \frac{3}{2} \left( Y^\dagger_{\rm d} \right)^{}_{\alpha\rho} \left( C^\dagger_5 \right)^{\beta\sigma} \left( C^{(1)}_{\ell e qu} - 12 C^{(3)}_{\ell e q u} \right)^{\sigma\lambda\rho\gamma} \right] \;,
	\nonumber
	\\
	\delta C^{\alpha\beta\gamma\lambda}_{\overline{q}u\ell\ell H} &=&  \loopf \left\{ - \left( C^{(1)}_{Hq} - C^{(3)}_{Hq} \right)^{\alpha\rho} \left( Y^{}_{\rm u} \right)^{}_{\rho \beta} \left( C^\dagger_5 \right)^{\gamma\lambda} + \left( Y^{}_{\rm u} \right)^{}_{\alpha\rho} C^{\rho\beta}_{Hu} \left( C^\dagger_5 \right)^{\gamma\lambda} \right.
	\nonumber
	\\
	&& - \left( Y^{}_{\rm u} \right)^{}_{\rho \beta} \left[ \left( C^{(1)}_{\ell q} + 5 C^{(3)}_{\ell q} \right)^{\sigma\gamma\alpha\rho} \left( C^\dagger_5 \right)^{\sigma\lambda} + \left( C^{(1)}_{\ell q} - C^{(3)}_{\ell q} \right)^{\sigma\lambda\alpha\rho} \left( C^\dagger_5 \right)^{\sigma\gamma}  \right]
	\nonumber
	\\
	&& + \left( Y^{}_{\rm u} \right)^{}_{\alpha\rho} \left[ \left( C^\dagger_5 \right)^{\lambda\sigma} C^{\sigma\gamma\rho\beta}_{\ell u} +   \left( C^\dagger_5 \right)^{\gamma\sigma} C^{\sigma\lambda\rho\beta}_{\ell u}  \right] + C^{(1)\sigma\rho\alpha\beta}_{\ell e q u} \left[ \left( C^\dagger_5 \right)^{\sigma\gamma} \left( Y^\dagger_l \right)^{}_{\rho\lambda} \right.
	\nonumber
	\\
	&& - \left. 2 \left( C^\dagger_5 \right)^{\sigma\lambda} \left( Y^\dagger_l \right)^{}_{\rho\gamma} \right]  - C^{\alpha\beta}_{uH} \left( C^\dagger_5 \right)^{\gamma\lambda}   + \frac{1}{2} C^{\ast\gamma\lambda}_5 \left[ 2 C^{(1)\rho\sigma\alpha\beta}_{\ell e qu} \left( Y^{\dagger}_l \right)^{}_{\sigma\rho}  \right.
	\nonumber
	\\
	&& +  4 \left( C^{(1)\alpha\rho\sigma\beta}_{qu} + \frac{4}{3} C^{(8)\alpha\rho\sigma\beta}_{qu} \right) \left( Y^{}_{\rm u} \right)^{}_{\rho\sigma}  - \left( Y^{\dagger}_{\rm d} \right)^{}_{\rho\sigma} \left( 6C^{(1)\alpha\beta\sigma\rho}_{quqd} + C^{(1)\sigma\beta\alpha\rho}_{quqd} \right.
	\nonumber
	\\
	&& + \left.\left. \frac{4}{3} C^{(8)\sigma\beta\alpha\rho}_{quqd} \right) \right] + \left( Y^{}_{\rm u} \right)^{}_{\alpha\beta} \left[ C^{\ast\gamma\lambda}_5 C^{}_{H\square} - \left( C^\dagger_5 C^{(1)}_{H\ell} \right)^{\gamma\lambda} - \left( C^\dagger_5 C^{(1)}_{H\ell} \right)^{\lambda\gamma} \right.
	\nonumber
	\\
	&&  + \left.\left. 2 \left( C^\dagger_5 C^{(3)}_{H\ell} \right)^{\gamma\lambda} + 2 \left( C^\dagger_5 C^{(3)}_{H\ell} \right)^{\lambda\gamma}  \right] \right\}  \;,
	\nonumber
	\\
	\delta C^{\alpha\beta\gamma\lambda}_{\overline{\ell}dud\widetilde{H}} &=& - \loopf 3 C^{\alpha\rho}_5 \left( Y^{}_{\rm d} \right)^{}_{\sigma\beta} C^{\lambda\gamma\sigma\rho}_{duq} \;,
	\nonumber
	\\
	\delta C^{\alpha\beta\gamma\lambda}_{\overline{\ell}dqq\widetilde{H}} &=&  \loopf \left[ - C^{\alpha\rho}_5 \left( Y^\dagger_{\rm u} \right)^{}_{\sigma\gamma} C^{\beta\sigma\lambda\rho}_{duq} - \frac{1}{2} C^{\alpha\rho}_5 \left( Y^\dagger_{\rm u} \right)^{}_{\sigma\lambda} C^{\beta\sigma\gamma\rho}_{duq} + \frac{1}{2} C^{\alpha\rho}_5 \left( Y^{}_{\rm d} \right)^{}_{\sigma\beta} \right.
	\nonumber
	\\
	&& \times \left. \left( 9 C^{(S)\gamma\lambda\sigma\rho}_{qqq} - 3 C^{(A)\gamma\lambda\sigma\rho}_{qqq} + 8 C^{(M)\gamma\lambda\sigma\rho}_{qqq} + 10 C^{(M)\sigma\gamma\lambda\rho}_{qqq}  \right) \vphantom{\frac{1}{2}}\right]  \;.
\end{eqnarray}

\end{appendices}


\begin{thebibliography}{99}

\bibitem{ParticleDataGroup:2022pth}
R.~L.~Workman \textit{et al.} [Particle Data Group],
``Review of Particle Physics,''
PTEP \textbf{2022} (2022), 083C01.

\bibitem{Xing:2020ijf}
Z.~z.~Xing,
``Flavor structures of charged fermions and massive neutrinos,''
Phys. Rept. \textbf{854} (2020), 1-147
[arXiv:1909.09610 [hep-ph]].

\bibitem{Buchmuller:1985jz}
W.~Buchmuller and D.~Wyler,
``Effective Lagrangian Analysis of New Interactions and Flavor Conservation,''
Nucl. Phys. B \textbf{268} (1986), 621-653.

\bibitem{Grzadkowski:2010es}
B.~Grzadkowski, M.~Iskrzynski, M.~Misiak and J.~Rosiek,
``Dimension-Six Terms in the Standard Model Lagrangian,''
JHEP \textbf{10} (2010), 085
[arXiv:1008.4884 [hep-ph]].

\bibitem{Brivio:2017vri}
I.~Brivio and M.~Trott,
``The Standard Model as an Effective Field Theory,''
Phys. Rept. \textbf{793} (2019), 1-98
[arXiv:1706.08945 [hep-ph]].

\bibitem{Isidori:2023pyp}
G.~Isidori, F.~Wilsch and D.~Wyler,
``The Standard Model effective field theory at work,''
[arXiv:2303.16922 [hep-ph]].

\bibitem{Henning:2014wua}
B.~Henning, X.~Lu and H.~Murayama,
``How to use the Standard Model effective field theory,''
JHEP \textbf{01} (2016), 023
[arXiv:1412.1837 [hep-ph]].

\bibitem{Weinberg:1980wa}
S.~Weinberg,
``Effective Gauge Theories,''
Phys. Lett. B \textbf{91} (1980), 51-55.

\bibitem{Jiang:2018pbd}
M.~Jiang, N.~Craig, Y.~Y.~Li and D.~Sutherland,
``Complete one-loop matching for a singlet scalar in the Standard Model EFT,''
JHEP \textbf{02} (2019), 031
[erratum: JHEP \textbf{01} (2021), 135]
[arXiv:1811.08878 [hep-ph]].

\bibitem{Criado:2018sdb}
J.~C.~Criado and M.~P\'erez-Victoria,
``Field redefinitions in effective theories at higher orders,''
JHEP \textbf{03} (2019), 038
[arXiv:1811.09413 [hep-ph]].

\bibitem{Weinberg:1979sa}
S.~Weinberg,
``Baryon and Lepton Nonconserving Processes,''
Phys. Rev. Lett. \textbf{43} (1979), 1566-1570.

\bibitem{Henning:2015alf}
B.~Henning, X.~Lu, T.~Melia and H.~Murayama,
``2, 84, 30, 993, 560, 15456, 11962, 261485, ...: Higher dimension operators in the SM EFT,''
JHEP \textbf{08} (2017), 016
[erratum: JHEP \textbf{09} (2019), 019]
[arXiv:1512.03433 [hep-ph]].

\bibitem{Criado:2019ugp}
J.~C.~Criado,
``BasisGen: automatic generation of operator bases,''
Eur. Phys. J. C \textbf{79} (2019) no.3, 256
[arXiv:1901.03501 [hep-ph]].

\bibitem{Fonseca:2019yya}
R.~M.~Fonseca,
``Enumerating the operators of an effective field theory,''
Phys. Rev. D \textbf{101} (2020) no.3, 035040
[arXiv:1907.12584 [hep-ph]].

\bibitem{Ma:2019gtx}
T.~Ma, J.~Shu and M.~L.~Xiao,
``Standard model effective field theory from on-shell amplitudes*,''
Chin. Phys. C \textbf{47} (2023) no.2, 023105
[arXiv:1902.06752 [hep-ph]].

\bibitem{Aoude:2019tzn}
R.~Aoude and C.~S.~Machado,
``The Rise of SMEFT On-shell Amplitudes,''
JHEP \textbf{12} (2019), 058
[arXiv:1905.11433 [hep-ph]].

\bibitem{Lehman:2014jma}
L.~Lehman,
``Extending the Standard Model Effective Field Theory with the Complete Set of Dimension-7 Operators,''
Phys. Rev. D \textbf{90} (2014) no.12, 125023
[arXiv:1410.4193 [hep-ph]].

\bibitem{Liao:2016hru}
Y.~Liao and X.~D.~Ma,
``Renormalization Group Evolution of Dimension-seven Baryon- and Lepton-number-violating Operators,''
JHEP \textbf{11} (2016), 043
[arXiv:1607.07309 [hep-ph]].

\bibitem{Liao:2019tep}
Y.~Liao and X.~D.~Ma,
``Renormalization Group Evolution of Dimension-seven Operators in Standard Model Effective Field Theory and Relevant Phenomenology,''
JHEP \textbf{03} (2019), 179
[arXiv:1901.10302 [hep-ph]].

\bibitem{Murphy:2020rsh}
C.~W.~Murphy,
``Dimension-8 operators in the Standard Model Eective Field Theory,''
JHEP \textbf{10} (2020), 174
[arXiv:2005.00059 [hep-ph]].

\bibitem{Li:2020gnx}
H.~L.~Li, Z.~Ren, J.~Shu, M.~L.~Xiao, J.~H.~Yu and Y.~H.~Zheng,
``Complete set of dimension-eight operators in the standard model effective field theory,''
Phys. Rev. D \textbf{104} (2021) no.1, 015026
[arXiv:2005.00008 [hep-ph]].

\bibitem{Durieux:2019eor}
G.~Durieux, T.~Kitahara, Y.~Shadmi and Y.~Weiss,
``The electroweak effective field theory from on-shell amplitudes,''
JHEP \textbf{01} (2020), 119
[arXiv:1909.10551 [hep-ph]].

\bibitem{AccettulliHuber:2021uoa}
M.~Accettulli Huber and S.~De Angelis,
``Standard Model EFTs via on-shell methods,''
JHEP \textbf{11} (2021), 221
[arXiv:2108.03669 [hep-th]].

\bibitem{Liao:2020jmn}
Y.~Liao and X.~D.~Ma,
``An explicit construction of the dimension-9 operator basis in the standard model effective field theory,''
JHEP \textbf{11} (2020), 152
[arXiv:2007.08125 [hep-ph]].

\bibitem{Li:2020xlh}
H.~L.~Li, Z.~Ren, M.~L.~Xiao, J.~H.~Yu and Y.~H.~Zheng,
``Complete set of dimension-nine operators in the standard model effective field theory,''
Phys. Rev. D \textbf{104} (2021) no.1, 015025
[arXiv:2007.07899 [hep-ph]].

\bibitem{Harlander:2023psl}
R.~V.~Harlander, T.~Kempkens and M.~C.~Schaaf,
``The Standard Model Effective Field Theory up to Mass Dimension 12,''
[arXiv:2305.06832 [hep-ph]].

\bibitem{Gherardi:2020det}
V.~Gherardi, D.~Marzocca and E.~Venturini,
``Matching scalar leptoquarks to the SMEFT at one loop,''
JHEP \textbf{07} (2020), 225
[erratum: JHEP \textbf{01} (2021), 006]
[arXiv:2003.12525 [hep-ph]].

\bibitem{Chala:2021cgt}
M.~Chala, \'A.~D\'\i{}az-Carmona and G.~Guedes,
``A Green\textquoteright{}s basis for the bosonic SMEFT to dimension 8,''
JHEP \textbf{05} (2022), 138
[arXiv:2112.12724 [hep-ph]].

\bibitem{Ren:2022tvi}
Z.~Ren and J.~H.~Yu,
``A Complete Set of the Dimension-8 Green's Basis Operators in the Standard Model Effective Field Theory,''
[arXiv:2211.01420 [hep-ph]].

\bibitem{Babu:1993qv}
K.~S.~Babu, C.~N.~Leung and J.~T.~Pantaleone,
``Renormalization of the neutrino mass operator,''
Phys. Lett. B \textbf{319} (1993), 191-198
[arXiv:hep-ph/9309223 [hep-ph]].

\bibitem{Chankowski:1993tx}
P.~H.~Chankowski and Z.~Pluciennik,
``Renormalization group equations for seesaw neutrino masses,''
Phys. Lett. B \textbf{316} (1993), 312-317
[arXiv:hep-ph/9306333 [hep-ph]].

\bibitem{Antusch:2001ck}
S.~Antusch, M.~Drees, J.~Kersten, M.~Lindner and M.~Ratz,
``Neutrino mass operator renormalization revisited,''
Phys. Lett. B \textbf{519} (2001), 238-242
[arXiv:hep-ph/0108005 [hep-ph]].

\bibitem{Jenkins:2013zja}
E.~E.~Jenkins, A.~V.~Manohar and M.~Trott,
``Renormalization Group Evolution of the Standard Model Dimension Six Operators I: Formalism and lambda Dependence,''
JHEP \textbf{10} (2013), 087
[arXiv:1308.2627 [hep-ph]].

\bibitem{Jenkins:2013wua}
E.~E.~Jenkins, A.~V.~Manohar and M.~Trott,
``Renormalization Group Evolution of the Standard Model Dimension Six Operators II: Yukawa Dependence,''
JHEP \textbf{01} (2014), 035
[arXiv:1310.4838 [hep-ph]].

\bibitem{Alonso:2013hga}
R.~Alonso, E.~E.~Jenkins, A.~V.~Manohar and M.~Trott,
``Renormalization Group Evolution of the Standard Model Dimension Six Operators III: Gauge Coupling Dependence and Phenomenology,''
JHEP \textbf{04} (2014), 159
[arXiv:1312.2014 [hep-ph]].

\bibitem{Alonso:2014zka}
R.~Alonso, H.~M.~Chang, E.~E.~Jenkins, A.~V.~Manohar and B.~Shotwell,
``Renormalization group evolution of dimension-six baryon number violating operators,''
Phys. Lett. B \textbf{734} (2014), 302-307
[arXiv:1405.0486 [hep-ph]].

\bibitem{Davidson:2018zuo}
S.~Davidson, M.~Gorbahn and M.~Leak,
``Majorana neutrino masses in the renormalization group equations for lepton flavor violation,''
Phys. Rev. D \textbf{98} (2018) no.9, 095014
[arXiv:1807.04283 [hep-ph]].

\bibitem{Wang:2023bdw}
Y.~Wang, D.~Zhang and S.~Zhou,
``Complete one-loop renormalization-group equations in the seesaw effective field theories,''
JHEP \textbf{05} (2023), 044
[arXiv:2302.08140 [hep-ph]].

\bibitem{Chala:2021juk}
M.~Chala and A.~Titov,
``Neutrino masses in the Standard Model effective field theory,''
Phys. Rev. D \textbf{104} (2021) no.3, 035002
[arXiv:2104.08248 [hep-ph]].

\bibitem{Chala:2021pll}
M.~Chala, G.~Guedes, M.~Ramos and J.~Santiago,
``Towards the renormalisation of the Standard Model effective field theory to dimension eight: Bosonic interactions I,''
SciPost Phys. \textbf{11} (2021), 065
[arXiv:2106.05291 [hep-ph]].

\bibitem{DasBakshi:2022mwk}
S.~Das Bakshi, M.~Chala, \'A.~D\'\i{}az-Carmona and G.~Guedes,
``Towards the renormalisation of the Standard Model effective field theory to dimension eight: bosonic interactions II,''
Eur. Phys. J. Plus \textbf{137} (2022) no.8, 973
[arXiv:2205.03301 [hep-ph]].

\bibitem{Helset:2022pde}
A.~Helset, E.~E.~Jenkins and A.~V.~Manohar,
``Renormalization of the Standard Model Effective Field Theory from geometry,''
JHEP \textbf{02} (2023), 063
[arXiv:2212.03253 [hep-ph]].

\bibitem{DasBakshi:2023htx}
S.~Das Bakshi and \'A.~D\'\i{}az-Carmona,
``Renormalisation of SMEFT bosonic interactions up to dimension eight by LNV operators,''
JHEP \textbf{06} (2023), 123
[arXiv:2301.07151 [hep-ph]].

\bibitem{Chala:2023jyx}
M.~Chala,
``Constraints on anomalous dimensions from the positivity of the S matrix,''
Phys. Rev. D \textbf{108} (2023) no.1, 015031
[arXiv:2301.09995 [hep-ph]].

\bibitem{Carmona:2021xtq}
A.~Carmona, A.~Lazopoulos, P.~Olgoso and J.~Santiago,
``Matchmakereft: automated tree-level and one-loop matching,''
SciPost Phys. \textbf{12} (2022) no.6, 198
[arXiv:2112.10787 [hep-ph]].

\bibitem{Barzinji:2018xvu}
A.~Barzinji, M.~Trott and A.~Vasudevan,
``Equations of Motion for the Standard Model Effective Field Theory: Theory and Applications,''
Phys. Rev. D \textbf{98} (2018) no.11, 116005
[arXiv:1806.06354 [hep-ph]].

\bibitem{Bilenky:1993bt}
M.~S.~Bilenky and A.~Santamaria,
``One loop effective Lagrangian for a standard model with a heavy charged scalar singlet,''
Nucl. Phys. B \textbf{420} (1994), 47-93
[arXiv:hep-ph/9310302 [hep-ph]].

\bibitem{Minkowski:1977sc}
P.~Minkowski,
``$\mu \to e\gamma$ at a Rate of One Out of $10^{9}$ Muon Decays?,''
Phys. Lett. B \textbf{67} (1977), 421-428.

\bibitem{Yanagida:1979as}
T.~Yanagida,
``Horizontal gauge symmetry and masses of neutrinos,''
Conf. Proc. C \textbf{7902131} (1979), 95-99
KEK-79-18-95.

\bibitem{Gell-Mann:1979vob}
M.~Gell-Mann, P.~Ramond and R.~Slansky,
``Complex Spinors and Unified Theories,''
Conf. Proc. C \textbf{790927} (1979), 315-321
[arXiv:1306.4669 [hep-th]].

\bibitem{Glashow:1979nm}
S.~L.~Glashow,
``The Future of Elementary Particle Physics,''
NATO Sci. Ser. B \textbf{61} (1980), 687.

\bibitem{Mohapatra:1979ia}
R.~N.~Mohapatra and G.~Senjanovic,
``Neutrino Mass and Spontaneous Parity Nonconservation,''
Phys. Rev. Lett. \textbf{44} (1980), 912.

\bibitem{Konetschny:1977bn}
W.~Konetschny and W.~Kummer,
``Nonconservation of Total Lepton Number with Scalar Bosons,''
Phys. Lett. B \textbf{70} (1977), 433-435.

\bibitem{Magg:1980ut}
M.~Magg and C.~Wetterich,
``Neutrino Mass Problem and Gauge Hierarchy,''
Phys. Lett. B \textbf{94} (1980), 61-64.

\bibitem{Cheng:1980qt}
T.~P.~Cheng and L.~F.~Li,
``Neutrino Masses, Mixings and Oscillations in SU(2) x U(1) Models of Electroweak Interactions,''
Phys. Rev. D \textbf{22} (1980), 2860.

\bibitem{Mohapatra:1980yp}
R.~N.~Mohapatra and G.~Senjanovic,
``Neutrino Masses and Mixings in Gauge Models with Spontaneous Parity Violation,''
Phys. Rev. D \textbf{23} (1981), 165.

\bibitem{Schechter:1980gr}
J.~Schechter and J.~W.~F.~Valle,
``Neutrino Masses in SU(2) x U(1) Theories,''
Phys. Rev. D \textbf{22} (1980), 2227.

\bibitem{Lazarides:1980nt}
G.~Lazarides, Q.~Shafi and C.~Wetterich,
``Proton Lifetime and Fermion Masses in an SO(10) Model,''
Nucl. Phys. B \textbf{181} (1981), 287-300.

\bibitem{Foot:1988aq}
R.~Foot, H.~Lew, X.~G.~He and G.~C.~Joshi,
``Seesaw Neutrino Masses Induced by a Triplet of Leptons,''
Z. Phys. C \textbf{44} (1989), 441.

\bibitem{Ma:1998dn}
E.~Ma,
``Pathways to naturally small neutrino masses,''
Phys. Rev. Lett. \textbf{81} (1998), 1171-1174
[arXiv:hep-ph/9805219 [hep-ph]].

\bibitem{Zhang:2021tsq}
D.~Zhang and S.~Zhou,
``Radiative decays of charged leptons in the seesaw effective field theory with one-loop matching,''
Phys. Lett. B \textbf{819} (2021), 136463
[arXiv:2102.04954 [hep-ph]].

\bibitem{Zhang:2021jdf}
D.~Zhang and S.~Zhou,
``Complete one-loop matching of the type-I seesaw model onto the Standard Model effective field theory,''
JHEP \textbf{09} (2021), 163
[arXiv:2107.12133 [hep-ph]].

\bibitem{Coy:2021hyr}
R.~Coy and M.~Frigerio,
``Effective comparison of neutrino-mass models,''
Phys. Rev. D \textbf{105} (2022) no.11, 115041
[arXiv:2110.09126 [hep-ph]].

\bibitem{Ohlsson:2022hfl}
T.~Ohlsson and M.~Pernow,
``One-loop matching conditions in neutrino effective theory,''
Nucl. Phys. B \textbf{978} (2022), 115729
[arXiv:2201.00840 [hep-ph]].

\bibitem{Li:2022ipc}
X.~Li, D.~Zhang and S.~Zhou,
``One-loop matching of the type-II seesaw model onto the Standard Model effective field theory,''
JHEP \textbf{04} (2022), 038
[arXiv:2201.05082 [hep-ph]].

\bibitem{Du:2022vso}
Y.~Du, X.~X.~Li and J.~H.~Yu,
``Neutrino seesaw models at one-loop matching: discrimination by effective operators,''
JHEP \textbf{09} (2022), 207
[arXiv:2201.04646 [hep-ph]].

\bibitem{Zhang:2022osj}
D.~Zhang,
``Complete one-loop structure of the type-(I+II) seesaw effective field theory,''
JHEP \textbf{03} (2023), 217
[arXiv:2208.07869 [hep-ph]].

\bibitem{Ma:2006km}
E.~Ma,
``Verifiable radiative seesaw mechanism of neutrino mass and dark matter,''
Phys. Rev. D \textbf{73} (2006), 077301
[arXiv:hep-ph/0601225 [hep-ph]].

\bibitem{Liao:2022cwh}
Y.~Liao and X.~D.~Ma,
``One-loop matching of scotogenic model onto standard model effective field theory up to dimension 7,''
JHEP \textbf{12} (2022), 053
[arXiv:2210.04270 [hep-ph]].

\bibitem{Christensen:2008py}
N.~D.~Christensen and C.~Duhr,
``FeynRules - Feynman rules made easy,''
Comput. Phys. Commun. \textbf{180} (2009), 1614-1641
[arXiv:0806.4194 [hep-ph]].

\bibitem{Alloul:2013bka}
A.~Alloul, N.~D.~Christensen, C.~Degrande, C.~Duhr and B.~Fuks,
``FeynRules  2.0 - A complete toolbox for tree-level phenomenology,''
Comput. Phys. Commun. \textbf{185} (2014), 2250-2300
[arXiv:1310.1921 [hep-ph]].

\bibitem{Hahn:2000kx}
T.~Hahn,
``Generating Feynman diagrams and amplitudes with FeynArts 3,''
Comput. Phys. Commun. \textbf{140} (2001), 418-431
[arXiv:hep-ph/0012260 [hep-ph]].

\bibitem{Shtabovenko:2016sxi}
V.~Shtabovenko, R.~Mertig and F.~Orellana,
``New Developments in FeynCalc 9.0,''
Comput. Phys. Commun. \textbf{207} (2016), 432-444
[arXiv:1601.01167 [hep-ph]].

\bibitem{Shtabovenko:2020gxv}
V.~Shtabovenko, R.~Mertig and F.~Orellana,
``FeynCalc 9.3: New features and improvements,''
Comput. Phys. Commun. \textbf{256} (2020), 107478
[arXiv:2001.04407 [hep-ph]].

\bibitem{Shtabovenko:2016whf}
V.~Shtabovenko,
``FeynHelpers: Connecting FeynCalc to FIRE and Package-X,''
Comput. Phys. Commun. \textbf{218} (2017), 48-65
[arXiv:1611.06793 [physics.comp-ph]].

\bibitem{Patel:2015tea}
H.~H.~Patel,
``Package-X: A Mathematica package for the analytic calculation of one-loop integrals,''
Comput. Phys. Commun. \textbf{197} (2015), 276-290
[arXiv:1503.01469 [hep-ph]].

\bibitem{Patel:2016fam}
H.~H.~Patel,
``Package-X 2.0: A Mathematica package for the analytic calculation of one-loop integrals,''
Comput. Phys. Commun. \textbf{218} (2017), 66-70
[arXiv:1612.00009 [hep-ph]].

\bibitem{Denner:1992me}
A.~Denner, H.~Eck, O.~Hahn and J.~Kublbeck,
``Compact Feynman rules for Majorana fermions,''
Phys. Lett. B \textbf{291} (1992), 278-280.

\bibitem{Denner:1992vza}
A.~Denner, H.~Eck, O.~Hahn and J.~Kublbeck,
``Feynman rules for fermion number violating interactions,''
Nucl. Phys. B \textbf{387} (1992), 467-481.

\bibitem{Elias-Miro:2014eia}
J.~Elias-Miro, J.~R.~Espinosa and A.~Pomarol,
``One-loop non-renormalization results in EFTs,''
Phys. Lett. B \textbf{747} (2015), 272-280
[arXiv:1412.7151 [hep-ph]].

\bibitem{Cheung:2015aba}
C.~Cheung and C.~H.~Shen,
``Nonrenormalization Theorems without Supersymmetry,''
Phys. Rev. Lett. \textbf{115} (2015) no.7, 071601
[arXiv:1505.01844 [hep-ph]].

\bibitem{Bern:2019wie}
Z.~Bern, J.~Parra-Martinez and E.~Sawyer,
``Nonrenormalization and Operator Mixing via On-Shell Methods,''
Phys. Rev. Lett. \textbf{124} (2020) no.5, 051601
[arXiv:1910.05831 [hep-ph]].

\bibitem{Cao:2023adc}
W.~Cao, F.~Herzog, T.~Melia and J.~Roosmale Nepveu,
``Non-linear non-renormalization theorems,''
JHEP \textbf{08} (2023), 080
[arXiv:2303.07391 [hep-ph]].

\bibitem{Babu:2001ex}
K.~S.~Babu and C.~N.~Leung,
``Classification of effective neutrino mass operators,''
Nucl. Phys. B \textbf{619} (2001), 667-689
[arXiv:hep-ph/0106054 [hep-ph]].

\bibitem{deGouvea:2007qla}
A.~de Gouvea and J.~Jenkins,
``A Survey of Lepton Number Violation Via Effective Operators,''
Phys. Rev. D \textbf{77} (2008), 013008
[arXiv:0708.1344 [hep-ph]].

\bibitem{Cai:2017jrq}
Y.~Cai, J.~Herrero-Garc\'\i{}a, M.~A.~Schmidt, A.~Vicente and R.~R.~Volkas,
``From the trees to the forest: a review of radiative neutrino mass models,''
Front. in Phys. \textbf{5} (2017), 63
[arXiv:1706.08524 [hep-ph]].

\bibitem{Cirigliano:2017djv}
V.~Cirigliano, W.~Dekens, J.~de Vries, M.~L.~Graesser and E.~Mereghetti,
``Neutrinoless double beta decay in chiral effective field theory: lepton number violation at dimension seven,''
JHEP \textbf{12} (2017), 082
[arXiv:1708.09390 [hep-ph]].

\bibitem{Cirigliano:2018yza}
V.~Cirigliano, W.~Dekens, J.~de Vries, M.~L.~Graesser and E.~Mereghetti,
``A neutrinoless double beta decay master formula from effective field theory,''
JHEP \textbf{12} (2018), 097
[arXiv:1806.02780 [hep-ph]].

\bibitem{Liao:2019gex}
Y.~Liao, X.~D.~Ma and H.~L.~Wang,
``Effective field theory approach to lepton number violating decays $K^\pm\rightarrow \pi^\mp l^{\pm}l^{\pm}$: short-distance contribution,''
JHEP \textbf{01} (2020), 127
[arXiv:1909.06272 [hep-ph]].

\bibitem{Liao:2020roy}
Y.~Liao, X.~D.~Ma and H.~L.~Wang,
``Effective field theory approach to lepton number violating decays $K^\pm\rightarrow \pi^\mp l^{\pm}_\alpha l^{\pm}_\beta$: long-distance contribution,''
JHEP \textbf{03} (2020), 120
[arXiv:2001.07378 [hep-ph]].

\bibitem{Fuentes-Martin:2022vvu}
J.~Fuentes-Mart\'\i{}n, M.~K\"onig, J.~Pag\`es, A.~E.~Thomsen and F.~Wilsch,
``Evanescent operators in one-loop matching computations,''
JHEP \textbf{02} (2023), 031
[arXiv:2211.09144 [hep-ph]].

\bibitem{Li:2023cwy}
X.~X.~Li, Z.~Ren and J.~H.~Yu,
``A complete tree-level dictionary between simplified BSM models and SMEFT (d $\leq$ 7) operators,''
[arXiv:2307.10380 [hep-ph]].



\end{thebibliography}
\end{document}